\documentclass[structabstract]{aa}  
                                   
\usepackage{txfonts}
\usepackage[hiresbb]{graphicx}
\usepackage[utf8]{inputenc}
\usepackage{cancel}
\usepackage[authoryear]{natbib}
\usepackage{color}
\usepackage{multirow,bigdelim}
\usepackage{mathrsfs}
\usepackage{rotating}
\usepackage{longtable,ltcaption}
\usepackage{pdflscape}
\usepackage{placeins}
\usepackage{txfonts}
\newcommand{\hel}[2] {He\,{\sc #1}~$\rm{\lambda}$#2}

\newcommand{\Nl}[2] {N\,{\sc #1}~$\lambda$#2}
\newcommand{\Nlmod}[2] {N\,{\sc #1}~#2}

\newcommand{\hhel}[1] {He\,{\sc #1}}
\newcommand{\SSil}[1] {Si\,{\sc #1}}

\newcommand{\msun}{\ensuremath{\mathit{M}_{\odot}}}
\newcommand{\mdot}{\ensuremath{\mathit{\dot{M}}}}
\newcommand{\msunyr}{\ensuremath{\mathit{M}_{\odot} {\rm yr}^{-1}}}
\newcommand{\lsun}{\ensuremath{\mathit{L}_{\odot}}}
\newcommand{\rsun}{\ensuremath{\mathit{R}_{\odot}}}
\newcommand{\vrot}{\ensuremath{\varv_{\rm e}\sin i}}
\newcommand{\veq}{\ensuremath{\varv_{\rm e}}}

\def\kms{\mbox{${\rm km}\:{\rm s}^{-1}$}}

\def\lesssim{\mathrel{\hbox{\rlap{\hbox{\lower4pt\hbox{$\sim$}}}\hbox{$<$}}}}
\let\la=\lesssim
\def\gtrsim{\mathrel{\hbox{\rlap{\hbox{\lower4pt\hbox{$\sim$}}}\hbox{$>$}}}}
\let\ga=\gtrsim

\defcitealias{evans}{Paper~I}
\defcitealias{dufton}{Paper~X}
\defcitealias{sana}{Paper~VIII}
\defcitealias{doran}{Paper~XI}
\defcitealias{ramirezagudelo}{Paper~XII}
\defcitealias{ramirezagudelo2015}{Paper~XXI}
\defcitealias{sabin_sanjulian2014}{Paper~XIII}
\defcitealias{walborn2014}{Paper~XIV}
\defcitealias{bestenlehner2014}{Paper~XVII}
\defcitealias{dunstall2015}{Paper~XXII}

\titlerunning{Stellar properties of the O-type giants and supergiants in 30 Doradus}
\begin{document}
  \title{The VLT-FLAMES Tarantula Survey\thanks{Based on observations
   collected at the European Southern Observatory under program ID 182.D-0222.} 
}

   \subtitle{ XXIV. Stellar properties of the O-type giants and supergiants in 30 Doradus}
   \author{
     O.H. Ram\'{i}rez-Agudelo   \inst{1,2,3}
    \and
     H. Sana      \inst{4}
     \and
     A. de Koter  \inst{1,4}
    \and
     F. Tramper  \inst{5}
    \and
    N.J. Grin    \inst{1,2}
     \and
     F.R.N. Schneider \inst{6}
    \and
     N. Langer    \inst{2}
    \and
    J. Puls \inst{7}    
     \and
     N. Markova  \inst{8}      
    \and
    J.M. Bestenlehner \inst{9}
    \and 
    N. Castro \inst{10}    
    \and 
    P.A. Crowther \inst{11}
    \and
     C.J. Evans  \inst{3}
    \and
     M. Garc\'ia  \inst{12}   
     \and	
      G.~Gr\"afener \inst{2}
     \and
     A. Herrero   \inst{13,14}
    \and	
     B. van Kempen \inst{1}
	\and	
	D.J. Lennon    \inst{5}
    \and
    J. Ma\'{i}z Apell\'aniz \inst{15}
     \and
     F. Najarro \inst{12} 
    \and
     C. Sab\'{i}n-Sanjuli\'{a}n \inst{16}
     \and
     S. Sim\'on-D\'{i}az   \inst{13,14}
    \and
     W.D. Taylor  \inst{3}  
    \and     
    J.S. Vink  \inst{17}      
}
\institute{ 
          Astronomical Institute Anton Pannekoek, 
          Amsterdam University,  
          Science Park 904, 1098~XH, 
          Amsterdam, The Netherlands\newline
          \email{oscar.ramirez@stfc.ac.uk}
\and 
           Argelander-Institut f\"ur Astronomie, 
           Universit\"at Bonn, 
           Auf dem H\"ugel 71, 
           53121 Bonn, Germany 
\and 
           UK Astronomy Technology Centre,
           Royal Observatory Edinburgh,
           Blackford Hill, Edinburgh, EH9 3HJ, United Kingdom
\and 
           Institute of Astrophysics,
           KU Leuven, 
           Celestijnenlaan 200D, 
           3001, Leuven, Belgium
\and 
           European Space Astronomy Centre (ESAC),
           Camino bajo del Castillo, s/n
           Urbanizacion Villafranca del Castillo,
           Villanueva de la Ca\~nada,
           E-28\,692 Madrid, Spain  
\and 
	       Department of Physics,
           University of Oxford, 
           Keble Road,
           Oxford OX1 3RH,
           United Kingdom 
\and 
             LMU Munich, Universit\"atssternwarte, 
	        Scheinerstrasse 1,
	        81679 M\"unchen, 
	        Germany           
\and  
            Institute of Astronomy with NAO,
            Bulgarian Academy of Sciences,
            PO Box 136, 4700 Smoljan, Bulgaria
\and  
            Max-Planck-Institut f\"ur Astronomie, 
            K\"onigstuhl 17, 
            69117 Heidelberg, 
            Germany	       
\and   
            Department of Astronomy, 
            University of Michigan, 
            1085 S. University Avenue, Ann Arbor, 
            MI 48109-1107, USA 
\and 
           Departament of Physic and Astronomy
           University of Sheffield,
           Sheffield,S3 7RH, 
           United Kingdom
\and 
        	Centro de Astrobiolog\'{i}a (CSIC-INTA),
	        Ctra. de Torrej\'on a Ajalvir km-4,
	        E-28850 Torrej\'on de Ardoz,
	        Madrid, Spain
\and 
           Departamento de Astrof\'{i}sica, 
           Universidad de La Laguna, 
           Avda. Astrof\'{i}sico Francisco S\'{a}nchez s/n, 
           E-38071 La Laguna, Tenerife, Spain
\and 
        Instituto de Astrof\'{i}sica de Canarias, 
        C/ V\'{i}a L\'{a}ctea s/n, E-38200 La Laguna, Tenerife,
         Spain
\and 
        Centro de Astrobiolog\'{i}a (CSIC-INTA), ESAC campus, 
        Camino bajo del castillo s/n, 
        Villanueva de la Ca{\~n}ada, 
        E-28\,692 Madrid, Spain.           
\and 
        Instituto de Investigaci\'on Multidisciplinar en Ciencia y Tecnolog\'ia, Universidad de La Serena, Ra\'ul Bitr\'an 1305, La Serena, Chile	    
\and 
        Armagh Observatory,
        College Hill,
        Armagh, BT61 9DG,
        Northern Ireland,
        United Kingdom 
}
             
   \date{Accepted ....}

 
 \abstract
   {The Tarantula region in the Large Magellanic Cloud contains the richest population of 
    spatially resolved massive O-type stars known so far. This unmatched sample 
    offers an opportunity to test models describing their main-sequence evolution
    and mass-loss properties.} 
   {Using ground-based optical spectroscopy obtained in the framework 
   of the VLT-FLAMES Tarantula Survey (VFTS), we aim to determine stellar, photospheric and wind 
   properties of 72 presumably single O-type giants, bright giants and supergiants and to
   confront them with predictions of stellar evolution and of line-driven mass-loss theories.}
  {We apply an automated method for quantitative spectroscopic analysis of O stars combining
   the non-LTE stellar atmosphere model {{\sc fastwind}} with the genetic fitting algorithm 
   {{\sc pikaia}} to determine the following stellar properties: effective temperature,
   surface gravity, mass-loss rate, helium abundance, and projected rotational velocity. The 
   latter has been constrained without taking into account the contribution from  macro-turbulent motions to the line broadening.}
   {We present empirical effective temperature versus spectral subtype calibrations at
   LMC-metallicity for giants and supergiants. The calibration for giants shows a +1kK offset compared to similar Galactic calibrations; a shift of the same magnitude has been reported for dwarfs. The supergiant calibrations, though only based on a handful of stars, do not seem to indicate such an offset. The presence of a  strong upturn at spectral type O3 and earlier can also not be confirmed by our data. 
   In the spectroscopic and classical
   Hertzsprung-Russell diagrams, our sample O stars are found to occupy the region  predicted to be the core hydrogen-burning phase 
   by \citeauthor{brott} (\citeyear{brott}) and \citeauthor{kohler2015} (\citeyear{kohler2015}). 
   For stars initially more massive than approximately 60\,\msun, the giant
   phase already appears relatively early on in the evolution; the supergiant phase develops
   later. Bright giants, however, are not systematically positioned between giants and supergiants 
   at $M_{\rm init} \ga 25$\, \msun. 
   At masses below 60\,\msun, the dwarf phase clearly precedes the giant and supergiant
   phases; however this behavior seems to break down at $M_{\rm init} \la 18$\,\msun. 
   Here, stars classified as late O\,III and II stars occupy the region where O9.5-9.7~V stars are expected, but 
   where few such late O\,V stars are actually seen. Though we can not exclude that these stars 
   represent a physically distinct group, this behaviour may reflect an
   intricacy in the luminosity classification at late O spectral subtype. Indeed, on the
   basis of a 
   secondary classification criterion, the relative strength of
   Si\,{\sc iv} to He\,{\sc i} absorption lines, these stars would have been
   assigned a luminosity class IV or V. %
    Except for five stars, the helium abundance of our sample stars is in agreement with the initial LMC composition. 
   This outcome is independent of their projected spin rates. The aforementioned five stars present moderate projected rotational velocities (i.e., $\vrot\,<\,$200\,\kms)
   and hence do not agree with current predictions of rotational mixing in main-sequence stars. 
    They may potentially reveal other physics not included in the models such as binary-interaction effects. %
   Adopting theoretical
   results for the wind velocity law, we find modified wind momenta
   for LMC stars that are $\sim$0.3 dex higher than earlier results. 
    For stars brighter than $10^{5}$\,\lsun, that is, in the regime of strong stellar winds, the measured (unclumped) mass-loss rates
    could be considered to be in agreement with line-driven wind predictions of \citeauthor{vink2001} (\citeyear{vink2001}) if the clump volume filling factors were  $f_{\rm V} \sim 1/8$ to $1/6$.
   }
   {}

   \keywords{
   			stars: early-type --
             stars:  evolution --
             stars: fundamental parameters -- 
             Magellanic Clouds --
             Galaxies: star clusters: individual: 30 Doradus
               }

   \maketitle
%

\section{Introduction}\label{sec:intro}

Bright, massive stars play an 
important role in the evolution of galaxies and 
of the universe as a whole. 
Nucleosynthesis in their interiors produces the bulk of the chemical elements \citep[e.g.,][]{prantzos2000,matteucci2008}, 
which are released into the interstellar medium through powerful stellar winds
\citep[e.g.,][]{puls2008} and supernova explosions.
The associated kinetic energy that is deposited in the ISM affects
the star-forming regions where massive stars reside 
\citep[e.g.,][]{beuther2008}. The radiation fields they emit add to this energy and supply copious amounts of hydrogen-ionizing photons 
and H$_{2}$ photo-dissociating photons. Massive stars that resulted from primordial star formation
\citep[e.g.,][]{hirano2014,hirano2015} are 
potential contributors to the re-ionization of the universe and 
have likely played a role in galaxy formation \citep[e.g.,][]{bromm2009}. 
Massive stars produce a variety of supernovae, including type Ib, Ic, Ic-BL, type IIP, IIL, IIb, IIn, and peculiar supernovae,
and gamma-ray bursts \citep[e.g.,][]{langer2012}, that can be seen up to high redshifts \citep[][]{zhang2009}.

Models of massive-star evolution predict the series of morphological states that these objects 
undergo before reaching their final fate \citep[e.g.,][]{brott,ekstrom2012,groh2014,kohler2015}.
Studying populations of massive stars spanning a range of metallicities is a proven way of 
testing and calibrating the assumptions of such calculations, and 
lends support to such
predictions at very low and zero metallicity. O-type
stars are of particular interest as they sample the main-sequence phase in the mass range
of 15 \msun\ to $\sim$70 \msun. 
They show a rich variety of spectral subtypes 
whether dwarfs, giants, or supergiants \cite[e.g.,][]{sota2011}, emphasizing the need for 
large samples to confront theoretical predictions.

To constrain the properties of massive stars, high-quality spectra and sophisticated modeling tools are required. In 
recent years, several tens of objects have been studied in the Large Magellanic Cloud (LMC) 
providing and initial basis to confront theory with observations. 
\citet{puls1996} included six LMC objects in
their larger sample of Galactic and Magellanic Cloud sources, pioneering the first large-scale quantitative spectroscopic
study of O stars.
\citet{crowther2002} presented an analysis of three LMC Oaf+ supergiants and one such object in the Small Magellanic Cloud (SMC) using 
far-ultraviolet FUSE, ultraviolet IUE/HST, and optical VLT ultraviolet-visual Echelle 
Spectrograph data. \citet{massey2004,massey2005} 
derived the properties 
of a total of 40 Magellanic Clouds stars, 24 of which are in the 
LMC (including 10 in R136) using data collected with Hubble Space Telescope (HST) and the 4m-CTIO telescope.  
\citet{mokiem2007} studied 23 LMC O stars 
using the VLT-FLAMES instrument, of which 17 are in the star-forming regions N11.
Expanding on their earlier work, \citet{massey2009} scrutinized another 23 Magellanic O-type stars, 11 of which being 
in the LMC, for which ultraviolet STIS spectra were available in the HST Archive and optical spectra were secured with 
the Boller \& Chivens Spectrograph at the Clay 6.5m (Magellan II) telescope at Las Campanas. Four of the LMC sources studied 
by these authors were included in a reanalysis, where results obtained with {\sc fastwind} \citep{puls2005,rivero_gonzalez2011} 
and {\sc cmfgen} \citep{hillier_miller_1998} were compared \citep{massey2013}.
Though this constitutes a promising start indeed, the morphological properties among O stars are so complex that still 
larger samples are required for robust tests of stellar evolution.

The Tarantula nebula in the LMC is particularly rich in O-type stars, 
containing hundreds of these objects. It has a well-constrained distance modulus of 
18.5\,mag \citep[][]{pietrzyski2013} and
only a modest foreground extinction, making it an ideal laboratory to study entire populations of massive stars. 
This motivated the VLT-FLAMES Tarantula Survey (VFTS), a multi-epoch
spectroscopic campaign that targeted 360 O-type and approximately 540 later-type stars across the
Tarantula region, spanning a field several hundred light years across 
\citep[][hereafter \citetalias{evans}]{evans}. 

In the present paper within the VFTS series, we analyze the properties of the 72 presumed single O-type giants,
bright giants, and supergiants in the VFTS sample. 
In all likelihood, not all of them  are truly single stars. Establishing the multiplicity properties of the targeted 
stars was an important component of the VFTS project 
\citep[][]{sana,dunstall2015} 
and the observing strategy was tuned 
to enable the detection of close pairs with periods up to $\sim$1000 days, that is, those that are expected to interact
during their evolution \citep[e.g.,][]{PJH1992}.  The finite number of epochs resulted in an average detection probability of approximately 70\%, 
implying that some of our targets may be binaries. 
Additionally, post-interaction systems may be disguised
as single stars by showing no or negligible radial velocity variations \citep{selma2014}.
By confronting the stellar characteristics with  evolutionary models 
for single stars we may not only test these models, 
but also identify possible post-interaction systems if their properties appear 
peculiar and contradict basic predictions from single-star models.

The outflows of O\,III to O\,I stars are dense and most of them feature signatures of mass-loss in
H$\alpha$ and He\,{\sc ii}\,$\lambda 4686$, allowing us to assess their wind strength. 
The stellar and wind properties of the dwarfs  are presented in \citet[][hereafter \citetalias{sabin_sanjulian2014}]{sabin_sanjulian2014} 
and Sab\'in-Sanjuli\'an et al. (subm.). Those of the most massive stars in the VFTS sample (the Of and WNh stars)
have been presented in \citet[][hereafter \citetalias{bestenlehner2014}]{bestenlehner2014}. 
Combining these results with those from this paper enables 
a confrontation with wind-strength predictions using a sample that is unprecedented in size.

The layout of the paper is as follows. Section \ref{sec:sample} describes the selection of 
our sample. The spectral analysis method is described in Section \ref{sec:method}. 
The results are presented and discussed in Section \ref{sec:results}. Finally, a summary is 
given in Section \ref{sec:conclusions}. 
 
\section{Sample and data preparation}
\label{sec:sample}

\begin{figure}
\centering
\includegraphics[width=\columnwidth]{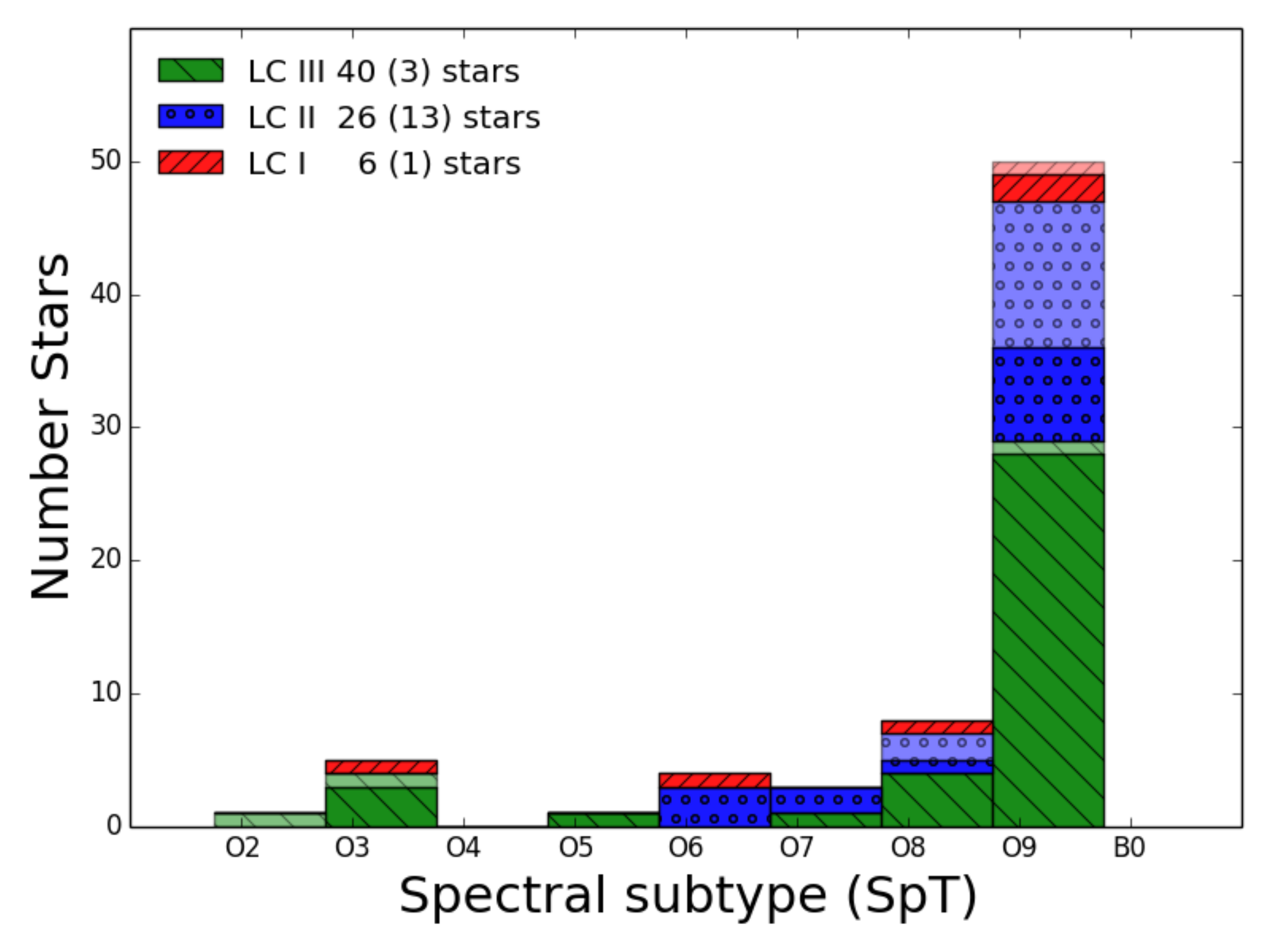}
\caption{Spectral-type distribution of the O-type stars in our sample, 
binned per spectral subtype (SpT). Different colors and shadings 
indicate different luminosity classes (LC); see legend.
The legend also gives the total number of stars in each LC class (e.g., 40 LC\,III).
In parentheses we provide the number of stars that were 
given an ambiguous LC classification within each category in 
\citet[][]{walborn2014} (e.g., 3 in LC\,III). They are plotted in their corresponding category 
with lower opacity (see main text for details).}
\label{fig:SpTLC}
\end{figure}

The VFTS project and the data have been described in \citetalias{evans}. 
Here we focus on a subset of the  O-type star sample that has been observed with 
the Medusa fibers of the VLT-FLAMES multi-object spectrograph:
the presumably single O stars with luminosity class (LC) III to I. 
The total Medusa sample contains 332 O-type objects observed with the Medusa fiber-fed Giraffe
spectrograph. Their spectral classification is available in \citet[][]{walborn2014}. 
Among that sample, \citet[][]{sana} have identified 116 spectroscopic 
binary (SB) systems from
significant radial velocity (RV) variations with a peak-to-peak amplitude ($\rm{\Delta RV}$) larger 
than 20~\kms. The remaining 216 objects either show no significant 
or significant but small RV variations  ($\rm{\Delta RV}\, \leq$ 20~\kms). 
They are presumed single stars although 
it is expected that up to 25\% of them are undetected binaries \citep[see][]{sana}. 
The rotational properties of the  O-type single and binary stars in the VFTS  
have been presented by \citet[][hereafter \citetalias{ramirezagudelo}]{ramirezagudelo} and \citet[][]{ramirezagudelo2015}.

Here we focus on the 72 presumably single O stars  with LC\,III to I. 
The remaining 31 spectroscopically single objects could not be assigned a LC 
classification \citep[see][]{walborn2014} and for that reason are discarded from the present analysis.
For completeness, we do however provide their parameters (see Sect.~\ref{sec:wind_ac}).

Our sample contains 
37 LC\,III, 13 LC\,II, and 5 LC\,I objects. In addition to these 55 stars,
there are 17 objects with a somewhat ambiguous LC, namely:
1 LC\,III-IV, 2 LC\,III-I, 10 LC\,II-III, 3 LC\,II-B0\,IV, and 1 LC\,I-II.
We adopted the first listed LC classification
bringing the total sample to 
40 giants, 26 bright giants, and 6 supergiants. 
Figure~\ref{fig:SpTLC} displays the distribution of spectral subtypes and shows that
69\%\ of the stars in the sample are O9-O9.7 stars. 
The lack of O\,4-5 III to I stars is in agreement with statistical fluctuations due to the sample size. The number
 of such objects in the full VFTS sample is comparable to that of 
 O3 stars; they are however almost all of LC\,V or IV  
(see Fig.~1 in \citetalias{ramirezagudelo} and Table~1 in \citealp{walborn2014}). 
There are only a few Of stars and no Wolf Rayet (WR) stars
in our sample. These extreme and very massive stars in the VFTS have been studied in
\citetalias{bestenlehner2014} (see Sect. \ref{sec:method}).

\begin{figure}
\centering
\includegraphics[width=\columnwidth]{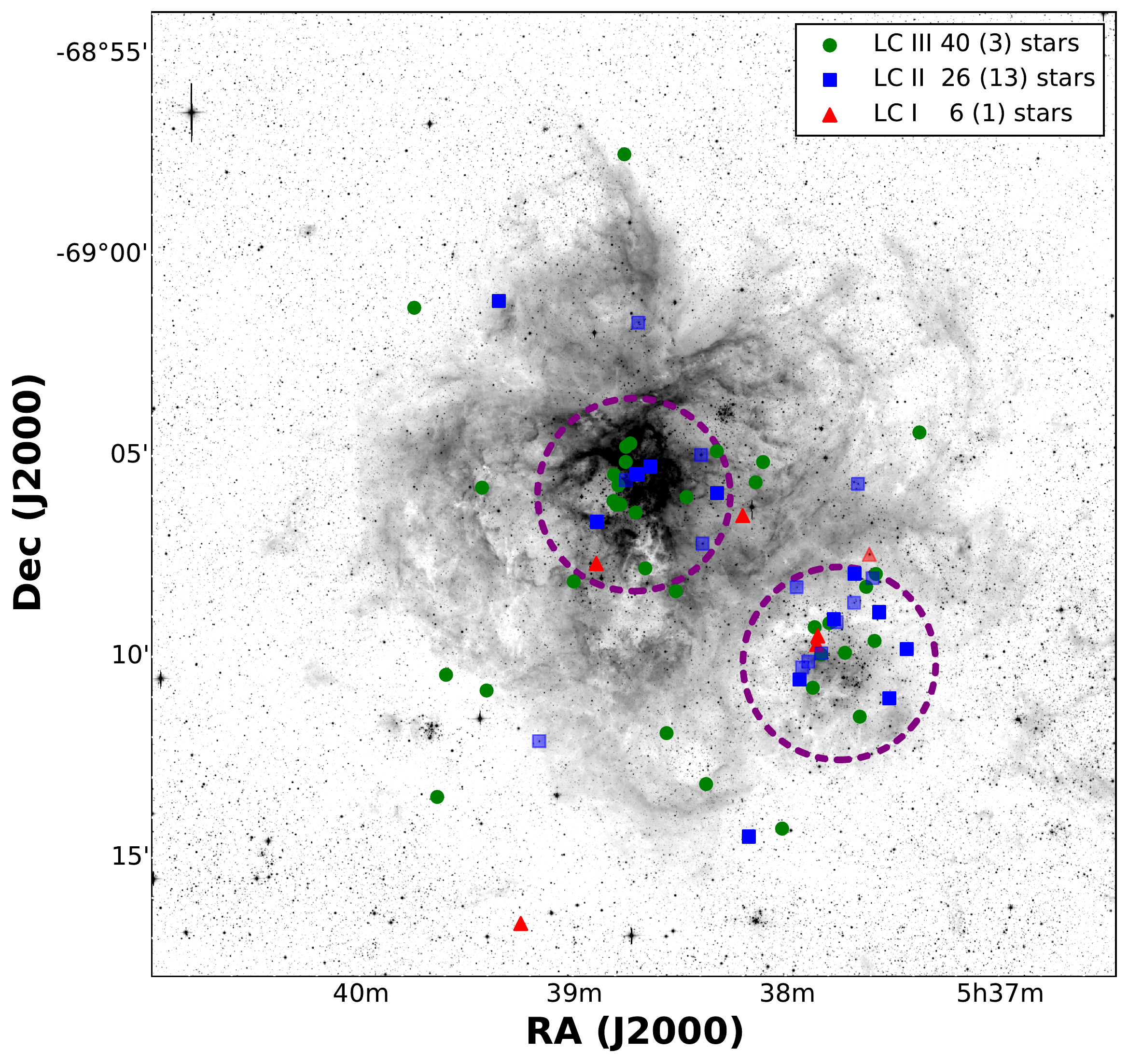}
\caption{Spatial distribution of the presumably single O-type stars as a function 
of LC. North is to the top; east is to the left. The circles define regions within 
2.4\arcmin\ of NGC\,2070 (central circle) and NGC\,2060 (south-west circle).
Different symbols indicate LC: III (circles), II (squares), I (triangles). 
Note that because of crowding a considerable fraction
of the sources overlap. Lower opacities again denote sources with
an ambiguous LC classification, similar as in Fig.~\ref{fig:SpTLC}.
}
\label{fig:SpTLC_spatial}
\end{figure}

The spatial distribution of our sample is shown in Fig.~\ref{fig:SpTLC_spatial}. 
The stars are concentrated in two 
associations, NGC\,2070 (in the centre of the image), and NGC\,2060 
(6.7\arcmin\ south-west of NGC\,2070), although a sizeable fraction are distributed throughout 
the field of view. For consistency with other VFTS papers,  
we refer to stars located further away than 2.4\arcmin\ from the 
centers of NGC\,2070 and NGC\,2060 as the stars \textit{outside of star-forming complexes}. 
These may originate from either NGC\,2070 or 2060 but may also have formed in other 
star-forming events in the 30\,Dor region at large. A circle of radius 2.4\arcmin\ (or 37~pc) around
NGC\,2070 contains 22 stars from our sample: 13 are of LC III, 
8 are of LC II and 1 is of 
LC I. NGC\,2060 contains 24 stars in a similar sized 
region and is believed to be somewhat older \citep{walborn}. Accordingly, it contains a larger fraction of LC~II and I stars (63\%; 15 out of 24) 
than NGC\,2070 (40\%). 

\subsection{Data preparation}
\label{subsec:data_prep}

The VFTS data are multi-epoch and multi-setting by nature. To increase the signal-to-noise of individual epochs and to simplify the atmosphere analysis process, we have combined, for each star,  the spectra from the various epochs and setups into a single normalized spectrum per object. 
We provide here a brief overview of the steps taken to reach that goal. We assumed that all stars are single by nature, that is, that no significant RV shift between the various observation epochs needs to be accounted for. This assumption is validated for our sample (see above), which selects either stars with no statistically significant RV variation, or significant but small RV shifts ($\Delta RV < 20$~\kms; hence less than half the resolution element of $\sim$~40\,\kms). 

For each object and setup, we started from the 
individual-epoch spectra normalized by \citet{sana} and first rejected 
the spectra of insufficient quality ($S/N < 5$). Subsequent steps are:
\begin{enumerate}
\item[{\em i.}] Rebinning to a common wavelength grid, using the largest common wavelength range. Step sizes of 0.2\,\AA\ and 0.05\,\AA\ are adopted for the LR and HR Medusa$-$Giraffe settings, respectively.
\item[{\em ii.}] Discarding spectra with a $S/N$ lower by a factor of three, or more,
compared to the median $S/N$ of the set of spectra for the considered object and setup;
\item[{\em iii.}] Computing the median spectrum;
\item[{\em iv.}] At each pixel, applying a 5$\sigma$-clipping around the median spectrum, using the individual error of each pixel;
\item[{\em v.}] Computing the weighted average spectrum, taking into account the individual pixel uncertainties and excluding the clipped pixels;
\item[{\em vi.}] Re-normalizing the resulting spectrum to correct for minor deviations that have become apparent thanks to the improved $S/N$ of the combined spectrum. 
The typical normalization error is better than 1\%\ \citep[see Appendix A in][]{sana}. The obtained spectrum is considered as the final spectrum for a given star and a given observing setup. 
\item[{\em vii.}]The error spectrum is computed through error-propagation throughout the described process.
\end{enumerate}

Once we have combined the individual epochs, we still have to merge the three observing setups. In particular, the averaged LR02 and LR03 spectra of each object are merged using a linear ramp between 4500 and 4525\AA. This implies that below 4500\AA\ the merged spectrum is from LR02 exclusively and that above 4525\AA\ it is from LR03. In particular, the information on the \ion{He}{ii}\,$\lambda$4541 line present in LR02 is discarded despite the fact that there are twice as many epochs of LR02 than of LR03. This is mainly due to  ({\em i}) a sometimes uncertain normalization of the \ion{He}{ii}\,$\lambda$4541 region in LR02, as the line lies very close to the edge of the LR02 wavelength range, and ({\em ii}) the fact that LR02 and LR03 setups yield  different spectral resolving power. Hence, we decided against the combination of data that differ in resolving power in such an important line for atmosphere fitting. While we might thus lose in $S/N$, we gain in robustness. In the 4500 -- 4525\AA\ transition region, we note that we did not correct for the difference in resolving power between LR02 and LR03. For most objects, no lines are visible there. Finally, HR15N was simply added as there is no overlap.

\section{Analysis method}
\label{sec:method}

To investigate the atmospheric properties of our sample stars, 
we obtained the stellar and wind parameters by fitting
synthetic spectra to the observed line profiles. 
This method is described in the following section.

\subsection{Atmosphere fitting}
\label{subsec:GA}

The stellar properties 
of the stars have been determined using an automated  method first developed by \citet{mokiem2005}. 
This method combines the non-LTE stellar atmosphere code 
{\sc fastwind} \citep{puls2005,rivero_gonzalez2011}
with the genetic fitting algorithm {\sc pikaia} \citep{charbonneau1995}. 
It allows for a standardized analysis of
the spectra of O-type stars by
a thorough exploration of the parameter space in affordable CPU time on a supercomputer 
\citep[see,][]{mokiem2006,mokiem2007,mokiem2007b,tramper2011,tramper2014}.

In the present study, we used {\sc fastwind} (version 10.1) with detailed
model atoms for hydrogen and helium \citep[described in][]{puls2005}, and in some
cases (see below) also for nitrogen \citep[][]{rivero_gonzalez2012b} and silicon
\citep[][]{trundle_et_al_2004} as `explicit' elements. 
Most of the other elements up to zinc were treated as
background elements. In brief, explicit elements are those used as
diagnostic tools and treated with high precision by detailed atomic
models and by means of a co-moving frame transport for all line
transitions. The background elements (i.e., the rest) are only needed
for the line-blocking/blanketing calculations, and are treated in a
more approximate way, though still solving the complete equations of
statistical equilibrium for most of them. In particular,
parametrized ionization cross sections following \citet[][]{seaton_et_al_1958} 
are used, and a co-moving frame transfer is applied only for the strongest lines,
whilst the weaker ones are calculated by means of the Sobolev approximation. 
For the abundances of these background 
elements we adopt solar values by \citet{asplund2005} scaled down by 0.3 dex to mimic the metal 
deficiency of the LMC \citep[e.g.,][]{rolleston2002}. The abundance of carbon is 
further adjusted by -1.1 dex (i.e., [C] = 7.0, where [X] is log (X/H)+12) 
and nitrogen by +0.35 dex (i.e., [N] = 7.7), 
characteristic for the surfacing of CN- and CNO-cycle products \citep{brott}.

The {\sc pikaia} algorithm was used to evolve a population of 79 
randomly drawn initial solutions (i.e., a population consisting of 79 individuals) 
over 300 generations. The population of each subsequent generation was based
on selection pressure (i.e., highest fitness) and random mutation of parameters.
Convergence was generally achieved after 30-50 generations, depending on the gravity, with 
lower-gravity objects requiring more generations to reach convergence. Computing a large number 
of generations beyond the convergence point allows us to fully explore the shape of the $\chi^{2}$ 
minimum while further ensuring that the absolute optimum has been identified.  
The population survival was based on the {\it fitness} ($F$) of the models, computed as:
\begin{equation}
F=\left(\frac{\sum_l w_{l} \cdot \chi^2_{l,{\rm red}}} {N_l} \right)^{-1},
\end{equation}
where $\chi^2_\mathrm{l,red}$ is the reduced $\chi^2$ between the data and the model for line $l$, 
$w_{l}$ is a weighting factor, and where the summation is carried out on the 
number of fitted lines $N_l$. We adopt unity for all weights, except for He\,{\sc ii}\,$\lambda$4200
($w = 0.5$) and the singlet transition
He\,{\sc i}\,$\lambda$4387 ($w = 0.25$), for reasons discussed in \citet{mokiem2005}.

While the exploration of the parameter space is based on the fitness to avoid a single line 
outweighing the others, the fit statistics -- and the error bars -- are however computed using the $\chi^2$ statistic, computed in the usual way.
\begin{equation}
\chi^2=\sum_l \chi^2_\mathrm{l}.
\end{equation}

The algorithm makes use of the normalized spectra to derive the effective temperature 
$(T_{\mathrm{eff}})$, the surface gravity $(\log\,g)$, the mass-loss rate $(\dot{M})$, the 
exponent of the $\beta$-type wind-velocity law $(\beta)$, 
the helium over hydrogen number density (later converted to surface helium abundance in mass fraction, $Y$, through the paper), 
the microturbulent velocity $(\varv_{\mathrm{turb}})$ and the projected
rotational velocity (\vrot). For additional notes on \vrot, we refer the reader to 
Sect.~\ref{sec:extra-broadening}. 

While the method, in principle, allows for the terminal 
wind velocity ($\varv_{\infty}$) to be a free parameter, this quantity cannot be 
constrained from the optical diagnostic lines. 
Instead, we adopted the empirical scaling of $\varv_{\infty}$ with the escape velocity
($\varv_{\mathrm{esc}}$) of \citet{Kudritzki_puls2000}, taking into account the 
metallicity ($Z$) dependence of \citet{LRD92}: $\varv_{\infty} = 2.65\,\varv_\mathrm{esc}\,Z^{0.13}$. In doing so, we 
corrected the Newtonian gravity as given by the spectroscopic mass for radiation 
pressure due to electron scattering. In units of the Newtonian gravity, this correction factor is $(1-\Gamma_{\rm e}$), where $\Gamma_{\rm e}$ is the
Eddington factor for Thomson scattering. 
This treatment of terminal velocity
ignores the large scatter that exists around the $\varv_{\infty}$ versus $\varv_{\mathrm{esc}}$ relation \citep[see discussions in][]{Kudritzki_puls2000,garcia2014}.
However, consistency checks performed in Sect.~\ref{sec:high_lum} indicate that this is not a major issue.

For each star in our sample, up to 12 diagnostic lines are adjusted: 
\hel{i+ii}{4026}, \hel{i}{4387}, 4471, 4713, 4922, \hel{ii}{4200}, 4541, 4686, 
H$\delta$, H$\gamma$, H$\beta$, and H$\alpha$. 
For a subset of stars (those with the earliest spectral subtypes, and mid- and late-O supergiants), our set of H and He diagnostic lines was not sufficient 
to accurately constrain their parameters. 
In these cases, we also adjusted nitrogen lines in the spectra and considered the nitrogen surface abundance to be a free parameter. 
Specifically, we included the following lines  in the list of diagnostic lines used: 
\Nl{ii}{3995}, \Nl{iii}{4097}, 4103, 4195, 4200, 4379, 4511, 4515, 
4518, 4523, 4634, 4640, 4641, \Nl{iv}{4058} and \Nl{v}{4603}, 4619.  
Tables~\ref{table:stats_lines_N}-\ref{table:stats_lines_pqf}
 summarize, for each star, the diagnostic lines that have simultaneously been considered in the  fit.
The fitting results for each object were visually inspected. Residuals of nebular correction were manually clipped, after which the fitting procedure was repeated. 
The best-fit model and the 
set of acceptable models, for every star, are displayed in Appendix~\ref{app:example} 
(see also Sect.~\ref{subsec:error_calculation}).

The de-reddened absolute magnitude and the RV of the star are needed as input parameters; the 
first is used to determine the object 
luminosity and hence the radius, 
while the second is used to shift the model and data to the same reference frame.
While both \citet{mokiem2005,mokiem2006,mokiem2007} and \citet{tramper2011,tramper2014} 
used the $V$-band magnitude as a photometric anchor, we choose to use the $K$-band magnitude
($M_{\mathrm{K}}$) to minimise the impact of uncertainties on the individual reddening of the
objects in our sample. We determined $M_{\mathrm{K}}$  using the VISTA observed $K$-band 
magnitude \citep{rubele2012}, adopting a distance modulus to the Tarantula nebula of 18.5\,mag 
\citepalias[see][]{evans} and an average $K$-band extinction 
$(A_{\mathrm{K}})$ of 0.21\,mag (Ma\'iz-Apell\'aniz et al., in prep.). The obtained $M_{\mathrm{K}}$ values are provided
in Table~\ref{table:new} and \ref{table:new_noLC}, for completeness. 
As for the RV values we used the measurements listed in \citet{sana}.

\subsection{Error calculation}
\label{subsec:error_calculation}

The parameter fitting uncertainties were estimated in the following way.
For each star and each model, we calculated the probability ($P$) 
that the $\chi^2$ value as large as the one that we observed is not a result of statistical fluctuation:
$P\,=\,1-\Gamma(\chi^2/2,\nu/2)$, where $\Gamma$ is the incomplete gamma function and $\nu$ the number of degrees of freedom.

Because $P$ is very sensitive to the $\chi^2$ value,
we re-normalized all $\chi^2$ values such that the best fitting model of a given star has a reduced $\chi^2$ ($\chi_{\mathrm{red}}^2$) equal to unity. We 
thus implicitly assumed that the model with the smallest $\chi^2$ 
 represents the data and that deviations of the best model's $\chi_{\mathrm{red}}^2$ from unity result from under- or overestimated error bars on the  normalized flux. 
This approach is valid if the best-fit model represents the data, which was visually checked for each star 
(see Sect.~\ref{sec:limitations} and Appendix~\ref{app:example}).
Finally, the 95\%\ confidence intervals  on the fitted parameters were obtained by considering the range of models that satisfy $P(\chi^2, \nu) > 0.05$. The latter can approximately be considered as $\pm2\sigma$ error estimates in cases where the
probability distributions follow a Gaussian distribution. 

The finite exploration of parameter space may however result in an underestimate of the confidence interval in the case of poor sampling near the borders $P(\chi^2, \nu) = 0.05$. As a first attempt to mitigate this situation, we adopt as boundaries of the 95\%\ confidence interval the first models 
that do
not satisfy $P(\chi^2, \nu) = 0.05$, hence making sure that the quoted confidence intervals are either identical or slightly larger than their exact 95\%\ counterparts. However, for approximately 10\%\ of the boundaries so determined, the results were still leading to unsatisfactory small, or large, upper and lower errors. We then turned to fitting the $\chi^2$ distribution envelopes.  The left- and right-hand part of the envelopes were fitted  separately for all quantities  using either a 3rd- or 4th degree polynomial or a Gaussian profile. The intersects of the fitted envelope with the critical $\chi^2$ threshold defined above ($P(\chi^2, \nu) = 0.05$) for the function that best represented the envelope were then adopted as upper- and lower-limit for the 95\%-confidence intervals.

\begin{figure*}[ht!]
\centering
\includegraphics[scale=0.45]{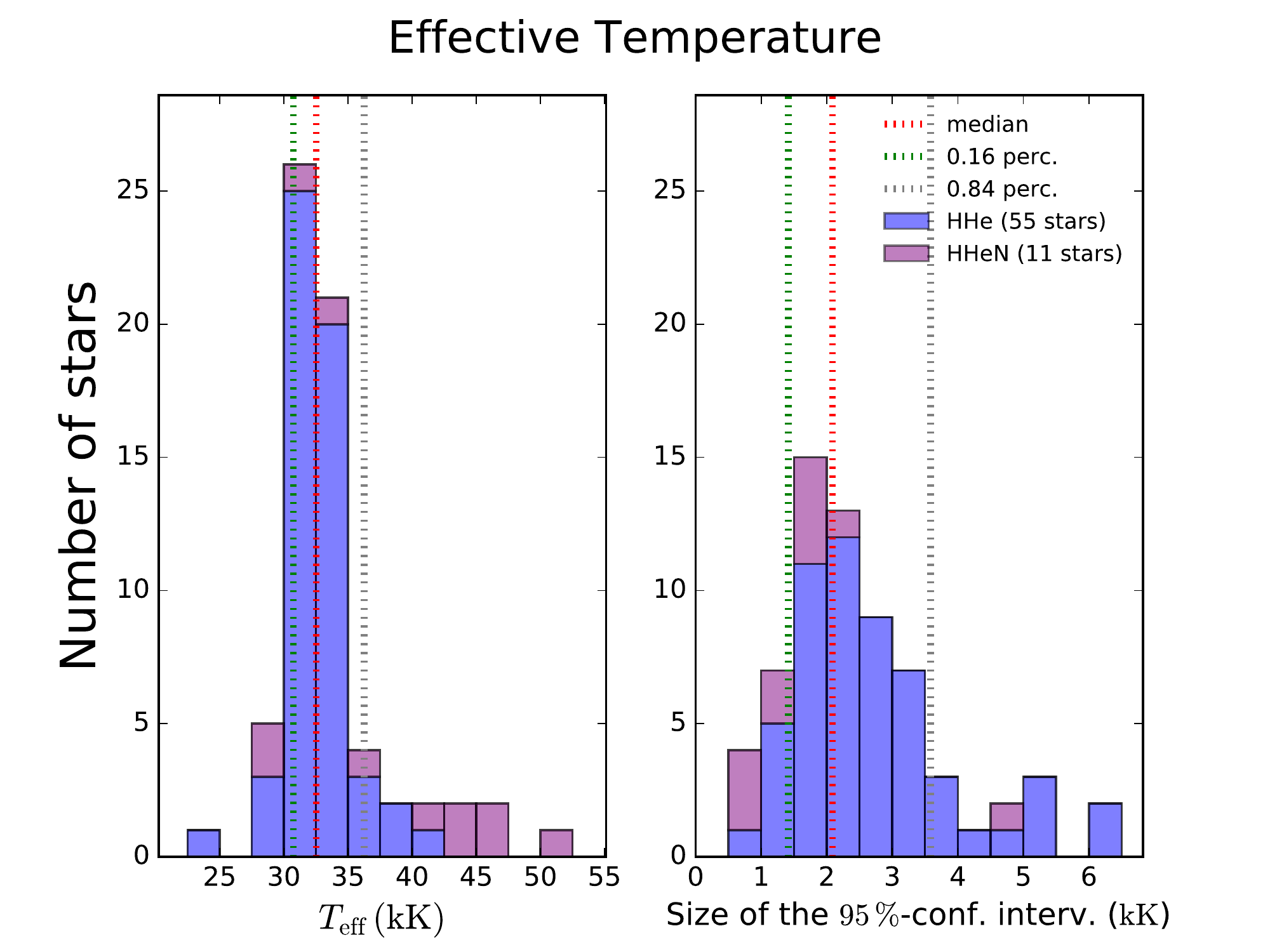}
\includegraphics[scale=0.45]{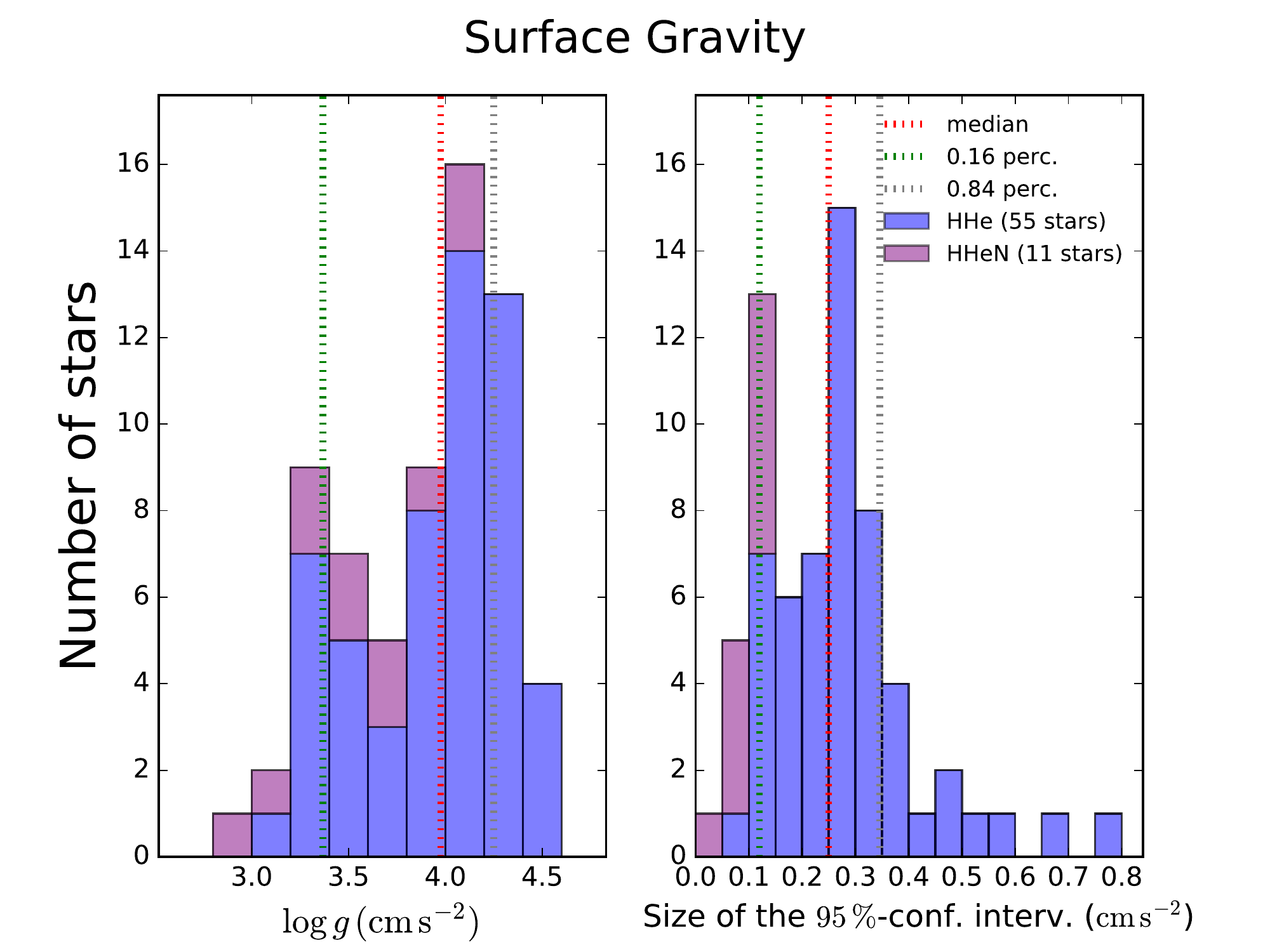}
\includegraphics[scale=0.45]{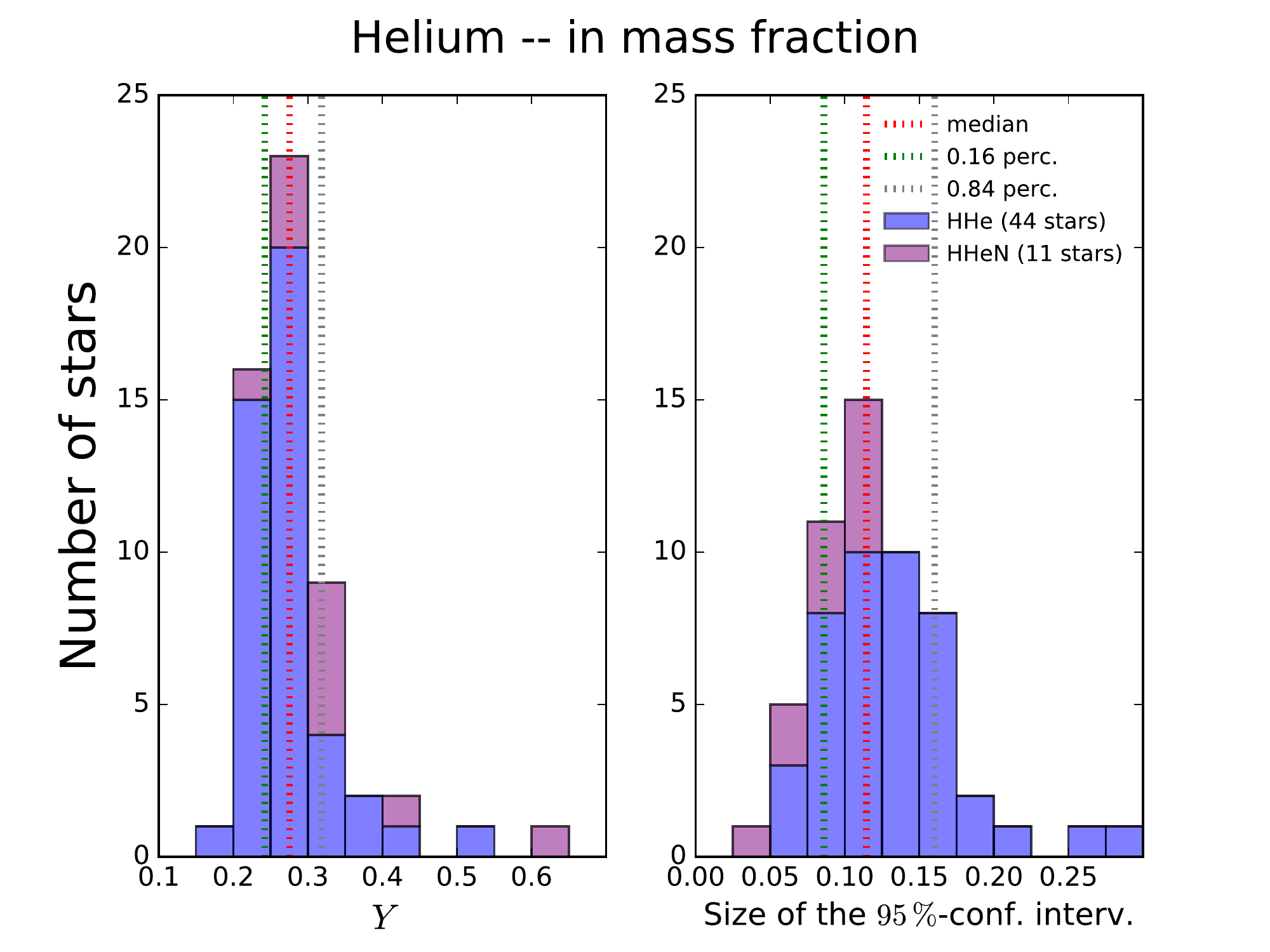}
\includegraphics[scale=0.45]{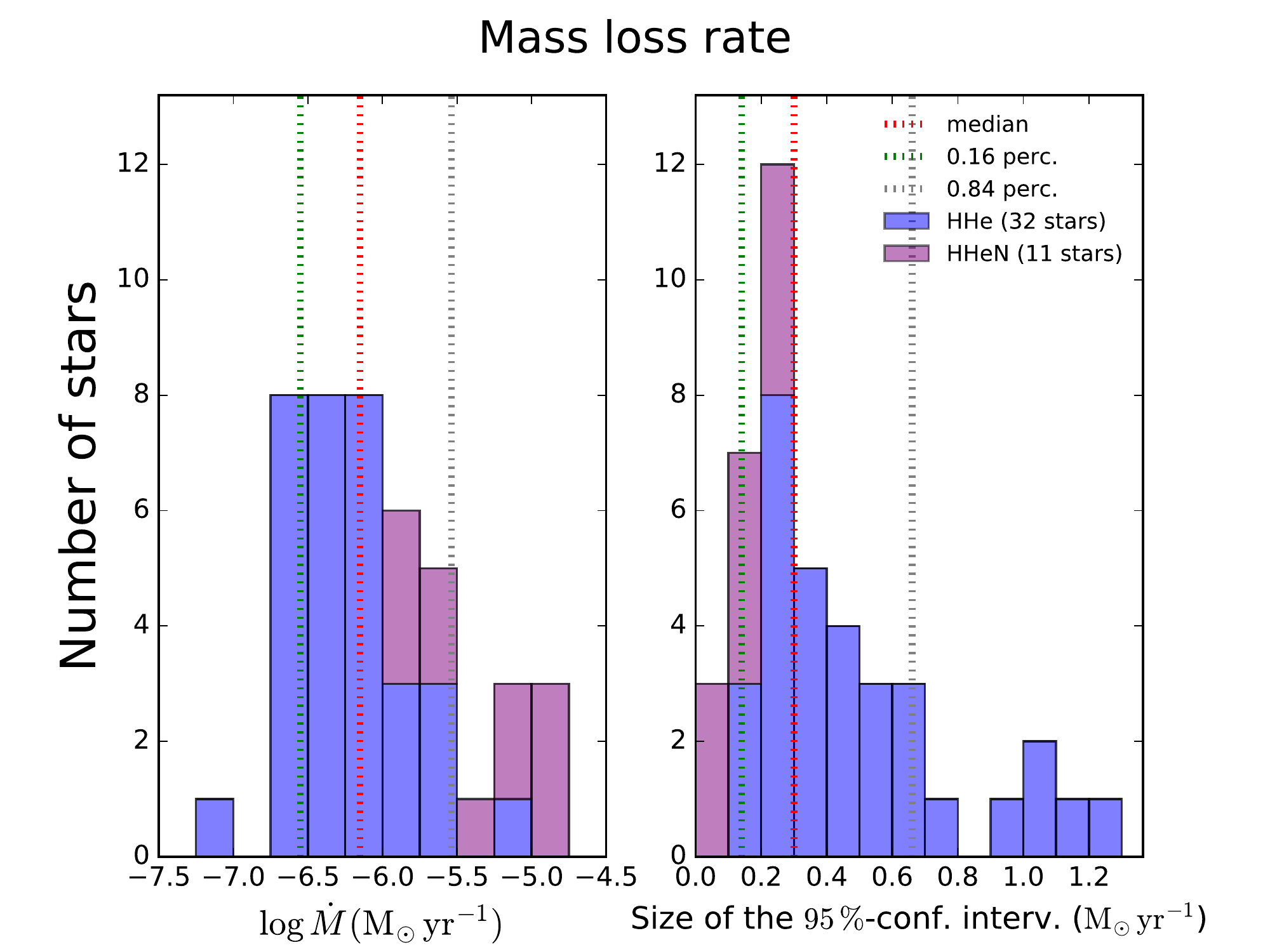}
\includegraphics[scale=0.45]{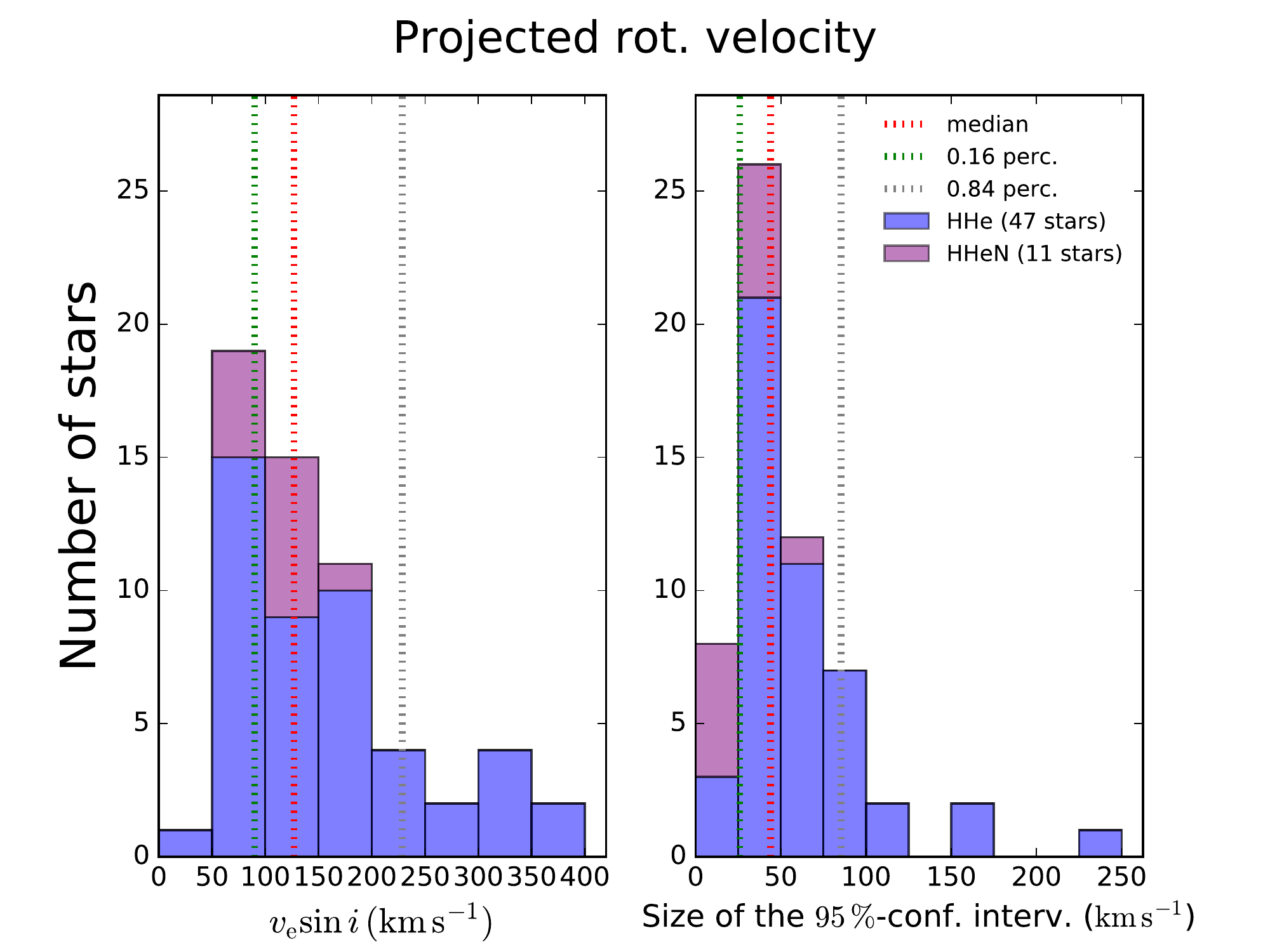}
\includegraphics[scale=0.45]{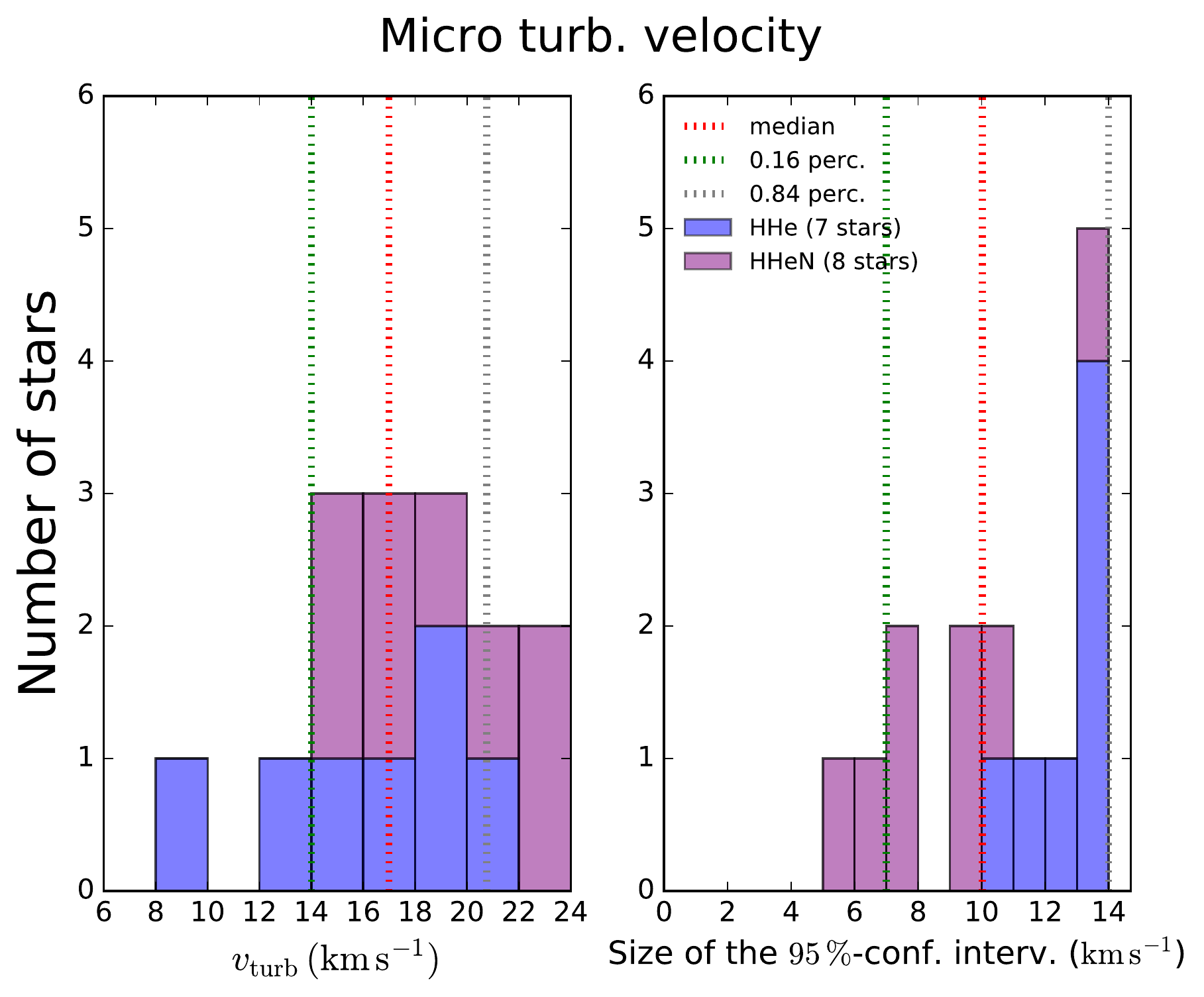}
\caption{Range of each fitted parameter (left panels of each set) and accompanying range of 95\%\ confidence
interval (right panel of each set) in the same unit.  Colors have been used to differentiate between stars
that have been analyzed using hydrogen and helium lines (HHe) and those for which also nitrogen lines (HHeN) were considered (see also Sect.~\ref{subsec:GA}).  
The distributions exclude stars for which only upper/lower limits could be determined, hence the number of stars shown in a panel depends on the parameter that is investigated. In each panel, the median value and the 16th and 84th percentiles are shown using vertical lines.}
\label{fig:distributions}
\end{figure*}

The obtained boundaries of the confidence intervals, relative to the
best-fit value, are provided in Table~\ref{table:new}.  For some quantities and for some stars, these boundaries are relatively asymmetric with respect to the best-fit values. Hence, the total range covered by the 95\%\ confidence intervals needs to be considered to understand the typical error budget in our sample stars, that is, not only the lower- or upper-boundaries. In Fig.~\ref{fig:distributions},   
we show the distribution of these widths for all model parameters that have their 
confidence interval constrained (i.e., excluding upper/lower limits). 
The median values of these uncertainties are 2090\,K for $T_{\mathrm{eff}}$, 0.25 dex 
for $\log\,g$, and 0.11 for Y. For those sources that have their mass-loss rates constrained, the median uncertainty in $\log\,\dot{M}$ is 0.3 dex. For the projected 
spin velocities it is 44\,\kms.  
We note that for some sources, the formal error estimates are very small.  This is particularly so in cases where nitrogen lines are used as 
diagnostics, which tend to place stringent limits on the  effective temperature, hence indirectly on the surface gravity, and the mass-loss rate. 
Results related to sources for which nitrogen was included in the analysis have been given a different color in 
Fig.~\ref{fig:distributions}. 

\subsection{Sources of systematic errors}
\label{subsec:systematics}

It is important to stress that the confidence intervals given in Table~\ref{table:new} represent the validity of the models as well as the
formal errors of the fits, that is, uncertainties measuring statistical variability.  They do not account for systematic uncertainties, 
which may be significant. 
Here we discuss possible sources of this type of uncertainty that may impact the accuracy of our results.

Systematic errors may relate to model assumptions, continuum placement biases, 
the assumed distance to the LMC, 
or an uncertain extinction, for example. 
Regarding the adopted model atmosphere, \citet{massey2013} performed a by-eye analysis of 
ten LMC O-type stars using both {\sc{cmfgen}} \citep[][]{hillier_miller_1998} and 
{\sc{fastwind}}. They report a systematic difference in the derived gravity 
of 0.12 dex, with {\sc{cmfgen}} values 
being higher. They argue that differences in the treatment of the electron scattering wings might explain the bulk of this difference, a treatment that is more refined in 
{\sc{cmfgen}}. A systematic error in the normalization of the local continuum may also impact the 
gravity estimate. If by-eye judgement would place it too high by 1\%\  
(where the typical normalization error is better than 1\%; see 
Sect.~\ref{subsec:data_prep}) for all relevant diagnostic lines, this would lead to a 
gravity that is higher by less than 0.1 dex. We do not, however, anticipate such a large 
systematic normalization error. 
The distance to the LMC is accurate to within 2\%\ \citep{pietrzyski2013}. 
We adopt a mean $K$-band extinction of 0.21\,mag (see 
Sect.~\ref{subsec:GA}). Typical deviations of this mean value are not larger than 0.1\,mag 
(Ma\'iz-Apell\'aniz et al. in prep.), hence correspond to an uncertainty in the luminosity 
of less than 10\%.  

Other systematic uncertainties may be present; for instance model assumptions that impact both a {\sc fastwind} and {\sc cmfgen} analysis. Examples 
are the neglect of macro-turbulence or the 
assumption of a spherical and constant mass-loss rate outflow.  

Systematic (quantifiable and non-quantifiable) errors will impact the formal confidence intervals discussed in Sect. 3.2.  In those cases where the quoted confidence intervals are approximately equal to their respective medians or larger, the systematic errors will likely contribute modestly to the total uncertainties. In cases where the formal errors are small, we caution the reader that systematic uncertainties may be larger than the statistical uncertainties presented in Table~\ref{table:new}.

\subsection{Consistency checks}

Here we compare aspects of the properties obtained for our sample stars to those of other
O-type sub-samples analyzed in the VFTS.

\subsubsection{O~V and IV stars}

To test the consistency of our results with the atmosphere fitting methods applied to 
O-type dwarfs within the VFTS, we selected a 
subset of 66 stars from \citetalias{sabin_sanjulian2014}. We computed the stellar properties by means of our atmosphere fitting approach. 
The fitting approach in \citetalias{sabin_sanjulian2014} also made use of {\sc fastwind} 
models, but applied a grid-based tool, where the absolute flux calibration relied on the $V$-band magnitude.
The values that we obtained are in agreement 
with those of \citetalias{sabin_sanjulian2014}. Specifically,
the weighted mean of the temperature
difference ($\Delta\,T_{\rm eff}$[\citetalias{sabin_sanjulian2014} $-$ this study]) and the 1$\sigma$ dispersion around the mean value are
$0.69\,\pm\,0.33$~kK and $1.21\,\pm\,0.37$~kK, respectively. 
 The weighted mean of the luminosity difference  
 ($\Delta\,\log L/\lsun\,$[\citetalias{sabin_sanjulian2014} $-$ this study]) and its
 $1\sigma$ dispersion are
 $0.08\,\pm\,0.02$ and $0.07\,\pm\,0.02$. 
Similarly, the weighted mean of the difference in gravity between both fitting approaches ($\Delta\,\log\,g\,$[\citetalias{sabin_sanjulian2014} $-$ this study]) and its $1\sigma$ dispersion are $0.16\,\pm\,0.04$ and $0.09\,\pm\,0.04$ (if $g$ is measured in units
of cm\,s$^{-2}$).
 Typical errors on the temperature, gravity, and luminosity in both methods are
1~kK, 0.1 dex, and 0.1 dex, respectively, that is, of the same order as the mean
differences. 
Hence, this comparison does not reveal conspicous discrepancies in temperature
and luminosity. Possibly, a modest discrepancy is present in gravity. 

\subsubsection{O9.7 stars}

The spectral analysis of the O9.7 subtype 
is complicated as the He\,{\sc ii} 
lines become very weak, hence small absolute uncertainties may have a larger impact on the 
determination of the effective temperature. For early-B stars,
one also relies on Si\,{\sc iii} and 
{\sc iv} lines as a temperature diagnostic 
\citep[see, e.g.,][]{mcevoy2015}. We performed several checks to assess the
reliability of our parameters for the late O stars. For 
four stars; VFTS\,035 (O9.5\,IIn), 235 (O9.7\,III), 253 (O9.7\,II), 
and 304 (O9.7\,III), we used {\sc fastwind} models that include
Si\,{\sc iii}\,$\lambda$4552, 4567, 4574, Si\,{\sc iv}\,$\lambda$4128, 4130, and
Si\,{\sc v}\,$\lambda$4089, 4116 as extra diagnostics. The temperatures and
gravities that we then obtain agree within 400\,K and 0.28 dex, respectively, that is, within typical uncertainties, 
suggesting that the lack of silicon lines in our automated {\sc fastwind} modeling does not 
introduce systematic effects.

One may also compare to atmosphere models that
assume hydrostatic equilibrium, that is, that neglect a stellar wind. This is 
done in \citet{mcevoy2015} for two O9.7 sources, VFTS\,087 (O9.7\,Ib-II) and
165 (O9.7\,Iab), where {\sc fastwind} analyses are compared to {\sc tlusty}
analyses \citep{hubeny1988,hubeny1995,lanz2007}. Here {\sc fastwind} 
settles on temperatures that are 1000\,K higher, which is within the
uncertainties quoted. This is accompanied by 0.07 dex higher gravities, which is well within the error range.  We also compared to preliminary  
 {\sc tlusty} results for some of the stars analyzed here  
(Dufton et al. in prep.); namely VFTS\,113, 192, 226, 607, 753, and 787, all 
O9.7\,II, II$-$III, III sources. The weighted mean of the temperature difference 
($\Delta\,T_{\rm eff}$[{\sc tlusty} $-$ this study]) and the associated $1\sigma$ 
dispersion are $-1.95\,\pm\,1.26$~kK and $0.81\,\pm\,4.15$~kK. This
offset is similar to that reported by \citet{massey2009}.
Similarly, the 
weighted mean of the difference in $\log\,g$ and the associated $1\sigma$ 
dispersion are $-0.29\,\pm\,0.07$, and $0.07\,\pm\,0.04$.
These differences are larger than one 
might expect and warrant caution. For LMC 
spectra of the quality studied here, systematic errors in $T_{\rm eff}$ and 
gravity between {\sc fastwind} and {\sc tlusty}, at spectral type O9.7, can not 
be excluded.

\subsubsection{The most luminous stars}\label{sec:high_lum}

Twelve objects in our sample are in common with  
\citetalias{bestenlehner2014}, which analyzed the stars with the highest masses and luminosities. 
These stars are VFTS\,016, 064, 171, 180, 259, 267, 333, 518, 566, 599, 664, and 669.
 VFTS\,064, 171, and 333 are, however, excluded from the present comparison because our obtained fits were rated as 
 poor quality (see Sect.~\ref{sec:poor_fits}).  For the remaining 
 nine stars, the values obtained in this paper agree well with those of \citetalias{bestenlehner2014}.
The weighted mean of the temperature difference ($\Delta\,T_{\rm eff}$[\citetalias{bestenlehner2014} $-$ this study]) and the associated $1\sigma$ 
dispersion of this set of nine stars are 
$1.52\,\pm\,0.18$~kK and $1.82\,\pm\,0.35$~kK. The weighted mean of the luminosity
difference ($\Delta\,\log L/\lsun\,$[\citetalias{bestenlehner2014} $-$ this study]) and its $1\sigma$ dispersion are $0.09\,\pm\,0.02$ and $0.06\,\pm\,0.01$.  
The gravities cannot be compared in a similar way as 
they were held constant in \citetalias{bestenlehner2014}. For this reason, the
present results are to be preferred for stars in common with
\citet{bestenlehner2014}. 

For the small number of stars for which $\varv_{\infty}$ could actually be measured in \citetalias{bestenlehner2014}, the terminal velocities are 
consistent with those that we estimated from the $\varv_{\infty}/\varv_{\rm esc}$ relation. Regarding the unclumped mass-loss rates, the weighted 
mean of the $\log\,\dot{M}$ differences ($\Delta \log\,\dot{M}\,$[\citetalias{bestenlehner2014}$-$ this study]), and its $1\sigma$ dispersion, 
are $-0.03\,\pm\,0.23\,M_{\odot}$/yr, and $0.21\,\pm\,0.19\,M_{\odot}$/yr, indicating the absence of  systematics between  the results of both studies. 
Finally, previous optical and ultraviolet analysis of  VFTS\,016 had constrained $\varv_{\infty}$ to $3450\pm50$~\kms\  \citep{evans2010}, 
in satisfactory agreement with the value of $3631^{+85}_{-122}$\,\kms\ that we derived from our best-fit parameters using the scaling with $\varv_{\rm esc}$.

\subsection{Derived properties}
\label{sec:derived_properties}

In addition to $\varv_{\infty}$ (see Sect.~\ref{subsec:GA}), several important quantities can be 
derived from the best-fit parameters: the bolometric luminosity $L$, the 
stellar radius $R$, the spectroscopic mass $M_{\mathrm{spec}}$, and the modified wind-momentum rate
$D_\mathrm{mom} = \dot{M} \, v_{\infty} \, \left(R/R_{\odot}\right)^{1/2}$. 
The latter quantity provides a convenient means to confront empirical with predicted wind strengths as
$D_\mathrm{mom}$ is expected to be almost independent of mass \citep[e.g.,][see  
Sect.~\ref{subsec:WLD}]{Kudritzki_puls2000}. 
To determine the radius, the
theoretical fluxes have been converted to $K$-band magnitude using the 2MASS filter response
function\footnote{2MASS filter response function are tabulated at http://www.ipac.caltech.edu/2mass/releases/allsky/doc /sec6\_4a.tbl3.html}
and the absolute flux calibration from \citet{cohen2003}.
The values of the parameters mentioned above for our 
sample stars together, with their corresponding 95\% confidence interval, are provided in Table~\ref{table:new} as well. 

\subsection{Binaries and poor quality fits}
\label{sec:poor_fits}

VFTS\,064, 093, 171, 332, 333, and 440 
have been subjected to further RV monitoring and are now 
confirmed to be spectroscopic binaries \citep[][see also Appendix~\ref{app:remarks}]{almeida_et_al_2016}. 
In VFTS, they all showed small but significant radial velocity variations 
\citep[$\rm{\Delta RV}\,\leq$\,20\,\kms;][]{sana}. 
 \citet{walborn2014} noted these six objects as having a somewhat problematic spectral classification.
Interestingly, our fits of these six stars often implied a helium
abundance significantly lower than the primordial value, which may be the result of line dilution by the 
continuum of the companion. We decided to discard these stars from our discussion, opting for a sample of 72$-$6 = 66
high-quality fits only and minimizing the risk of misinterpretations. We do provide the obtained 
parameters and formal uncertainties of these six stars in Table~\ref{table:new} but warn against possible systematic biases.

All other spectral fits were screened by eye to assess their quality. We concluded that all fits were 
acceptable within the range of models that pass our statistical criteria except those of six objects without 
LC (VFTS\,145, 360, 400, 446, 451, and 565).  
We also provide the obtained 
parameters of these six stars in Table~\ref{table:new_noLC} but warn that they may not be representative of the stars 
physical parameters as their fits have limited quality.

\subsection{Limitations of the method}
\label{sec:limitations}

We discuss two limitations of the method in more detail, that is, the
neglect of macro-turbulence and the lack of a diagnostics that allows us to constrain
the spatial velocity gradient of the outflow.

\subsubsection{Extra line-broadening due to macro-turbulent motions}
\label{sec:extra-broadening}

When comparing the models with the data, we take into account several sources of 
spectral line broadening:  intrinsic broadening, rotational broadening, and 
broadening due to the instrumental profile. 
However, we do not take into account the possibility of extra-broadening as a result of
macro-turbulent motions in the stellar photosphere \citep[e.g.,][]{Gray}.
This approach is somewhat different to that of \citetalias{ramirezagudelo}, 
in which a Fourier transform method was used to help differentiate between rotation 
and macro-turbulent broadening, neglecting intrinsic line broadening and given 
a model for the behavior of macro-turbulence (we refer to \citetalias{ramirezagudelo} for a discussion).

Appendix~\ref{subsec:vsini_paperXII_GA} compares the rotation rates 
of the sample of 66 stars obtained 
through both methods. The systematic difference 
($\Delta$\,\vrot\,[this study - \citetalias{ramirezagudelo}]) is approximately 7 \kms\ with a 
standard deviation of approximately 21 \kms. This is within the 
uncertainties discussed in  \citetalias{ramirezagudelo}. 
At projected spin velocities below 160\,\kms\ the present measurements may
overestimate \vrot\ by up to several tens of \kms\ in cases where macro-turbulence 
is prominent. Though, the best of our knowledge, there are no theoretical assessments of the impact
of macro-turbulence on the determination of the stellar 
parameters, we do not expect the differences in \vrot\  to affect the determination of other stellar properties in any significant way as the \vrot\ measurements of both methods are within uncertainties.

\subsubsection{Wind-velocity law}
\label{sec:wind_ac}

The spatial velocity gradient,
measured by the exponent $\beta$ of the wind-velocity law 
becomes unconstrained 
if the diagnostic lines which are sensitive to mass-loss rate (H$\alpha$ and \hel{ii}{4686}) are 
formed close to the photosphere. In such cases, the flow velocity 
is indeed still low compared to $\varv_{\infty}$. In an initial
determination of the parameters,
we let $\beta$ be a free parameter in the interval [0.8, 2.0]. In approximately half of the cases,
the fit returned a central value for $\beta$ larger than 1.2 and with large uncertainties. 
From theoretical computations, such a large
acceleration parameter is not expected for normal O stars  and we
identified these sources as having an unconstrained $\beta$. 
Given the large percentage of stars that fell in this category and the potential impact 
of $\beta$ on the derived mass-loss rate, we decided to adopt 
$\beta = 0.9$ for giants and $0.95$ for bright giants and supergiants, 
following theoretical predictions by \citet{muijres2012}. 
For the 31 O stars that could not be assigned a LC, we adopted 
the canonical value $\beta = 1$ (see Table~\ref{table:new_noLC}). 
We will discuss the impact of this assumption in Sect.~\ref{subsec:WLD}.

\section{Results and discussion}\label{sec:results}

We discuss our findings for the effective temperature, gravity, helium abundance,
mass loss, and mass, and place these results in the broader context of 
stellar evolution, mass-loss behavior, and mass discrepancy.

\subsection{Effective temperature vs. spectral subtype calibrations}\label{sec:teff_spt}

\begin{figure}
\centering
\includegraphics[width=\columnwidth]{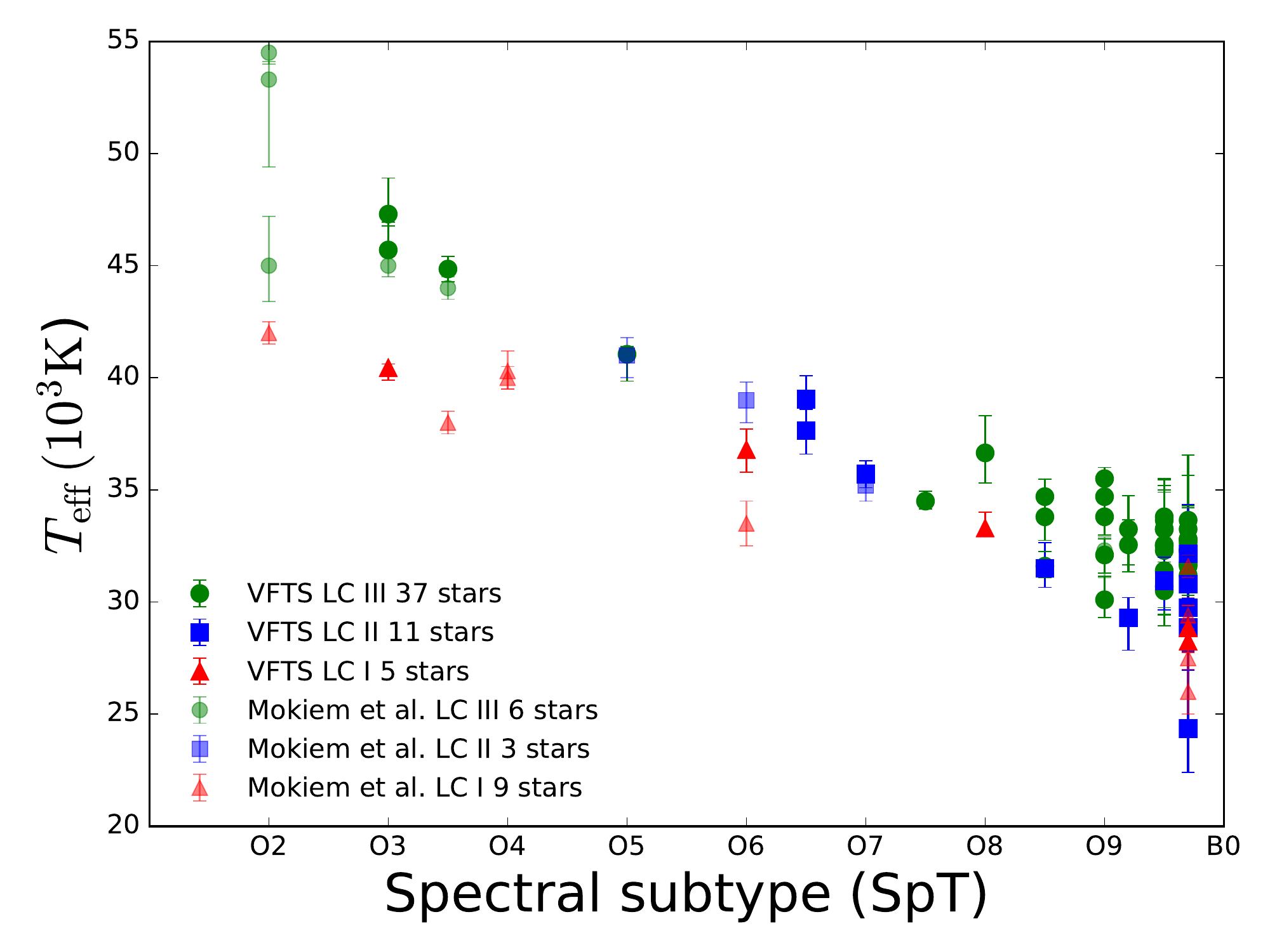}
\caption{
Effective temperature {\it vs.} spectral subtype for the O-type with well defined LC (see main text).
The lower-opacity symbols give the results for the sample of
LMC stars investigated by \citet{mokiem2007b}.
}
\label{fig:spt_teff}
\end{figure}

Figure~\ref{fig:spt_teff} plots the derived effective temperature for
53 giants, bright giants and supergiants as a function of spectral subtype. This sample of 
53 stars corresponds to the high-quality fits (66 stars) minus the stars that
have a somewhat ambiguous luminosity classification (17 minus the newly confirmed binaries VFTS 093, 171, 332, and 333 fits, hence 13 stars; 
we refer to Sects.~\ref{sec:sample} and~\ref{sec:poor_fits}).
For LC~III and LC~II stars, the scatter at late spectral type is too large to be solely explained by measurement errors
and may thus also reflect intrinsic differences in gravity, hence in evolutionary
state \citep[for a discussion, see][]{simon2014}.
Added to the figure are
results for 18 LC\,III to I LMC stars by \citet{mokiem2007b}. These were analyzed using the 
same fitting technique, save that these authors did not use nitrogen lines 
in cases where either He\,{\sc i} or He\,{\sc ii} lines were absent (see  
Sect.~\ref{subsec:GA}). Both our sample and that of \citeauthor{mokiem2007b} yield results that are compatible with each other, therefore we 
combine both samples in the remainder of this section.

Though the overall trend in Fig.~\ref{fig:spt_teff} is clearly that of
a monotonically decreasing temperature with spectral subtype, such a 
trend need not necessarily reflect a linear relationship. Work by
\citet{rivero_gonzalez2012b,rivero_gonzalez2012a} for early-O dwarfs in the
LMC, for instance, suggests a steeper slope at the earliest subtypes (O2-O3).
This seems to be supported by first estimates of the properties of 
O2 dwarfs in the VFTS by \citet{sabin_sanjulian2014}. The presence of such an upturn starting at  spectral sub-type O4 is not confirmed in  Fig.~\ref{fig:spt_teff}. The three O2~III stars (all from \citeauthor{mokiem2007b}) do show a spread that may be compatible with a steeper slope for giants but such an increased slope is not yet needed at subtype O3. Furthermore, the only two O2~I and O3~I stars in Fig.~\ref{fig:spt_teff} are  perfectly compatible with a constant slope down to the earliest spectral sub-types for the supergiants. In regards to the insufficient number of stars, to fully test for the presence of an upturn at subtype O2, we limit our  $T_{\rm eff}$-SpT calibrations to subtypes O3 and later.

\begin{figure}[t!]
\centering
\includegraphics[width=\columnwidth]{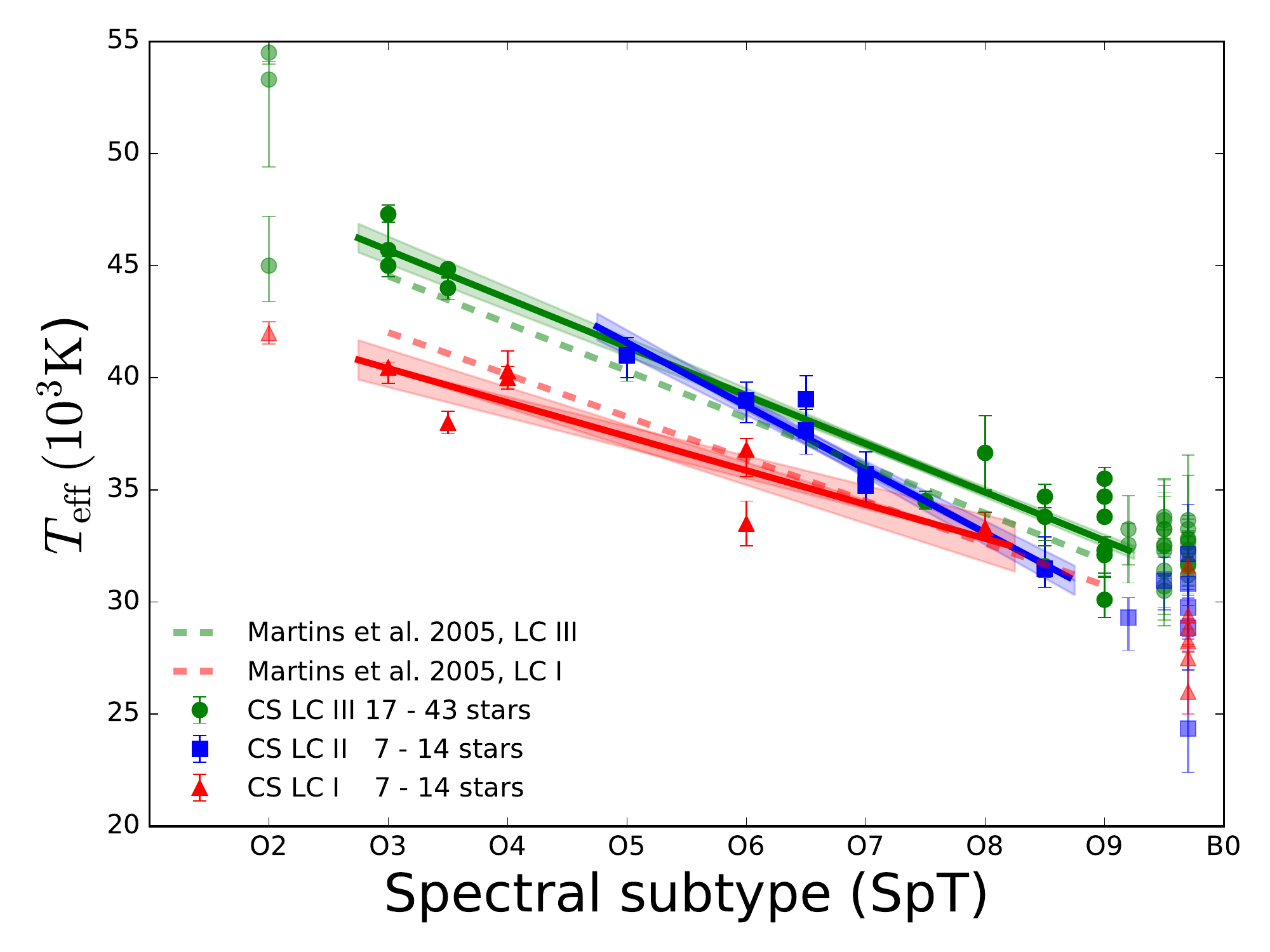}
\caption{Effective temperature {\it vs.} spectral subtype 
but now displaying fits that combine our sample 
with that of \citet{mokiem2007b}: CS\,= Combined sample (see main text). 
Stars with spectral subtype O2-3 or later than O9
have been plotted with lower opacity.
The dashed lines give the theoretical calibrations of \citet{martins} for Galactic class 
III and I stars. The leading number in the legend refers to the 
total number of O3-O9 stars for which the fit has been derived.
The trailing number refers to the total number of stars in each sample.}
\label{fig:spt_teff_combine}
\end{figure}

A shallowing of the $T_{\rm eff}$-SpT relation at  subtypes later than O9 
(relative to the O3-O9 regime) is also relatively conspicuous in Fig.~\ref{fig:spt_teff}. We too exclude this regime
from the relations given below, also because the luminosity classification
of this group in particular may be debated (see Sect.~\ref{subsec:gravities}).
We thus aim to derive $T_{\rm eff}$-SpT relations for LMC
O-type stars in the regime O3-O9. To do so, we used a weighted least-square 
linear fit to adjust the relation
\begin{equation}
     T_\mathrm{eff} = a + b \times \mathrm{SpT},
\label{eq:1}
\end{equation}
where the spectral subtype is represented by a real number, for example, 
SpT = 6.5 for an O6.5 star. Figure~\ref{fig:spt_teff_combine} shows these
linear fits for our sample and that of \citet{mokiem2007} combined, for each 
luminosity class separately. The fit coefficients and their uncertainties
are provided in Table~\ref{table:teff}. 

A comparison of our combined LC\,III, II and I relations with theoretical results
for a LMC metallicity is not feasible as, to our knowledge, such 
predictions are not yet available.
One may anticipate that a LMC calibration would be shifted up to higher 
temperatures, as, in a lower metallicity environment, the effects of
line blocking/blanketing are less important than in a high-metallicity 
environment. Thus, fewer photons are scattered back, contributing less to
the mean intensity in those regions where the He\,{\sc i} lines are formed. Consequently,
a higher $T_{\rm eff}$ is needed to reach the same degree of ionization 
for stars in the LMC compared to those with a higher metal abundance 
\citep[which have stronger blocking/blanketing, see][]{repolust2004}. 
In Fig.~\ref{fig:spt_teff_combine}, we compare our results to the LC\,III and I empirical calibrations of 
\citet[][their equation 2]{martins} for Galactic stars. Below we discuss the results for LC III, II, and I separately:

\begin{enumerate}
\item[-] {\it Giants (LC\,III):} 
The slope of the $T_{\rm eff}$-SpT relation for giants is in excellent agreement with
the (observational) \citet{martins} calibration, though an upward shift of 
approximately 1\,kK is required to  account for the lower metallicity. 
\citet{doran} report that a +1\,kK shift is required to match the LMC dwarfs,
but that no shift seems required for O-giants. 
Our results suggest that this upward shift should be applied to this category as well.

\item[-] {\it Bright giants  (LC\,II):} 
The $T_{\rm eff}$-SpT relation for the bright giants is relatively steep,
and crosses the relations for the giants and supergiants. As explained in
the notes for individual stars (Appendix~\ref{app:remarks}), the spectra of
some of these stars are peculiar. We also note that
in the Hertzsprung-Russell diagram the O\,II stars do not appear to constitute a 
distinct group intermediate between the giants and supergiants (see 
Figs.~\ref{fig:logg_logT} and \ref{fig:HR}); rather, they mingle between the O\,V and I stars. This might 
explain their behavior in Fig.~\ref{fig:spt_teff_combine} and implies that
one should be cautious in using this relation as a calibration. We
recommend to refrain from doing so and to wait until more data become available.

\item[-] {\it Supergiants  (LC\,I):} 
Our supergiant sample is smaller than that of the giants and some O\,I stars have peculiar spectra (Appendix~\ref{app:remarks}), yet it is the largest LMC supergiants sample assembled so far and hence worthy of some in-depth discussion.  As also observed at Galactic metallicity \citep{martins}, our derived $T_{\rm eff}$-SpT relation for  supergiants is  shallower than that for giants. 
The slope for the supergiants is even more shallow at LMC than obtained by \citeauthor{martins} in the Milky Way. Furthermore, the upward shift measured for LMC V and III stars compared to those in the Galaxy is not seen for the supergiants. If anything, a downward shift is present at the earliest spectral types. Within uncertainties however, one may still accept the Galactic-metallicity relation derived for LC~I by \citeauthor{martins} as a reasonable representation of the LMC supergiants. A larger sample would be desirable to confirm or discard these preliminary conclusions as well as to investigate the physical origin of the different metallicity effects for LC~I objects compared to LC V and III stars.
\end{enumerate}

\begin{table}
\caption{$T_{\rm eff}-$SpT linear-fit parameters and their 1$\sigma$ error bars derived for stars with spectral subtypes O3 to O9 in the
combined sample (see text).}             
\label{table:teff}      
\centering                          
\begin{tabular}{lccc}        
\hline\hline\\[-8pt]                 
Sample &  \# stars &$a$\,(kK) & $b$\,(kK) \\ 
\hline\\[-8pt] 
LC III &17   &   52.17 $\pm$ 1.03  &    $-$2.15 $\pm$ 0.14    \\     
LC II  & 7   &   55.71 $\pm$ 2.07  &    $-$2.83 $\pm$ 0.31     \\     
LC I   & 7   &   44.97 $\pm$ 1.87  &    $-$1.52 $\pm$ 0.36    \\     
\hline                        
\end{tabular}
\end{table}

\subsection{Gravities and luminosity classification}
\label{subsec:gravities}

\begin{figure*}
\centering
\includegraphics[scale=0.70]{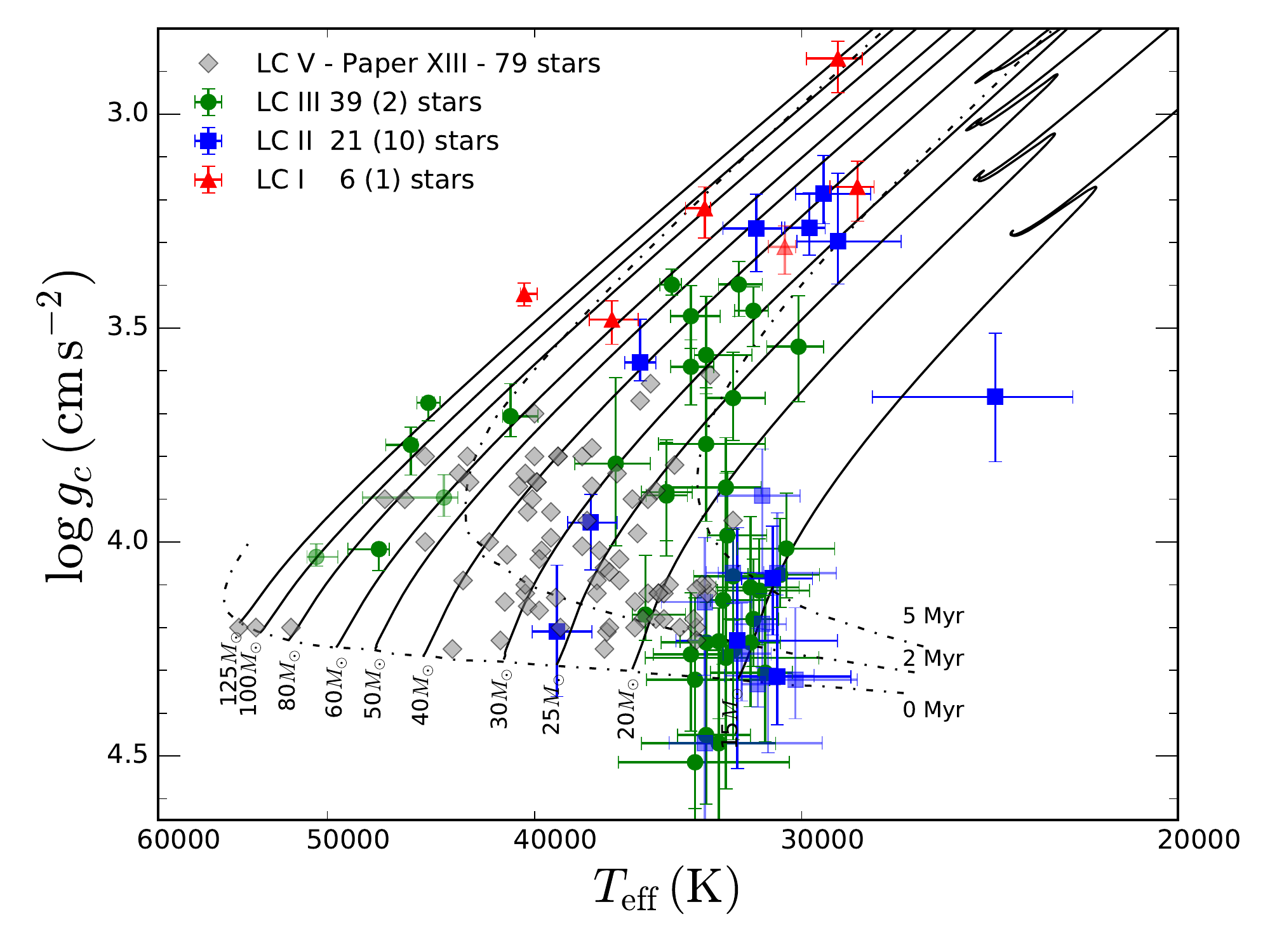}
\includegraphics[scale=0.70]{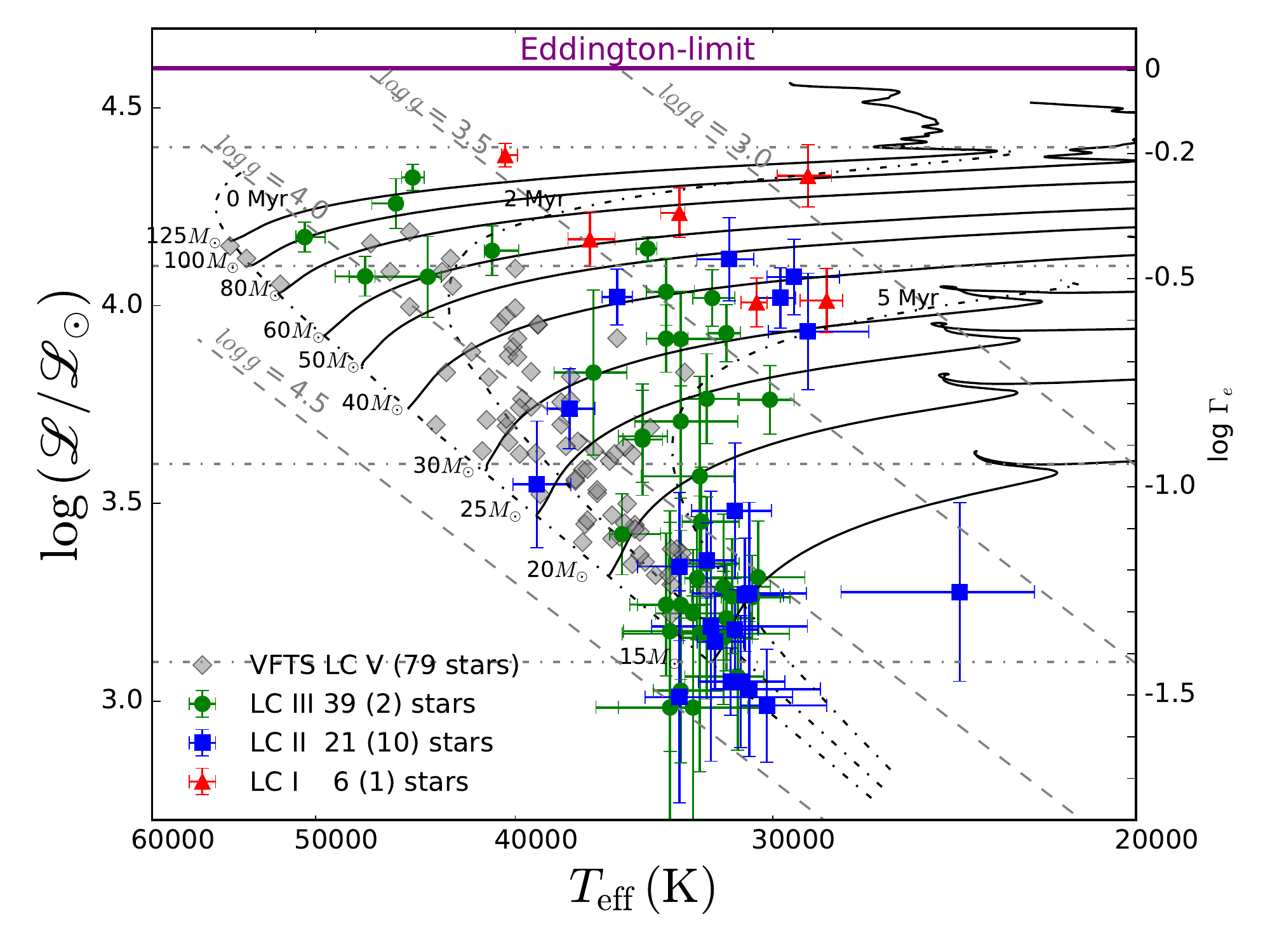}
\caption{
$\log\,g_c$ vs. $\log\,T_{\rm eff}$ (upper panel) and spectroscopic
Hertzsprung-Russell (lower panel) diagrams of the O-type giants, bright giants, and 
supergiants, where $\mathscr{L} \equiv T_{\mathrm{eff}}^{4}/g_{\rm c}$ (see Sect.~\ref{subsec:gravities}). 
Symbols and colors have the same meaning as in Fig.~\ref{fig:spt_teff_combine}.
Evolutionary tracks and isochrones are for models that have an initial rotational velocity
of approximately 200\,\kms\  \citep[][]{brott,kohler2015}. In the lower panel, 
the right-hand axis gives the classical 
Eddington factor $\Gamma_{\rm e}$ for the opacity of free electrons in a fully ionized plasma with solar helium abundance
\citep[c.f.][]{langer_kudritzki2014}. 
The horizontal line at  $\log\,\mathscr{L}/\mathscr{L_{\sun}}\, =\, 4.6$ indicates the location of the corresponding Eddington limit. 
The dashed straight lines are lines of constant $\log\,g$ as indicated.
Lower opacities of the green and blue symbols and the numbers in parentheses in 
the legend have the same meaning as in Fig.~\ref{fig:SpTLC}.
[Color version available online].
}
\label{fig:logg_logT}
\end{figure*}

We present the Newtonian gravities graphically using the $\log\,g_{\rm c}$\,$-$\,$\log\,T_{\mathrm{eff}}$ 
diagram and the spectroscopic Hertzsprung-Russell (sHR) diagram (Fig.~\ref{fig:logg_logT}). In doing so, 
the gravities were corrected for centrifugal acceleration using $\log\,g_{\rm c} = \log\,[g+(\vrot)^2/R]$ 
\citep[see also][]{herrero,repolust2004}.
The sHR diagram shows $\mathscr{L}$ versus $T_\mathrm{eff}$.  $\mathscr{L} \equiv T_\mathrm{eff}^4/g_c$ 
is proportional to $L/M$, thus to $\Gamma_{\rm e} / \kappa$, 
where $\kappa$ is the flux-mean opacity \citep[see][]{langer_kudritzki2014}. For a fixed $\kappa$, 
the vertical axis of this diagram thus sorts the stars according to their proximity to
the Eddington limit: the higher up in the diagram the closer their atmospheres are to 
zero effective gravity \citep[see also][]{castro2014}.

Figure~\ref{fig:logg_logT} shows both diagrams for our stars. We have supplemented them 
with VFTS LC\,V stars analyzed in \citetalias{sabin_sanjulian2014}. 
Stars that evolve away from the ZAMS increase their radii, and hence
decrease their surface gravity. Therefore, it is expected that the different
luminosity classes are separated in these diagrams, that is, stars assigned a 
lower roman numeral are located further from the ZAMS.
This behavior is clearly visible for the supergiants that seem to be the most 
evolved stars along the main sequence. The bright giants mingle 
with the supergiants, though some, at 25$-$30\,\msun\ reside where the dwarf 
stars dominate. They do not appear to form a well defined regime intermediate between
giants and supergiants, though it should be mentioned that the sample size of these
stars is small.

At initial masses of approximately 60\,\msun\ and higher,  
giants and bright giants appear closer to the ZAMS.
This is the result of a relatively high mass-loss rate, as the morphology of 
He\,{\sc ii}\,$\lambda$4686 -- the main diagnostic used to assign luminosity class -- traces wind density. 
At initial masses in-between approximately 18\,\msun\ and 60\,\msun, the dwarf phase clearly precedes
the giant and bright giant phase. However, at lower initial masses the picture
is more complicated. Here a group of late-O III and II stars populate the regime
relatively close to the ZAMS, where dwarf stars are expected. The properties of 
these stars are indeed more characteristic for LC V objects; they have gravities 
$\log\,g_{\rm c}$ between 4.0 and 4.5 and radii of approximately 5$-$8\,\rsun. Consequently,
their absolute visual magnitudes are fainter than calibrations suggest 
\citep{walborn1973}. In addition, these objects display higher spectroscopic masses than 
evolutionary masses (see Sect.~\ref{subsec:mass_disc} and Fig.~\ref{fig:discrep_mass_He}).

What could explain this peculiar group of stars? Though we do not want
to exclude the possibility that these objects belong to a separate physical group, 
we do find that they populate a part of the HRD where few dwarf O stars are actually
seen (see Sect.~\ref{ref:hrd}). A simple explanation may thus be an 
intricacy with the LC classification.

The spectral classification in the VFTS is described in \citet{walborn2014}. For the late-O stars, following, for example 
\citet{sota2011}, it relies  on the equivalent width ratio 
\hel{ii}{4686} / \hel{i}{4713} as its primary luminosity criterion. The relative strength 
of \SSil{iv} to \hhel{i} absorption lines may serve as a secondary criteria, a 
measure that is somewhat susceptible to metallicity effects \citep[see][]{walborn2014}.
The \SSil{iv} / \hhel{i} ratio is however the primary classifier in 
early-B stars.

Though \hel{ii}{4686} / \hel{i}{4713} is the primary criterion in the VFTS, the group of 
problematic stars being discussed here have \SSil{iv} weaker than expected for LC III,
favoring a dwarf or sub-giant classification. Indeed, it is often for this reason that 
the classification of these stars is lower rated in \citet{walborn2014}. Other reasons for the problematic
classification of these stars may be relatively poor quality spectra and an inconspicuous binary nature. 
Regarding the latter possibility, 
we mention that a similar behavior is seen in some Galactic O stars, 
as discussed in \citet{sota2014} and in the third paper of the Galactic O-Star 
Spectroscopic Survey series \citep[][]{jesus_apellaniz_et_al_2016}. More specifically,  
it is seen in the A component of $\sigma$\,Ori\,AB, that this star itself is a spectroscopic 
binary. For this system, \citet{simon-diaz2011,simon-diaz2015} find that the spectrum is 
the composite of that of an O9.5\,V and B0.5\,V star. 
Further RV monitoring has indeed revealed 
that some of these late-O\,III and II stars are genuine spectroscopic binaries (see annotations in Appendix~\ref{app:remarks}).

\subsection{Helium abundance}
\label{subsec:yhe} 

\begin{figure}
\centering
\includegraphics[width=\columnwidth]{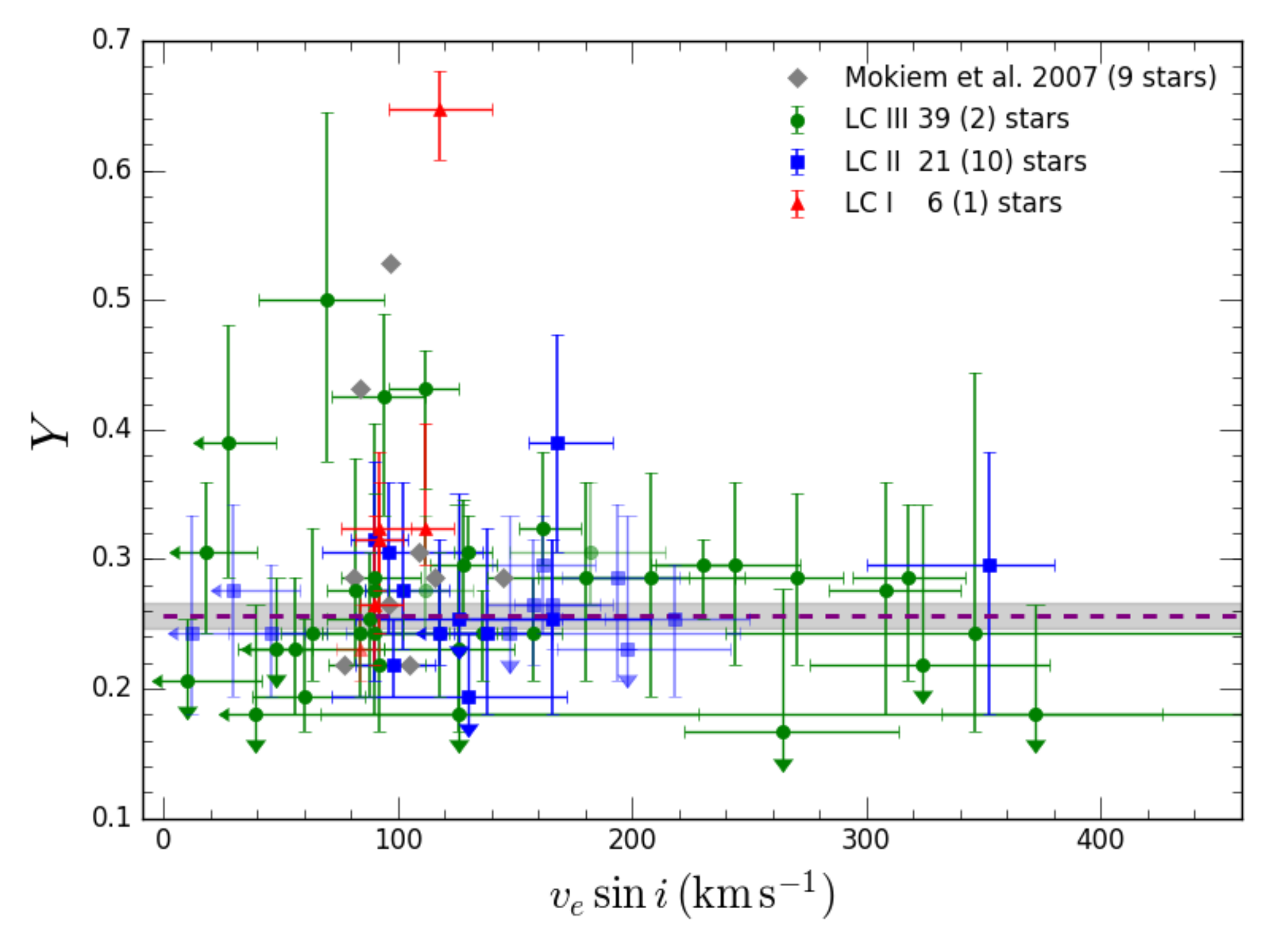}
\includegraphics[width=\columnwidth]{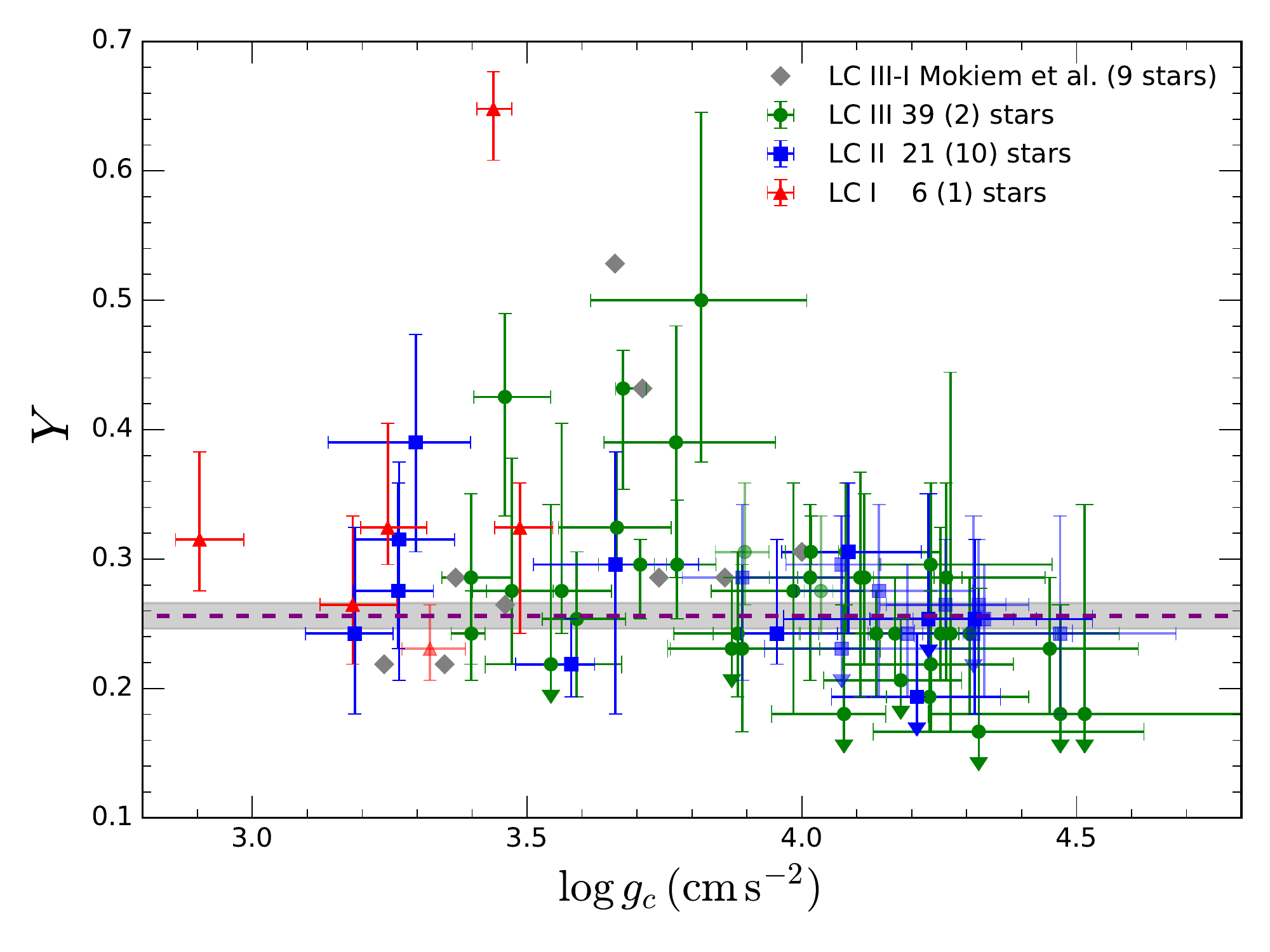}
\caption{ 
Helium mass fraction $Y$ versus \vrot\ (upper panel) and $\log\,g_c$ (lower panel). Symbols and colors have the same meaning as in Fig.~\ref{fig:spt_teff_combine}. 
Gray diamonds denote stars with LC\,III to I from \citet{mokiem2007}.
The purple dashed line at $Y\,=\,0.255$ defines the initial composition for LMC stars; the gray bar
is the 3$\sigma$ uncertainty in this number.} 
\label{fig:heliumabund}
\end{figure}

Figure~\ref{fig:heliumabund} shows the helium mass fraction $Y$ 
as a function of \vrot\ (top panel) and $\log\,g_c$ (bottom panel).  Most of the stars in our sample
agree within their 95\%\ confidence intervals with the initial composition of the LMC, $Y\,=\,0.255 \pm 0.003$,
which has been derived by scaling the primordial value \citep{peimbert2007} linearly with
metallicity \citep{brott}.

Table~\ref{table:yhe} summarizes the frequency of stars in the total sample and in
given sub-populations that 
have $Y$ larger, by at least $2\sigma_Y$, than a specified limit. We find 
that 92.4\%\ of our 66-star sample does not show 
a clear signature of enrichment given the uncertainties, that is, \ has $Y-2\sigma_Y\,\leq\,0.30$.
 Five stars (VFTS\,046, 180,  518,  546, and  819), hence 7.6\%\ of our sample, meet the requirement of $Y-2\sigma_Y > 0.30$ for a clear signature of enrichment. Interestingly, all these sources have
a projected spin velocity less than 200\,\kms\ 
(see upper panel Fig.~\ref{fig:heliumabund}).
The lower panel of Fig.~\ref{fig:heliumabund} plots helium abundance as
a function of surface gravity. All sources with $Y-2\sigma_Y\,>\,0.35$ have
gravities less than or equal to 3.83 dex, though not all sources 
that have such low gravities have $Y-2\sigma_Y\,>\,0.35$. 
This conclusion does not change if we take $\log\,g_c$ instead.

\begin{table}
\caption{Frequency of stars from different sub-samples that display a helium abundance by mass ($Y$) larger 
than the specified limit by at least $2\sigma_Y$. The sample consists of 66 sources. We provide the number of stars with ambiguous LC in parentheses. 
The error bars indicate the 68\%-confidence intervals on the given fractions and were computed using simulated samples and binomial statistics.}             
\label{table:yhe}      
\centering                          
\begin{tabular}{clll}        
\hline\hline\\[-8pt]                 
 &  \multicolumn{3}{c}{f($Y$)}  \\[1pt] \\[-8pt]  \hline  
Sample  &  $>$ 0.30 &  $>$ 0.35 & $>$ 0.40  \\[1pt]  \hline\\[-8pt]  
LC\,III 39 (2)  stars & 0.08 $\pm$ 0.04 & 0.05 $\pm$ 0.04      & 0.00 $\pm$ {\it n/a} \\
LC\,II  21 (10) stars & 0.05 $\pm$ 0.05 & 0.00 $\pm$ {\it n/a} & 0.00 $\pm$ {\it n/a} \\
LC\,I    6  (1) stars & 0.17 $\pm$ 0.15 & 0.17 $\pm$ 0.15      & 0.17 $\pm$ 0.15 \\ 
LC\,III to I 66 (13) stars & 0.08 $\pm$ 0.03 & 0.05 $\pm$ 0.03      & 0.02 $\pm$ 0.02 \\
\hline                       
\end{tabular}
\end{table}

We ran Monte-Carlo simulations to estimate the number of spuriously detected He-rich stars in our
sample, that is, the number of stars that have normal He-abundance but for which the high $Y$ value 
obtained  may purely result from statistical fluctuations in the measurement process. Given our sample size and 
measurement errors, we obtained a median number of two spurious detections. Within a 90\%\ 
confidence interval, this number varies between zero and three. 
While some detections of He-rich stars in our sample may thus result from 
statistical fluctuations, it is unlikely that all 
detections are spurious.

Further, some of the stars appear to have a sub-primordial helium abundance. This is
thought to be unphysical, possibly indicating an issue with the analysis 
such as continuum dilution. Continuum dilution may be caused by multiplicity
(either through physical companions or additional members of an unresolved stellar association) and nebular continuum emission, 
contributing extra flux in the Medusa fiber. In the former case, the extra continuum flux of the 
companion may weaken the lines, essentially mimicking
an unrealistically low helium content.
Alternative explanations may be linked to effects of magnetic fields and of (non-radial)
pulsations, though, at the present time, little is known about the impact of these processes
on the (apparent) surface helium abundance.

\subsubsection{The dependence of $Y$ on the mass-loss rate and rotation rate}

\begin{figure}
\centering
\includegraphics[width=\columnwidth]{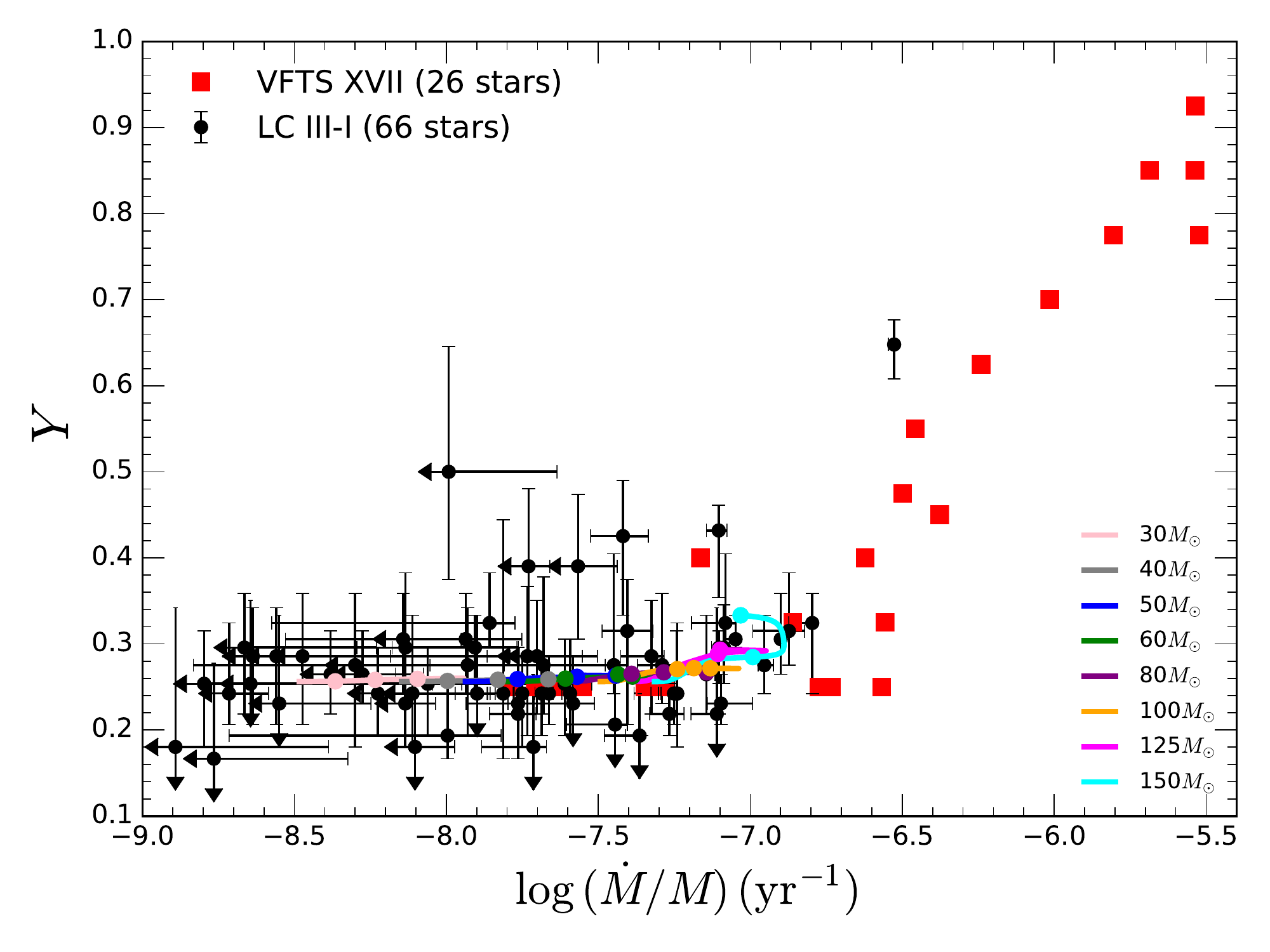}
\includegraphics[width=\columnwidth]{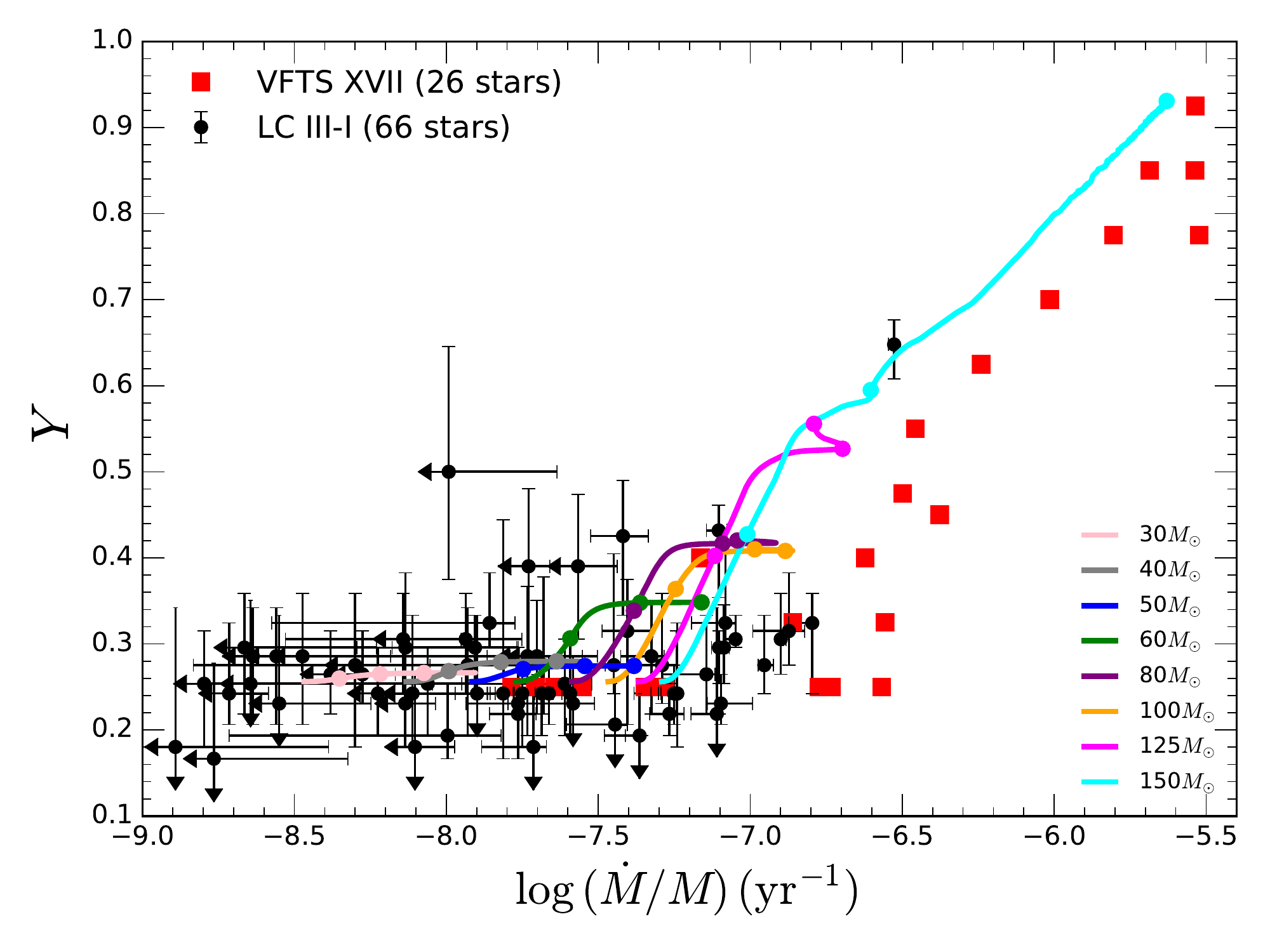}
\caption{ 
Helium mass fraction $Y$ versus the empirical (unclumped) mass-loss rate
relative to the stellar mass $(\dot{M}/M)$ for our sample stars 
with their respective 95\% confidence intervals. Added to this is the 
set of very massive luminous O, Of, Of/WN and WNh stars analyzed in 
\citetalias{bestenlehner2014}, excluding the nine stars in common with this
paper. Also shown are evolutionary tracks by \citet{brott} and \citet{kohler2015}
for stars with initial spin rates of approximately 200\,\kms\  (upper panel) and 300\,\kms\  (lower panel) with dots every 1\,Myr of evolution. 
These tracks are truncated at 30\,kK, which is approximately 
the temperature where the stars evolve into B-type objects and thus are no
longer part of our observational sample.
} 
\label{fig:heliumabund_massloss}
\end{figure}

As a relatively low surface gravity ($\log\,g_{\rm c} \leq 3.83$) seems a prerequisite 
for surface helium 
enrichment, envelope stripping through stellar winds 
may be responsible for the high $Y$. 
To investigate this possibility we plot $Y$ versus the mass-loss rate relative to the mass
of the star ($\dot{M}/M$) in Fig.~\ref{fig:heliumabund_massloss}. Here, we adopt $M_\mathrm{spec}$ as a proxy for the mass; using the evolutionary mass $M_{\rm evol}$ yields similar results. The  quantity $\dot{M}/M_\mathrm{spec}$
is the reciprocal of the momentary stellar evaporation timescale. 
Also plotted are the set of 26 very massive O, Of, Of/WN, and WNh stars (VMS) analyzed  
in \citetalias{bestenlehner2014}. At $\log\,(\dot{M} / M)$\,$\ga$\,$-7$,  
these stars display a clear correlation with helium abundance.
This led \citet{bestenlehner2014} to hypothesize that, in this regime, 
mass loss is exposing helium enriched layers. 

To explore this further, we compare the data with the main-sequence predictions for $Y$ versus $\dot{M}/M$ 
by \citet{brott} and \citet{kohler2015} for massive stars in the range 
of 30$-$150 \msun. So far, this is the only set of tracks at LMC metallicity that includes 
rotation and that covers a wide range of initial spin rates.
The plotted tracks have been truncated at 30\,kK, that is, 
approximately where the stars evolve into B-type (super)giants and thus leave our observational sample.

The empirical mass-loss rates used to construct this diagram (i.e., the data points)
assume a homogeneous outflow. In Sect.~\ref{subsec:WLD} we discuss wind clumping, there we point
out that for the stars studied here our optical wind diagnostics can be 
reconciled with wind-strength predictions as used in the evolutionary calculations 
if the empirical $\log\,\dot{M}$ values are reduced by $\sim$0.4\,dex. Hence, in Fig.~\ref{fig:heliumabund_massloss}, the empirical measurements of
$\log\,(\dot{M}/M)$ should also be reduced by this amount. 
Regarding the $\log\,(\dot{M}/M)$ measurements 
of \citetalias{bestenlehner2014} (the red squares in Fig.~\ref{fig:heliumabund_massloss}), 
these should also be shifted to lower values. Yet, as the mass estimates 
obtained in \citetalias{bestenlehner2014} were upper limits and not actual measurements, the 
reduction in $\log\,(\dot{M}/M)$ of these stars may be
limited to $\sim$0.2$-$0.4 dex assuming similar clumping properties in Of, Of/WN and WNh stars as 
applied for O stars.

The upper panel in Fig.~\ref{fig:heliumabund_massloss} shows tracks for initial spin velocities close to 200\,\kms. Within the 
framework of the current models, no significant enrichment is expected in the O or WNh 
phase, with the possible exception of stars initially more massive than $\sim$150\,\msun.
We add that mass-loss prescriptions adopted in the evolutionary tracks discussed here account
for a bi-stability jump at spectral type B1.5, where the mass-loss rate is predicted to strongly increase 
\citep{vink1999}. Beyond the bi-stability jump stars initially more massive than 
$\sim$60$-$80\,\msun\ do show strong helium enrichment but, by then, the stars have already left our O\,III-I sample. 

The lower panel in Fig.~\ref{fig:heliumabund_massloss} shows O-star tracks for an initial
spin rate of approximately 300\,\kms. In this case, the \citet{kohler2015} models do predict an
increase in $Y$ during the O star phase for initial masses $\sim$60\,\msun\ and up. 
Initially, they spin so fast that rotationally-induced mixing prevents the 
build-up of a steep chemical gradient at the core boundary. 
The lack of such a barrier explains the initial rise in $Y$. However, as a result of loss of angular momentum via the stellar wind and the associated spin-down of the star, a chemical gradient barrier may develop during its main-sequence evolution. Such a gradient effectively acts as a `wall' inhibiting the transport of helium to the surface. This can be seen in Fig.~\ref{fig:heliumabund_massloss}  as 
a flattening of the Y increase with time. 
Once such a barrier develops, the star starts to evolve to cooler temperatures, an
evolution that was prohibited in the preceding phase of quasi-chemically homogeneous evolution.
Once redward evolution commences, stripping of the envelope by mass loss may aid
in increasing the surface helium abundance. In our tracks this is only significant
for initial masses 125\,\msun\ and up.

Finally, our findings might indicate that the current implementation of 
rotational mixing and wind stripping in single-star models is not able to justify the $Y$ abundances of most of the 
helium enriched stars in our sample. 
In the following subsection we combine the constraints on the helium abundance with 
the projected spin rate of the star and its position 
in the Hertzsprung-Russell diagram to further scrutinize the evolutionary models.

\vspace{6mm}

\subsection{Hertzsprung-Russell diagram}
\label{ref:hrd}

\begin{figure}
\centering
\includegraphics[width=\columnwidth]{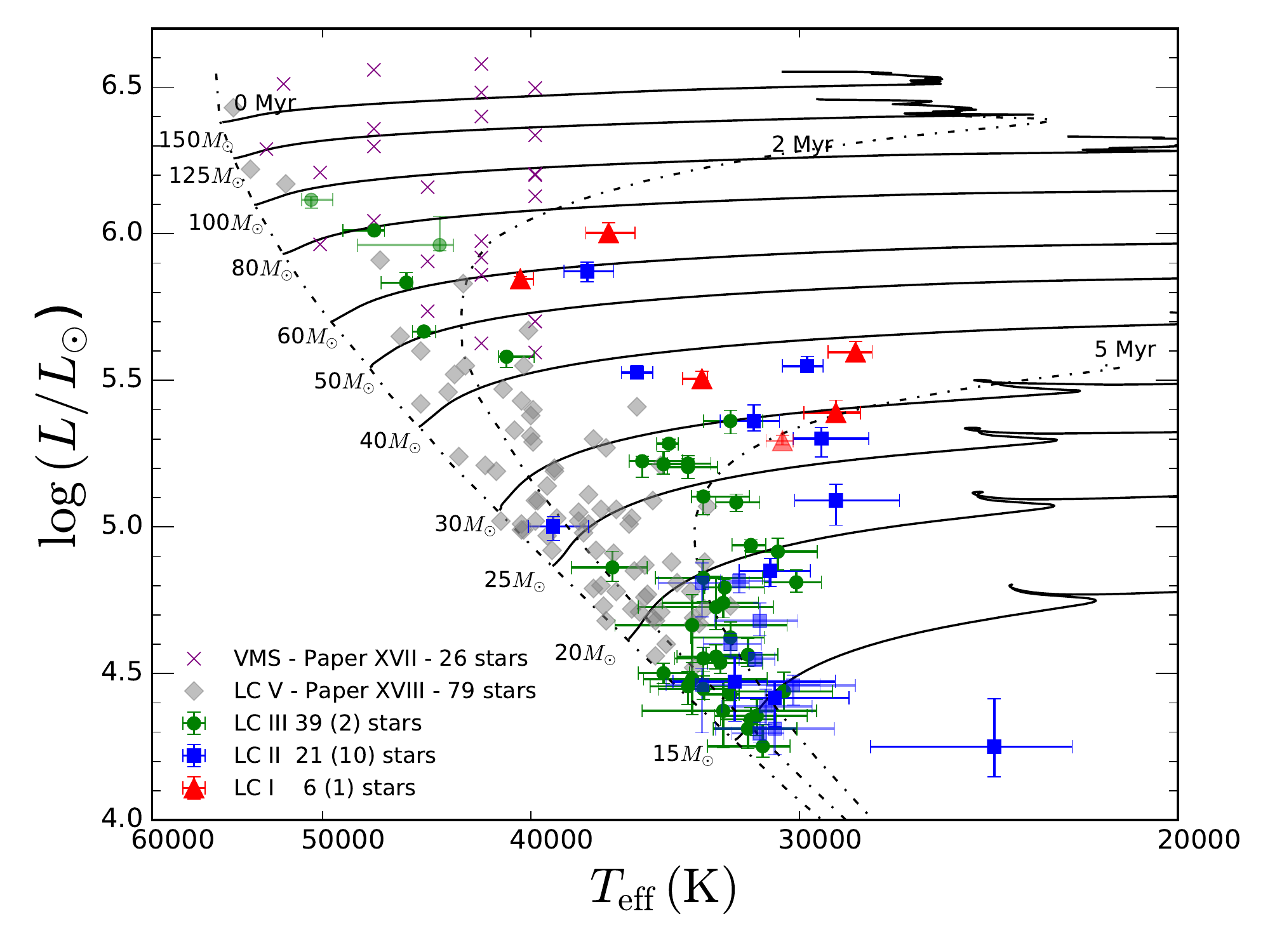}
\includegraphics[scale=0.47]{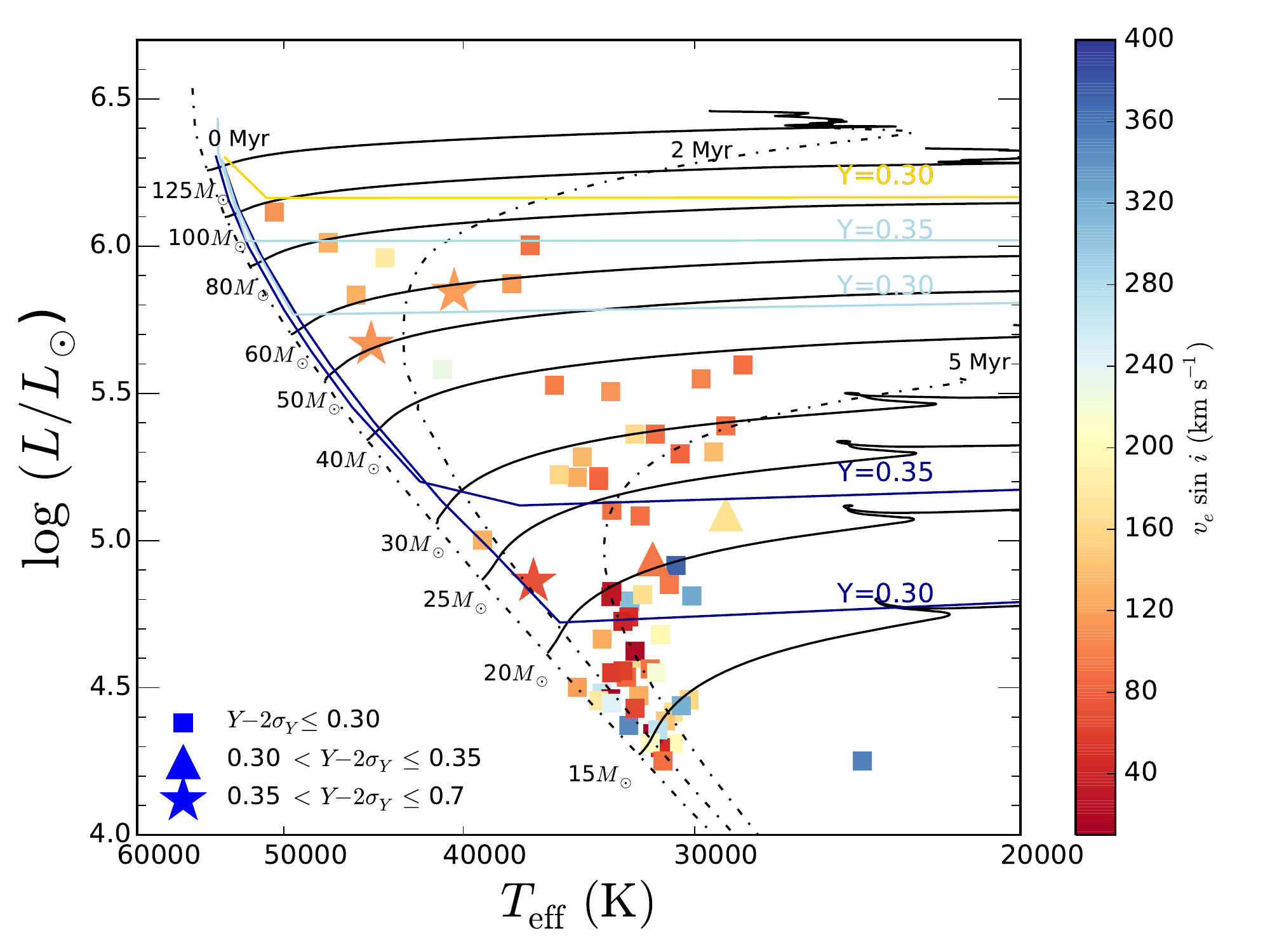}
\caption{Two versions of the Hertzsprung-Russell diagram. In the top panel our sample of single O-type
giants and supergiants is supplemented with the dwarf O star sample of \citetalias{sabin_sanjulian2014} and
the sample of very massive Of and WNh stars by \citetalias{bestenlehner2014}.
We exclude the results of \citetalias{bestenlehner2014} for the nine stars in common with this paper and adopted our results (see Sect.~\ref{sec:high_lum}).
Symbols and colors have the same meaning as in Fig.~\ref{fig:logg_logT}. 
The lower panel only contains the sample studied here. The symbol shapes in the lower panel show three categories of helium mass
fraction, that is, not enriched (squares), moderately enriched (triangles), and enriched (stars). 
The symbol colors refer to their projected rotational velocity (see color bar on the right). 
Evolutionary tracks and isochrones are for models that have an initial rotational velocity
of approximately 200\,\kms\  \citep[][]{brott,kohler2015}. 
Iso-helium lines for different rotational velocities
are from \citet[][]{kohler2015} and are color-coded using the color bar.
}
\label{fig:HR}
\end{figure}

In this section, we explore the evolutionary status of our sample stars 
by means of the Hertzsprung-Russell diagram (see Fig.~\ref{fig:HR}).  
Two versions of the HRD are shown in Fig.~\ref{fig:HR}. In the top panel, our sample of 
giants, bright giants, 
and supergiants is complemented with the VFTS samples of very massive stars (VMS) from
\citetalias[][]{bestenlehner2014} and of LC\,V stars from \citetalias[][]{sabin_sanjulian2014}.
VMS populate the upper left part of the HRD.
Giants, bright giants, and supergiants are predominantly located in between 
the 2 and 5\,Myrs isochrones while dwarfs are found closer to the ZAMS.
The location of LC\,V stars compared to III, II and I stars reflects their higher surface gravities
as shown in Fig.~\ref{fig:logg_logT}. At the lowest luminosities, we note a predominance of LC III and II stars and an
absence of LC V stars. As discussed in Sect.~\ref{subsec:gravities}
this may reflect a classification issue.

The positions of the O stars in the HRD do not reveal an obvious preferred age but 
rather show a spread of ages, supporting findings of \citet[][]{demarchi2011}, \citet{Cioni2015a}, and \citet{sabbi2015b}. 
HRDs of each of the spatial sub-populations defined in Sect.~\ref{sec:sample} do not
point to preferred ages either (see Appendix~\ref{sec:hrd_clusters} and Fig.~\ref{fig:HR_clusters}), suggesting 
that  star formation has been sustained for the last 5~Myr at least throughout the Tarantula region.
We stress that
the central 15" of Radcliffe 136, the core cluster of NGC\,2070, is excluded from the VFTS sample.
The age distribution of the Tarantula massive stars will be investigated in detail in a subsequent
paper in the VFTS series (Schneider et al., in prep.).

In the lower panel of Fig.~\ref{fig:HR}, we include information on $Y$ and \vrot\ for our sample stars.
We also include iso-helium lines for $Y$\,=\,0.30 and 0.35 as a function of initial rotational 
velocity \citep[see figure 10 of][]{kohler2015}. 
According to these tracks, main-sequence stars initially less massive than $\sim$100\,\msun\ with 
initial rotation rates of 200\,\kms\ or less are not 
expected to show significant helium surface enrichment, that is, $Y\,<\,0.30$. 
Stars with an initial rotation rate of 300\,\kms\ are only supposed to reach detectable helium enrichment in
the O star phase if they are
initially at least 60\,\msun. 
Helium enrichment is common for 20\,\msun\ stars and up if they spin extremely fast at birth
(\veq\,$ > 400$~\kms). Below we discuss how this compares with our sample stars.

First, our finding that all helium enriched stars have a present day projected spin 
rate of less than 200\,\kms\ (see also Fig.~\ref{fig:heliumabund}) appears at odds with 
the predictions of the tracks referred to above. 
In the LMC, significant spin-down due to angular-momentum loss 
through the stellar wind and/or secular expansion is only expected
by \citet{brott} and \citet{kohler2015} for stars 
initially more massive than $\sim$40 \msun, once these objects 
evolve into early-B supergiants \citep{vink2010}.
Only for much higher initial mass are the winds 
sufficiently strong to cause rotational braking during the O-star phase. 
This could perhaps help explain 
the two highest-luminosity He-enriched objects, VFTS\,180 and 518,
though in the context of our models this requires an initial spin of 400\,\kms\
and wind strengths typical for at least $\sim$125\,\msun\ stars. 
Their evolutionary masses are at most 50\,\msun. It is furthermore extremely 
unlikely that the remaining three 
He-enriched stars at lower luminosity (having initial masses $< 40\,\msun$) 
spin at 400\,\kms\ and are all seen almost pole-on. For the two hot He-enriched stars 
VFTS\,180 and 518 we included a set of nitrogen diagnostic lines (see Sect.~\ref{subsec:GA}). Interestingly, we find that they are
nitrogen enriched as well (i.e., [N] $>$ 8.5). A thorough nitrogen analysis of the full sample is presented by 
\citet[][see also Summary]{grin_et_al_2016}.

If indeed these are main-sequence (core H-burning) stars that live their life in isolation,
rotational mixing, as implemented in the evolutionary predictions 
employed here, cannot explain the surface helium mass fraction in this particular subset of stars. 
This would point to deficiencies 
in the physical treatment of mixing processes in the stellar interior.

Alternatively, the high helium
abundances could point to a binary history 
\citep[e.g., mass transfer or even merger events; see e.g.,][]{selma2014,bestenlehner2014}
 or post-red supergiant (post-RSG) evolution.
Concerning the former option, one of these sources is VFTS\,399, which has been 
identified as an X-ray binary by \citet{clark2015}. 
Concerning the latter option, LMC evolutionary tracks that account for 
rotation and that cover the core-He burning phase have been computed by
\citet{meynet2005}. These tracks indicate that a brief part of the evolution
of stars initially more massive than 25\,\msun\ may be spent as post-RSG stars
hotter than 30\,000\,K. However, these exceptional 
stars would be close to the
end of core-helium burning and feature much higher 
helium (and nitrogen) surface abundances. 

Second, while we have only a few fast rotators, these stars do not seem to be helium enriched 
(see again Fig.~\ref{fig:heliumabund}). All of them have masses below 20~\msun, therefore 
no significant helium enrichment is expected, in agreement with our measurements.
If such fast rotators are spun-up secondaries resulting from binary interaction 
\citep[e.g.,][]{ramirezagudelo,selma}, then the interaction process should have been 
helium neutral. Some of the stars appear to have
sub-primordial helium abundances. This could also be an indication of present-day binarity 
(see Sect.~\ref{subsec:yhe}). Among them are some of the fastest spinning objects, 
consistent with the latter conjecture.

\subsection{Mass loss and modified wind momentum}
\label{subsec:WLD}

In the optical, the mass-loss rate determination relies on wind infilling in H$\alpha$ and 
\ion{He}{ii}\,$\lambda4686$. These recombination lines are indeed sensitive to the invariant wind-strength 
parameter $Q\,=\,\dot{M}/(R \varv_{\infty})^{3/2}$ that is inferred from the spectral analysis
\citep[see, e.g.,][]{puls1996,dekoter1998}. For approximately 40\%\ of our sample 
only upper limits on $\dot{M}$ can be determined. 
These stars mostly have $\dot{M} < 10^{-7}$\,\msunyr\  and 
$\log\,\left(L/L_{\sun}\right) < 5.0$. 
This group of relatively modest-mass stars
($M_{\rm spec}\,\leq\,25\,M_{\odot}$) 
is excluded from the analysis presented in this section.

\begin{figure}[t!]
\centering
\includegraphics[width=\columnwidth]{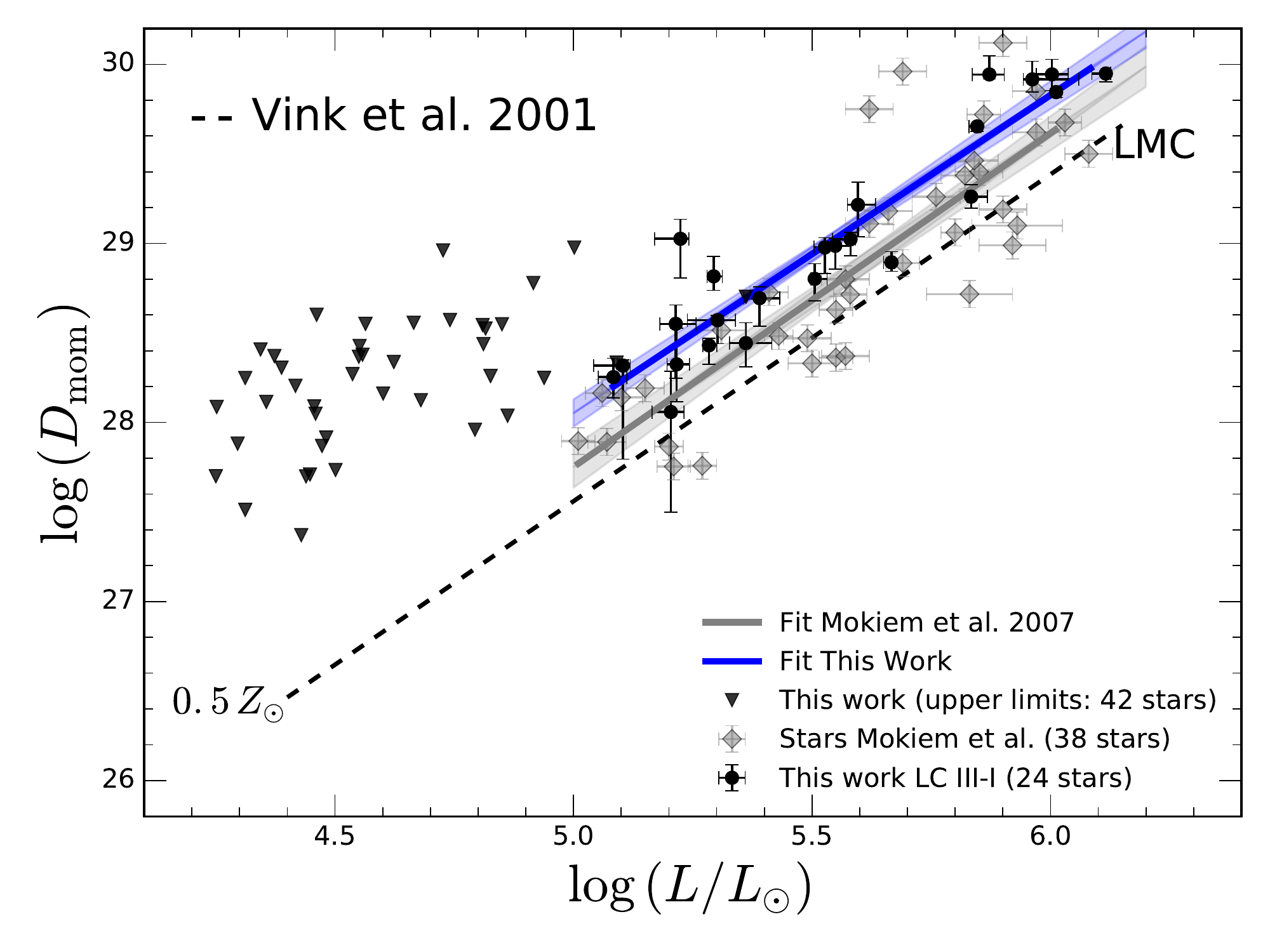}
\includegraphics[width=\columnwidth]{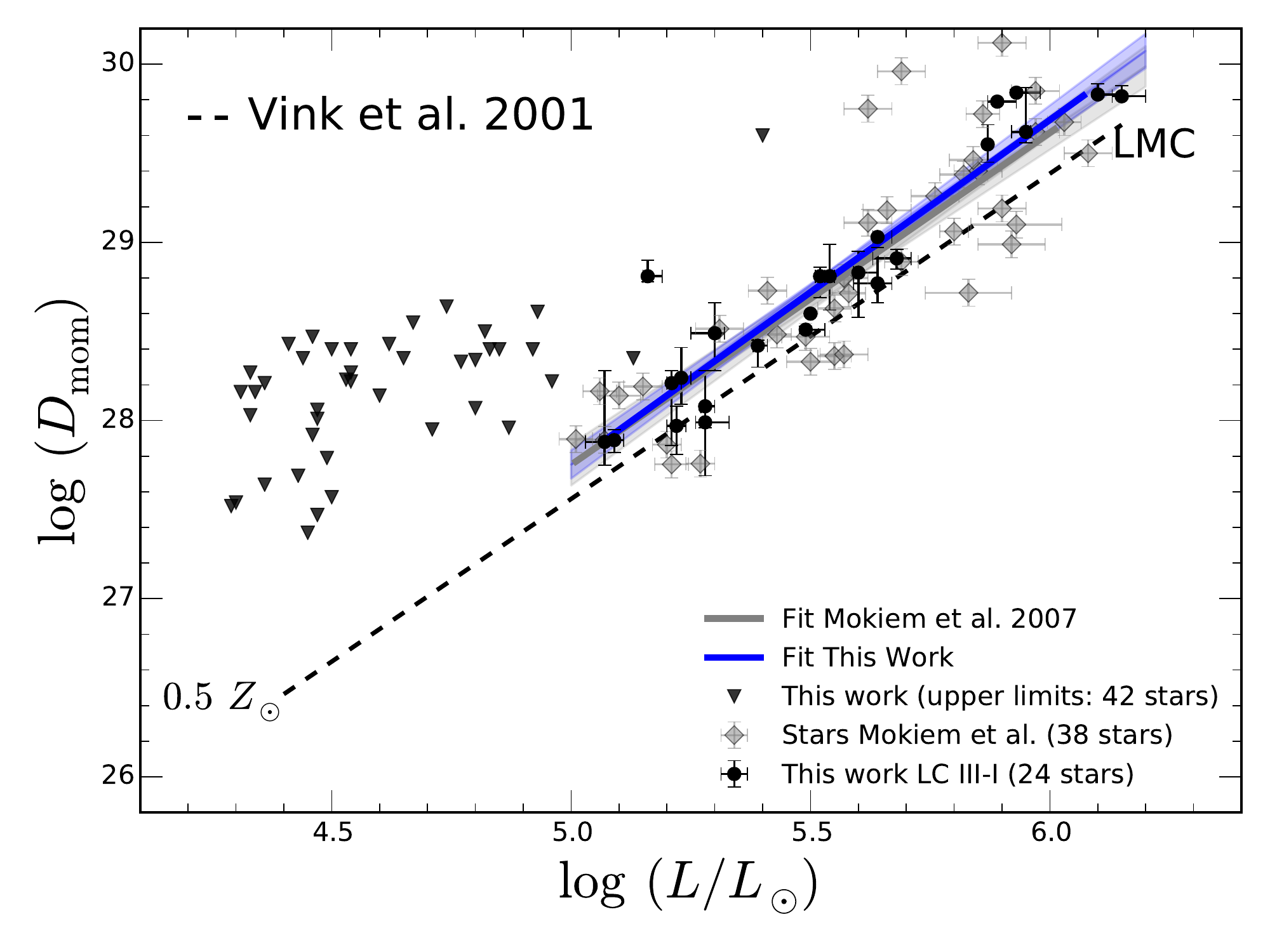}
\caption{Modified wind momentum ($D_{\rm mom}$) vs. luminosity diagram. 
The dashed lines indicate the theoretical predictions of \citet{vink2001} 
for homogeneous winds. {\em Top panel}: the empirical 
fit for this work and \citet{mokiem2007} (both for $L/L_{\sun} > 5.0$) in shaded blue
and gray bars, respectively. For stars with $L/L_{\sun} \leq\ 5.0$, only
upper limits could be constrained. These stars are not considered in the analysis. 
{\em Bottom panel}: same as top panel but now for an analysis in which
the acceleration of the wind flow, $\beta$, is a free parameter and in which the analysis does not
include nitrogen lines, but relies on hydrogen and helium lines only.
[Color version available online].  
}
\label{fig:wld}
\end{figure}

To facilitate a comparison of the mass-loss rates of the remaining stars with theoretical results, we use
 the modified wind momentum luminosity diagram (WLD; Fig.~\ref{fig:wld}).
The modified wind momentum 
$D_{\rm mom}$ is defined in Sect.~\ref{sec:derived_properties}.
For a given metallicity, $D_{\rm mom}$ is predicted to be a 
power-law of the stellar luminosity, that is,
\begin{equation}
\log\,D_{\mathrm{mom}} = x\log\left(L_{*}/L_{\sun}\right) + \log\,D_{0},
\end{equation}
where $x$ is the inverse of the slope of the line-strength distribution function corrected 
for ionization effects \citep{puls2000}. For a metal content of solar 
down to $\sim$1/5th solar, $x$ and $D_{0}$ 
do not depend on spectral type for the parameter range 
considered here, which allows for a simple (i.e., power-law) prescription of 
the mass-loss metallicity dependence.

The top panel of Fig.~\ref{fig:wld} shows the WLD for our sample, where $D_{\rm mom}$ is in 
the usual units of g\,cm\,s$^{-2}$. 
Upper limits for the weak-wind stars are also shown (see legend).
A linear fit to the stars for which we have a constraint 
on the mass-loss rate is given in blue, with the shaded blue area representing the uncertainty as
a result of errors in $D_{\rm mom}$.
Also plotted in the figure are the results of \citet{mokiem2007b} for 38 stars
in total in the LMC (sub-)sample, 16 of which are in N11. 
Our results exhibit somewhat higher $D_{\mathrm{mom}}$ values 
than those of \citet{mokiem2007b}.
The reason for this discrepancy 
is illustrated in the lower panel, where we have repeated our analysis applying 
identical fitting constraints as \citeauthor{mokiem2007b} 
This implies that we let $\beta$ be a free parameter and have removed the 
nitrogen lines from our set of diagnostics. In that case, we 
recover essentially the same result. As pointed out in
Sect.~\ref{sec:limitations}, allowing the  method to constrain the 
slope of the velocity law yields higher $\beta$ values compared to 
the adopted values, that is, those based on theoretical
considerations, for a substantial fraction of the stars. 
A shallower velocity stratification 
(that is, a higher $\beta$) in the H$\alpha$ and He\,{\sc ii}\,$\lambda$4686 forming regions 
corresponds to a higher density for the same $\dot{M}$. As the recombination
lines depend on the square of the density,  
the emission will be stronger (at least in the central
regions of the profile). Hence, to fit the profiles compared to a lower $\beta$, one
needs to reduce the mass loss in the models. 

When compared to the theoretical predictions of \citet{vink2001}, who apply the
same prescription to estimate $\varv_{\infty}$ as used here (see Sect.~\ref{subsec:GA}), our 
strong-wind stars
show higher $D_{\rm mom}$ values. This is interpreted as being due to inhomogeneities
in the outflow, usually referred to as clumping. Empirical evidence for clumpy outflows
has been presented by \citet{eversberg1998}, \citet{lepine2008}, and \citet{prinja2010}, for example. 
If the winds are clumped, disregarding this effect would lead to an 
overestimation of the 
empirical H$\alpha$ or He\,{\sc ii}\,$\lambda$4686-based mass-loss rate by a factor of
$f_{\rm V}^{-1/2}$, where $f_{\rm V}$ is the clump 
volume-filling factor. This assumes 
the clumps to be optically thin for the considered diagnostic lines, and the inter-clump medium to be void. To reconcile our 
results with theory would require $f_{\rm V} \sim$ 1/8 to 1/6, reducing
the mass-loss rate by a factor of 2.8. This is a somewhat stronger reduction than implied by the volume-filling factors
$f_{\rm V} \sim$ 1/2 to 1/3 found by \citet{mokiem2007b}, which correspond to a reduction in $\dot{M}$ relative to a homogeneous
outflow of approximately a factor of 1.5. 

\begin{table}[t!]
\caption{Coefficients describing empirical and theoretical modified-wind momentum relations.}           
\label{table:2new}      
\centering                          
\begin{tabular}{lcc}        
\hline\hline\\[-8pt]                 
 Sample                           &  Slope                       & Intercept  \\[1pt]  \hline\\[-9pt]    
 {\em Empirical} & & \\
 This work                       & 1.78 $\pm$ 0.14     & 19.17 $\pm$ 0.79 \\ 
 \citet{mokiem2007b}             & 1.87 $\pm$ 0.19     & 18.30 $\pm$ 1.04 \\[2pt]
\hline\\[-9pt]
 {\em Theoretical LMC relation} & & \\
\citet{vink2001}             & 1.83 $\pm$ 0.04     & 18.43 $\pm$ 0.26 \\ 
\hline
\end{tabular}
\end{table}

Placing constraints on the properties of the clumps
in the H$\alpha$ and He\,{\sc ii}\,$\lambda$4686-forming region relies 
on the accuracy of the theoretical
mass-loss rates but does not imply in any way that the \citet{vink2001} predictions are correct.
Critical assumptions in these theoretical results are that $\dot{M}$ relies on a
global energy conservation argument \citep[see][]{abbott1985,dekoter1997} and that the
outflow is homogeneous.
For the strong-wind stars investigated here, \citet{muijres2012} showed that 
wind solutions based on a detailed treatment of the line force yielded mass-loss 
rates to within 0.1 dex when adopting the same 
terminal flow velocities, supporting the reliability of the global energy conservation assumption
applied by \citet{vink2001}. If the material in the outflow would be concentrated
in relatively few and strongly over-dense clumps, porosity effects may cause photons
to escape `in-between the clumps' reducing the line-driving force and hence the
mass-loss rate \citep[for corresponding scaling relations, see][]{sundqvist_et_al_2014}. 
However, \citet{muijres2011} demonstrated that for clumps that are
smaller than 1/100th of the local density scale height, thought to represent physically
realistic situations, such effects are not significant
for volume-filling factors as low as approximately 1/30. 

As for the empirically derived filling factors, \citet{massa2003} and \citet{fullerton2006},
by analyzing the P\,{\sc v}\,$\lambda 1118,1128$ resonance line doublet, find cases where
$f_{\rm V}$ reaches values as low as 1/100, as does \citet{najarro2011}. 
\citet{bouret2003,bouret2005,bouret2013} derive clumping 
properties from O\,{\sc vi}\,$\lambda1371$, with a mode of 1/10 but also reporting extremely low 
volume-filling factors in some cases. Extending the original work from \citet{oskinova2007} to 3D
simulations, \citet{surlan2013} point out that the assumption of
optically thin clumps breaks down for the phosphorous lines, 
showing that for a distribution
of clump optical depths, a match to both the strength of P\,{\sc v} and H$\alpha$ is found
for much larger $f_{\rm V}$. These authors present such matches for an assumed $f_{\rm V} = 1/10$,
but we note that simultaneous fits may also be realised for somewhat larger filling factors.
\citet{sundqvist2010,sundqvist2011} compute stochastic wind models, allowing also for porosity in 
velocity space and a non-void interclump medium. For the case of $\lambda$\,Cep their results imply a mass-loss rate
that is half of that predicted by \citet{vink2001} and $f_{\rm{V}}$ values
larger than 1/30. Finally, for O stars brighter than $L = 10^{5}\,L_{\odot}$ a model independent mass-loss
constraint that can be obtained from stars that have spectral morphologies in transition from
Of to Wolf-Rayet type, that is, Of/WNh stars, points to volume-filling factor  
$f_{\rm V} \sim 1/10$ \citep{vink2012}.

We conclude that the mass-loss rate predictions of \cite{vink2001} for LMC
metallicity are consistent with H$\alpha$ and He\,{\sc ii}\,$\lambda$4686-based 
wind volume-filling factors of $f_{\rm V} \sim$ 1/8 to 1/6 and that such
volume-filling factors appear to be in reasonable 
agreement with empirical constraints that rely on models that account for
optical depth effects in the clumps and porosity of the wind medium. 

\subsection{Mass discrepancy}
\label{subsec:mass_disc}

The discourse on the mass discrepancy in massive 
stars, triggered by the work on Galactic stars by \citet{groenewegen_and_lamers1989} and \citet{herrero}, 
is extensive and a general consensus on the topic is yet to be reached. If present, the 
discrepancy usually implies that evolutionary masses are found to be larger than 
spectroscopic masses.
Limiting ourselves to presumed-single giant and supergiant 
LMC stars, \citet{massey2005} and \citet{mokiem2007} do not find a conspicuous mass 
discrepancy for O stars based on samples of 10 and 14 stars, respectively. 
For B supergiants, \citet{trundle2005} (18 stars) and \citet{mcevoy2015} 
(34 stars) study somewhat larger samples and
report a tentative mass discrepancy that is decreasing with luminosity.
We too aim to investigate this issue and determine the  spectroscopic  ($M_\mathrm{spec}$)  and evolutionary  ($M_\mathrm{evol}$) masses as outlined in the following paragraph.

The {\it spectroscopic mass} can be derived from the spectroscopically determined
gravity and the $K$-band magnitude constrained radius.
The gravities were corrected for the (small) contribution by the centrifugal acceleration
(see Sect.~\ref{subsec:gravities}). We derive the current {\it evolutionary mass}
of our stars by comparison with the single-star evolutionary tracks of \citet{brott} and \citet{kohler2015}. For this purpose,  
we used {\sc bonnsai}\footnote{The {\sc bonnsai} web-service is available at 
https://www.astro.uni-bonn.de/stars/bonnsai/}, a bayesian method, to constrain the 
evolutionary state of stars \citep{schneider2014}.
As independent prior functions, we adopt a \citet{salpeter1955}  initial mass function, an initial rotational velocity distribution 
from \citetalias{ramirezagudelo},
a random orientation of rotation
axes, and a uniform age distribution (equivalent to a past constant star-formation 
rate). As for the observables, 
we used the derived effective temperature, luminosity, 
and projected spin velocity. 
Because of the limited resolution of the model grid, we impose minimum error bars of 500\,K in $T_{\rm eff}$ and 0.1 dex in $\log\,L$.
On the basis of these constraints, we computed  the posterior probability distribution  
of the present-day mass for each star, yielding its mean mass and associated
68\% confidence intervals.  The evolutionary tracks
that were used  are limited to the main-sequence, that is, evolved stars that are moving blueward in the HRD are not
considered. In all cases (66 stars)
the probability distribution of current evolutionary masses
$M_{\rm evol}$ yielded a single, well defined peak. Both spectroscopic and evolutionary 
mass estimates are given in Table~\ref{table:new}. 

\begin{figure}
\centering
\includegraphics[width=\columnwidth]{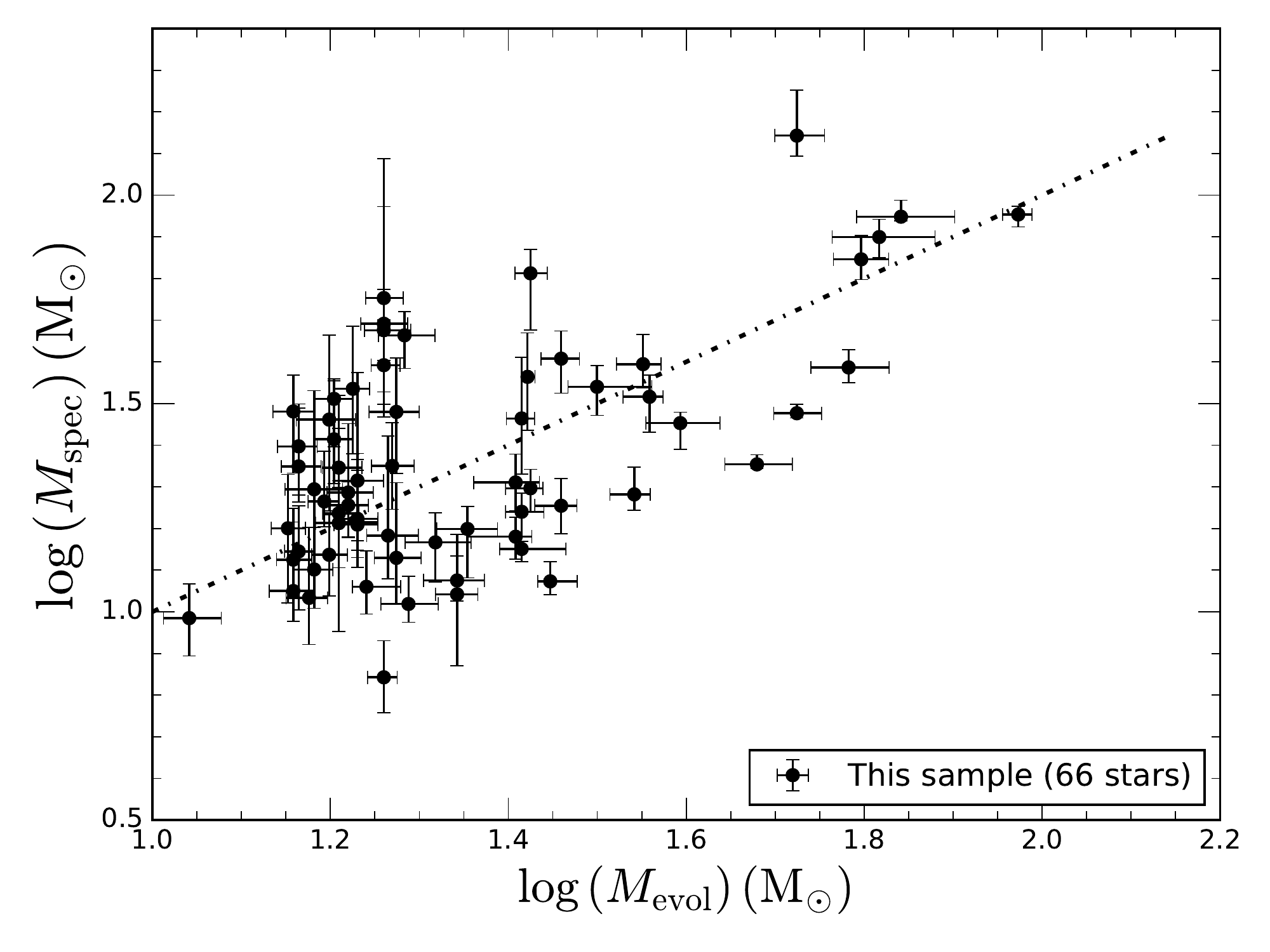}
\includegraphics[width=\columnwidth]{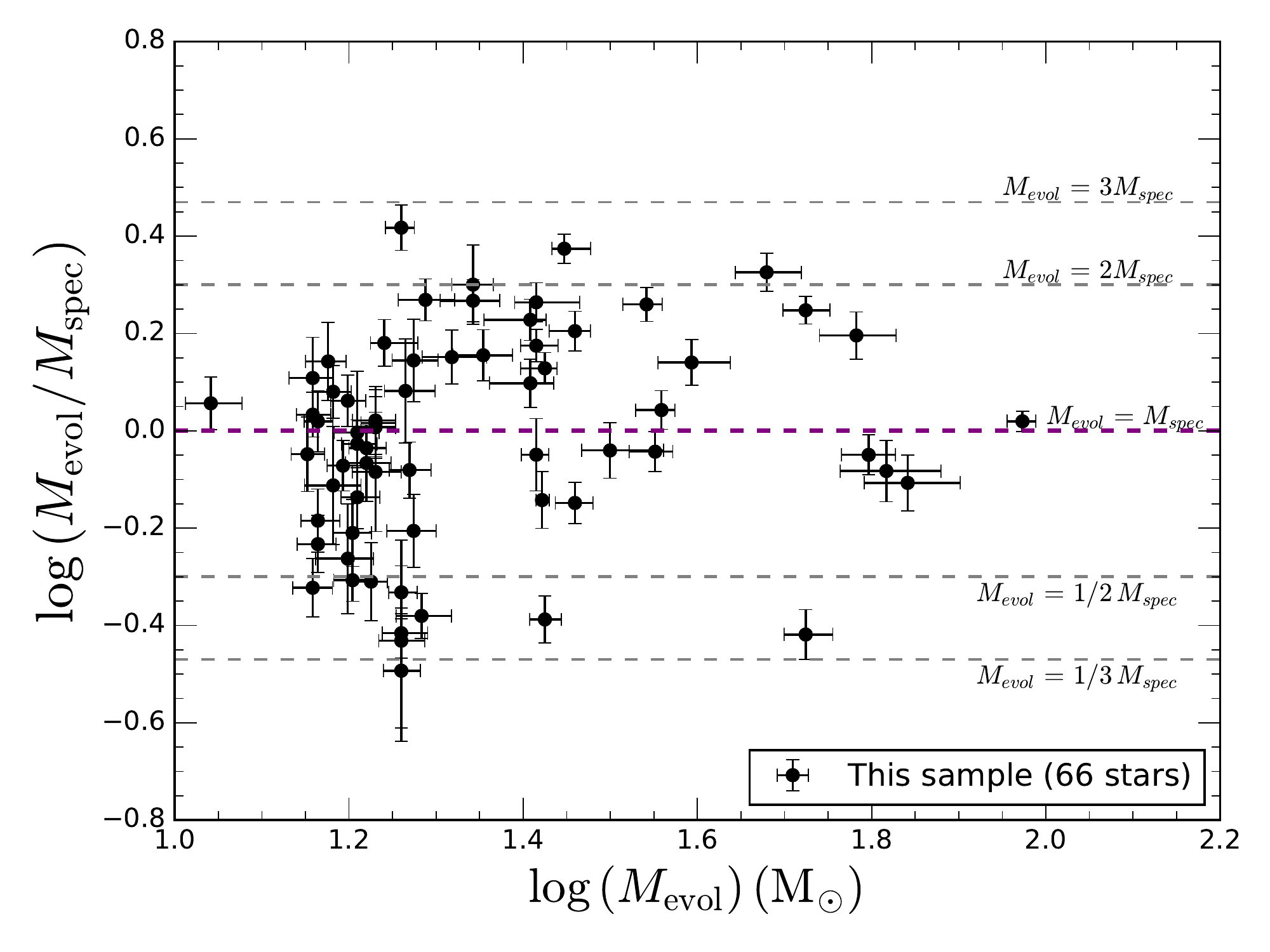}
\caption{Comparison of spectroscopic masses and evolutionary masses for
the 66 sources that have $M_{\rm evol}$ constrained by {\sc bonnsai}. 
The upper panel shows a one-to-one comparison; the lower panel shows 
the ratio of the masses.}
\label{fig:discrep_mass}
\end{figure}

The spectroscopic  and evolutionary masses are
compared in Fig.~\ref{fig:discrep_mass}. For relatively small masses, the uncertainty in the spectroscopic mass is often larger than the uncertainty in the evolutionary mass. For the high-mass sources, the uncertainties in the evolutionary
masses become larger as the observables span a larger mass range per unit temperature and luminosity.
For the sample as a whole, we find a weighted mean in $\log\,(M_\mathrm{evol}/M_\mathrm{spec})$ of $0.081\,\pm\,0.009$, that is, small but significant. However, the scatter is sizeable ($1\sigma$ dispersion of $0.201\,\pm\,0.010$), 
which precludes confirmation of a systematic mass discrepancy.

\begin{figure}
\centering
\includegraphics[scale=0.48]{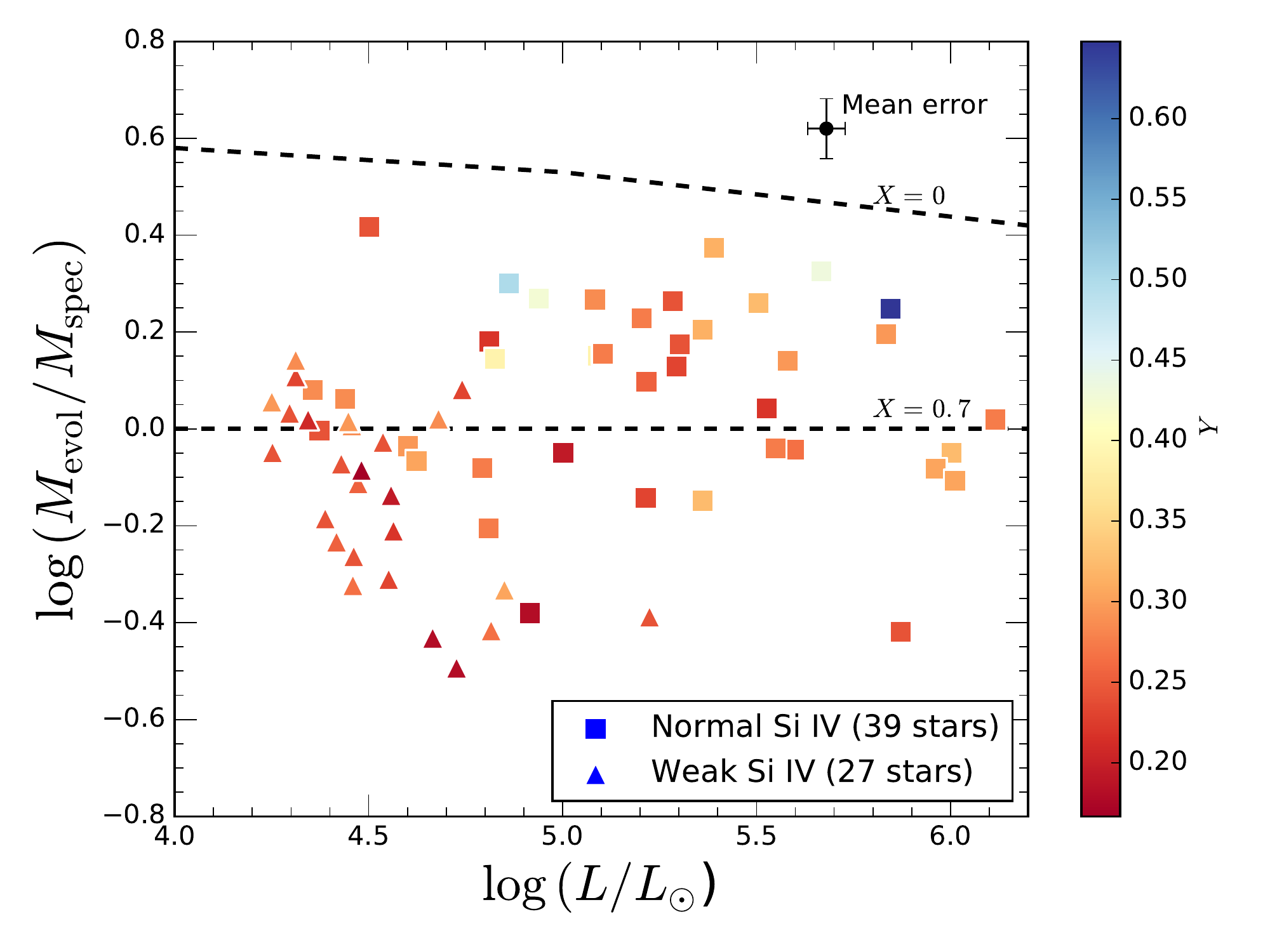}
\caption{Mass discrepancy plotted against luminosity for the 66 sources that 
have $M_{\rm evol}$ constrained by {\sc bonnsai}. The upper dashed line is from the 
$L(M)$ relation of helium stars; see main text for discussion.
The lower line is $M_{\rm spec}\,=\,M_{\rm evol}$, and is a lower limit as stars cannot 
be under-luminous. The color  
denotes the surface helium abundance. Squares denote stars for which the LC is certain
and triangles refer to the sources with weak Si\,{\sc iv} relative to
He\,{\sc i} and for which the luminosity class may be debated (they may be dwarf 
stars). Helium enriched sources are all positioned above the one-to-one relation.}
\label{fig:discrep_mass_He}
\end{figure}

Following earlier work \citep[e.g.,][]{mcevoy2015}, Fig.~\ref{fig:discrep_mass_He} shows the
mass discrepancy, in terms of $\log\,(M_\mathrm{evol}/M_\mathrm{spec})$, as a function of luminosity. 
According to stellar models, a He-enriched star of given
mass is expected to be more luminous than its He-normal
equivalent \citep{langer_1992}, that is, yield a higher $M_\mathrm{evol}$. This effect is illustrated by the upper dashed line in Fig.~\ref{fig:discrep_mass_He}, which represents the mass discrepancy
that could arise if the source were fitted with a baseline helium abundance $Y=0.255$ while
in reality it is a pure helium star ($Y=1.0$), adopting the $L(M)$ relation for helium
stars of \citet{grafener2011}. The He-enriched stars do seem to systematically show higher evolutionary masses than spectroscopic masses. However, we do not find a clear trend of the mass discrepancy with respect to the helium abundances of the stars.

It is interesting to note the subgroup of late O\,III and II stars in our sample (those that have weak Si\,{\sc iv}
features, see Sect.~\ref{subsec:gravities} of this paper and Table A.1 and 2 in \citet{walborn2014}; labeled as triangles in Fig.~\ref{fig:discrep_mass_He}). These sources systematically show larger spectroscopic than evolutionary masses (weighted mean $\log\,(M_\mathrm{evol}/M_\mathrm{spec})\,=\,-0.148\,\pm\,0.012$), which is the opposite to what is usually reported in the literature \citep[e.g.,][]{herrero}. Whether this is related to their nature remains unclear. We do note that they occupy a region in the HRD relatively devoid of O-dwarfs, which supports the hypothesis that they are regular main-sequence O-type stars (see Fig.~\ref{fig:HR} and the corresponding discussion in Sect.~\ref{ref:hrd}).

In view of the potential luminosity classification intricacy of the aforementioned  group, we also assess the presence or lack of a discrepancy excluding these sources. Hence for the remaining stars (labelled as squares in Fig.~\ref{fig:discrep_mass_He}), we find a weighted mean
of $\log\,(M_\mathrm{evol}/M_\mathrm{spec})\,=\,0.106\,\pm\,0.007$.
This suggests the presence of a modest (systematic) mass discrepancy for this subset. However, again, scatter is sizeable with a $1\sigma$ dispersion of $0.184\,\pm\,0.007$.

Again excluding the sources for which the luminosity class is debated 
(see above and Sect.~\ref{subsec:gravities}) one might perceive by eye 
a trend similar to that reported by \citet{trundle2005} and \citet{mcevoy2015}. However, 
the Pearson correlation coefficient does not allow us to accept the presence of such a linear trend at the 5\%\ significance level. We also computed 
the Spearman's and Kendall's rank correlations and reached the same conclusions that our data does not allow us to 
establish the significance of any trend between the degree of mass discrepancy ($\log\,(M_\mathrm{evol}/M_\mathrm{spec})$) and the stellar luminosity.

Finally, though certain individuals show worrying inconsistencies, the data does not allow us to confirm nor to reject the presence of a systematic mass discrepancy.

\section{Summary}
\label{sec:conclusions}

We have determined the stellar and wind properties of the 72 presumably single O-type 
giants, bright giants, and supergiants observed
in the context of the VLT-FLAMES Tarantula Survey. Our main findings can be summarized as follows:

\begin{itemize}
\item   We use our sample of LMC stars and the sample analyzed by \citet{mokiem2007b} to
        calibrate the spectral type versus $T_{\rm eff}$ dependence for luminosity class 
        III and I in the spectral range O3-O9. Linear relations for these calibrations 
        are provided in Table~\ref{table:teff}.

\item   Supergiants appear to be
        more evolved than the other luminosity classes and are among the brightest objects
        featuring the lowest surface gravities. Stars initially more massive than 60\,\msun\
        show giant characteristics already relatively early on in their evolution. Bright 
        giants with $M \ga 25$\,\msun\ do not constitute a conspicuous group intermediate 
        to giants and supergiants in the sHRD and HRD
        but instead reside in a large area of these two diagrams, ranging from the dwarf 
        (LC\,V) domain to the supergiants (LC\,I).            

\item   The group of late-O III and II stars with $\log\,g_c$ in between 4.0 and 4.5 reside
        in a region relatively close to the ZAMS that seems devoid of late dwarf O stars. This
        could point to a spectral classification issue. Indeed, though the primary
        luminosity criterion (the He\,{\sc ii}\,$\lambda$4686 / He\,{\sc i}\,$\lambda$4713
        ratio) indicates a LC III or II, the secondary criterion (the ratio of Si\,{\sc iv}
        to He\,{\sc i} lines) is more in line with a LC IV or V classification.    
            
\item   The positions of the O giants to supergiants in the HRD do not point to a preferred
        age but rather seem indicative of a continuum of ages. The sub-populations centered on 
        the associations NGC2070 (excluding its
        core cluster R136, not covered by the VFTS) and NGC2060 (6.7\arcmin\ south-west
        of R136) do not show preferred ages either, neither relative to each other nor relative
        to the remaining field population.

\item   The sample of presumed single stars contains a handful of helium enriched 
        stars (five stars with $Y > 0.30$), which all have $\vrot \leq 200$\,\kms, and unenriched helium stars 
        that spin in excess of 300\,\kms. This is not in accordance with expectations of rotational mixing in 
        main-sequence stars as computed by the evolutionary tracks of \citet{brott} and \cite{kohler2015}. 
        While it is very unlikely that these stars are post-RSGs, we cannot exclude the possibility 
        that they are post-interaction
        binaries. For four out of the six stars spinning in excess of 300\,\kms, we find a helium content 
        below the primordial value. We consider this to be a spurious result that may indicate a present-day binary nature of these systems.

\item   The $\beta$ parameter of the wind acceleration law cannot be reliably constrained with our data. 
	   Its value however significantly impacts the derived mass-loss rates. Adopting theoretical $\beta$ values 
           from \citet{muijres2012} yields a $\log\,D_{\rm mom}$ $-$ $\log\,L/\lsun$ relation that is shifted upwards by 
           $\sim$0.3 dex compared to earlier LMC results from \citet{mokiem2007}. The latter results can be recovered if we treat 
           the wind acceleration $\beta$ as a free parameter.  The  $\log\,D_{\rm mom}$ $-$ $\log\,L/\lsun$ relation that we obtained can be reconciled with the mass-loss predictions of 	\citet{vink2001} if the wind is clumped with a clump volume filling factor $f_{\rm V} \sim 1/8-1/6$.

\item   The current masses derived from the spectroscopic analysis are in fair agreement with those 
        derived from a comparison with evolutionary tracks, though the scatter is sizeable. 
        We do not detect a conspicuous systematic mass discrepancy.
\end{itemize}

The analysis presented here is part of a project that aims to establish the properties of the
bulk of the hot massive stars in the Tarantula Nebula. The aim is 
to better constrain the physics governing their
evolution, specifically the role of rotational mixing, mass loss, and binarity. In a follow-up study
we use the results obtained here to study the efficiency of rotational mixing in 
O III$-$I stars in more detail, using the surface nitrogen abundances as a probe \citep[][]{grin_et_al_2016}.
Of all massive stars that feature strong winds, the wind-driving mechanism of the group of O III$-$I stars
is thought to be best understood. However, even for these objects, the intricacies of accurately establishing
their wind properties remain challenging. Here, this is exemplified by a discussion
of the degeneracy of the wind acceleration $\beta$ and the mass-loss rate \mdot\
if only optical spectra are analyzed \cite[see also e.g.][]{markova2004}. Firm 
constraints on both parameters, as well as an independent measure of the 
clumping properties of the outflowing gas, can be obtained from far-ultraviolet spectra
and we signal the need to obtain such spectra to further our understanding of the 
mass-loss mechanism of the most massive stars. Finally, an in-depth 
reassessment of the luminosity class assignment of the group of late-O giants and 
bright giants featuring Si\,{\sc iv} lines that are weak relative to He\,{\sc i} lines 
is warranted. This should establish whether or not these stars
form a separate physical group.

\begin{acknowledgements}
O.H.R.A. acknowledges funding from the European Union's Horizon 2020 research and 
innovation programme under the Marie Sk{\l}odowska-Curie grant agreement No 665593   
awarded to the Science and Technology Facilities Council. N.J.G. is part of the International Max Planck
Research School (IMPRS) for Astronomy and Astrophysics at the Universities of Bonn
and Cologne. S.S.D. and A.H. acknowledge funding by the Spanish Ministry of
Economy and Competitiveness (MINECO) under grants AYA2010-21697-C05-
04, Consolider-Ingenio 2010 CSD2006-00070, and Severo Ochoa SEV-2011-
0187, and by the Canary Islands Government under grant PID2010119.
M.G. and F.N. acknowledge MINECO grants FIS2012-39162-C06-01,
ESP2013-47809-C3-1-R  and   ESP2015-65597-C4-1-R. 
C.S.S. acknowledges support from the Joint Committee ESO-Government of Chile and DIDULS programme from Universidad
de La Serena under grant DIDULS Regular PR16145. N.M. acknowledges the financial support of the Bulgarian NSF under grant DN08/1/13.12.2016.
This work was carried out on the Dutch national e-infrastructure with the support of SURF Cooperative.
\end{acknowledgements}

\bibliographystyle{aa}	

\begin{thebibliography}{113}
\expandafter\ifx\csname natexlab\endcsname\relax\def\natexlab#1{#1}\fi

\bibitem[{{Abbott} \& {Lucy}(1985)}]{abbott1985}
{Abbott}, D.~C. \& {Lucy}, L.~B. 1985, \apj, 288, 679

\bibitem[{{Almeida} {et~al.}(2017){Almeida}, {Sana}, {Taylor}, {Barb{\'a}},
  {Bonanos}, {Crowther}, {Damineli}, {de Koter}, {de Mink}, {Evans}, {Gieles},
  {Grin}, {H{\'e}nault-Brunet}, {Langer}, {Lennon}, {Lockwood}, {Ma{\'{\i}}z
  Apell{\'a}niz}, {Moffat}, {Neijssel}, {Norman}, {Ram{\'{\i}}rez-Agudelo},
  {Richardson}, {Schootemeijer}, {Shenar}, {Soszy{\'n}ski}, {Tramper}, \&
  {Vink}}]{almeida_et_al_2016}
{Almeida}, L.~A., {Sana}, H., {Taylor}, W., {et~al.} 2017, \aap\ in press
  (ArXiv e-prints 1610.03500)

\bibitem[{{Asplund} {et~al.}(2005){Asplund}, {Grevesse}, \&
  {Sauval}}]{asplund2005}
{Asplund}, M., {Grevesse}, N., \& {Sauval}, A.~J. 2005, in Astronomical Society
  of the Pacific Conference Series, Vol. 336, Cosmic Abundances as Records of
  Stellar Evolution and Nucleosynthesis, ed. T.~G. {Barnes}, III \& F.~N.
  {Bash}, 25

\bibitem[{{Bestenlehner} {et~al.}(2014){Bestenlehner}, {Gr{\"a}fener}, {Vink},
  {Najarro}, {de Koter}, {Sana}, {Evans}, {Crowther}, {H{\'e}nault-Brunet},
  {Herrero}, {Langer}, {Schneider}, {Sim{\'o}n-D{\'{\i}}az}, {Taylor}, \&
  {Walborn}}]{bestenlehner2014}
{Bestenlehner}, J.~M., {Gr{\"a}fener}, G., {Vink}, J.~S., {et~al.} 2014, \aap,
  570, A38 (\citetalias{bestenlehner2014})

\bibitem[{{Beuther} {et~al.}(2008){Beuther}, {Linz}, \&
  {Henning}}]{beuther2008}
{Beuther}, H., {Linz}, H., \& {Henning}, T., eds. 2008, Astronomical Society of
  the Pacific Conference Series, Vol. 387, {Massive Star Formation:
  Observations Confront Theory}

\bibitem[{{Bouret} {et~al.}(2003){Bouret}, {Lanz}, {Heap}, {Hubeny}, {Hillier},
  {Lennon}, {Evans}, \& {Smith}}]{bouret2003}
{Bouret}, J.-C., {Lanz}, T., {Heap}, S., {et~al.} 2003, in SF2A-2003: Semaine
  de l'Astrophysique Francaise, ed. F.~{Combes}, D.~{Barret}, T.~{Contini}, \&
  L.~{Pagani}, 499

\bibitem[{{Bouret} {et~al.}(2005){Bouret}, {Lanz}, \& {Hillier}}]{bouret2005}
{Bouret}, J.-C., {Lanz}, T., \& {Hillier}, D.~J. 2005, \aap, 438, 301

\bibitem[{{Bouret} {et~al.}(2013){Bouret}, {Lanz}, {Martins}, {Marcolino},
  {Hillier}, {Depagne}, \& {Hubeny}}]{bouret2013}
{Bouret}, J.-C., {Lanz}, T., {Martins}, F., {et~al.} 2013, \aap, 555, A1

\bibitem[{{Bromm} {et~al.}(2009){Bromm}, {Yoshida}, {Hernquist}, \&
  {McKee}}]{bromm2009}
{Bromm}, V., {Yoshida}, N., {Hernquist}, L., \& {McKee}, C.~F. 2009, \nat, 459,
  49

\bibitem[{{Brott} {et~al.}(2011){Brott}, {de Mink}, {Cantiello}, {Langer}, {de
  Koter}, {Evans}, {Hunter}, {Trundle}, \& {Vink}}]{brott}
{Brott}, I., {de Mink}, S.~E., {Cantiello}, M., {et~al.} 2011, \aap, 530, A115

\bibitem[{{Campbell} {et~al.}(2010){Campbell}, {Evans}, {Mackey}, {Gieles},
  {Alves}, {Ascenso}, {Bastian}, \& {Longmore}}]{campbell2010}
{Campbell}, M.~A., {Evans}, C.~J., {Mackey}, A.~D., {et~al.} 2010, \mnras, 405,
  421

\bibitem[{{Castro} {et~al.}(2014){Castro}, {Fossati}, {Langer},
  {Sim{\'o}n-D{\'{\i}}az}, {Schneider}, \& {Izzard}}]{castro2014}
{Castro}, N., {Fossati}, L., {Langer}, N., {et~al.} 2014, \aap, 570, L13

\bibitem[{{Charbonneau}(1995)}]{charbonneau1995}
{Charbonneau}, P. 1995, \apjs, 101, 309

\bibitem[{{Cioni} \& {the VMC team}(2015)}]{Cioni2015a}
{Cioni}, M.-R.~L. \& {the VMC team}. 2015, ArXiv e-prints

\bibitem[{{Clark} {et~al.}(2015){Clark}, {Bartlett}, {Broos}, {Townsley},
  {Taylor}, {Walborn}, {Bird}, {Sana}, {de Mink}, {Dufton}, {Evans}, {Langer},
  {Ma{\'{\i}}z Apell{\'a}niz}, {Schneider}, \& {Soszy{\'n}ski}}]{clark2015}
{Clark}, J.~S., {Bartlett}, E.~S., {Broos}, P.~S., {et~al.} 2015, \aap, 579,
  A131

\bibitem[{{Cohen} {et~al.}(2003){Cohen}, {Wheaton}, \& {Megeath}}]{cohen2003}
{Cohen}, M., {Wheaton}, W.~A., \& {Megeath}, S.~T. 2003, \aj, 126, 1090

\bibitem[{{Crowther} {et~al.}(2002){Crowther}, {Hillier}, {Evans}, {Fullerton},
  {De Marco}, \& {Willis}}]{crowther2002}
{Crowther}, P.~A., {Hillier}, D.~J., {Evans}, C.~J., {et~al.} 2002, \apj, 579,
  774

\bibitem[{{de Koter} {et~al.}(1997){de Koter}, {Heap}, \&
  {Hubeny}}]{dekoter1997}
{de Koter}, A., {Heap}, S.~R., \& {Hubeny}, I. 1997, \apj, 477, 792

\bibitem[{{de Koter} {et~al.}(1998){de Koter}, {Heap}, \&
  {Hubeny}}]{dekoter1998}
{de Koter}, A., {Heap}, S.~R., \& {Hubeny}, I. 1998, \apj, 509, 879

\bibitem[{{De Marchi} {et~al.}(2011){De Marchi}, {Paresce}, {Panagia},
  {Beccari}, {Spezzi}, {Sirianni}, {Andersen}, {Mutchler}, {Balick}, {Dopita},
  {Frogel}, {Whitmore}, {Bond}, {Calzetti}, {Carollo}, {Disney}, {Hall},
  {Holtzman}, {Kimble}, {McCarthy}, {O'Connell}, {Saha}, {Silk}, {Trauger},
  {Walker}, {Windhorst}, \& {Young}}]{demarchi2011}
{De Marchi}, G., {Paresce}, F., {Panagia}, N., {et~al.} 2011, \apj, 739, 27

\bibitem[{{de Mink} {et~al.}(2013){de Mink}, {Langer}, {Izzard}, {Sana}, \& {de
  Koter}}]{selma}
{de Mink}, S.~E., {Langer}, N., {Izzard}, R.~G., {Sana}, H., \& {de Koter}, A.
  2013, \apj, 764, 166

\bibitem[{{de Mink} {et~al.}(2014){de Mink}, {Sana}, {Langer}, {Izzard}, \&
  {Schneider}}]{selma2014}
{de Mink}, S.~E., {Sana}, H., {Langer}, N., {Izzard}, R.~G., \& {Schneider},
  F.~R.~N. 2014, \apj, 782, 7

\bibitem[{{Doran} {et~al.}(2013){Doran}, {Crowther}, {de Koter}, {Evans},
  {McEvoy}, {Walborn}, {Bastian}, {Bestenlehner}, {Gr{\"a}fener}, {Herrero},
  {K{\"o}hler}, {Ma{\'{\i}}z Apell{\'a}niz}, {Najarro}, {Puls}, {Sana},
  {Schneider}, {Taylor}, {van Loon}, \& {Vink}}]{doran}
{Doran}, E.~I., {Crowther}, P.~A., {de Koter}, A., {et~al.} 2013, \aap, 558,
  A134

\bibitem[{{Dunstall} {et~al.}(2015){Dunstall}, {Dufton}, {Sana}, {Evans},
  {Howarth}, {Sim{\'o}n-D{\'{\i}}az}, {de Mink}, {Langer}, {Ma{\'{\i}}z
  Apell{\'a}niz}, \& {Taylor}}]{dunstall2015}
{Dunstall}, P.~R., {Dufton}, P.~L., {Sana}, H., {et~al.} 2015, \aap, 580, A93

\bibitem[{{Ekstr{\"o}m} {et~al.}(2012){Ekstr{\"o}m}, {Georgy}, {Eggenberger},
  {Meynet}, {Mowlavi}, {Wyttenbach}, {Granada}, {Decressin}, {Hirschi},
  {Frischknecht}, {Charbonnel}, \& {Maeder}}]{ekstrom2012}
{Ekstr{\"o}m}, S., {Georgy}, C., {Eggenberger}, P., {et~al.} 2012, \aap, 537,
  A146

\bibitem[{{Evans} {et~al.}(2011){Evans}, {Taylor}, {H{\'e}nault-Brunet},
  {Sana}, {de Koter}, {Sim{\'o}n-D{\'{\i}}az}, {Carraro}, {Bagnoli}, {Bastian},
  {Bestenlehner}, {Bonanos}, {Bressert}, {Brott}, {Campbell}, {Cantiello},
  {Clark}, {Costa}, {Crowther}, {de Mink}, {Doran}, {Dufton}, {Dunstall},
  {Friedrich}, {Garcia}, {Gieles}, {Gr{\"a}fener}, {Herrero}, {Howarth},
  {Izzard}, {Langer}, {Lennon}, {Ma{\'{\i}}z Apell{\'a}niz}, {Markova},
  {Najarro}, {Puls}, {Ramirez}, {Sab{\'{\i}}n-Sanjuli{\'a}n}, {Smartt},
  {Stroud}, {van Loon}, {Vink}, \& {Walborn}}]{evans}
{Evans}, C.~J., {Taylor}, W.~D., {H{\'e}nault-Brunet}, V., {et~al.} 2011, \aap,
  530, A108 (\citetalias{evans})

\bibitem[{{Evans} {et~al.}(2010){Evans}, {Walborn}, {Crowther},
  {H{\'e}nault-Brunet}, {Massa}, {Taylor}, {Howarth}, {Sana}, {Lennon}, \& {van
  Loon}}]{evans2010}
{Evans}, C.~J., {Walborn}, N.~R., {Crowther}, P.~A., {et~al.} 2010, \apjl, 715,
  L74

\bibitem[{{Eversberg} {et~al.}(1998){Eversberg}, {L{\'e}pine}, \&
  {Moffat}}]{eversberg1998}
{Eversberg}, T., {L{\'e}pine}, S., \& {Moffat}, A.~F.~J. 1998, \apj, 494, 799

\bibitem[{{Fullerton} {et~al.}(2006){Fullerton}, {Massa}, \&
  {Prinja}}]{fullerton2006}
{Fullerton}, A.~W., {Massa}, D.~L., \& {Prinja}, R.~K. 2006, \apj, 637, 1025

\bibitem[{{Garcia} {et~al.}(2014){Garcia}, {Herrero}, {Najarro}, {Lennon}, \&
  {Alejandro Urbaneja}}]{garcia2014}
{Garcia}, M., {Herrero}, A., {Najarro}, F., {Lennon}, D.~J., \& {Alejandro
  Urbaneja}, M. 2014, \apj, 788, 64

\bibitem[{{Gr{\"a}fener} {et~al.}(2011){Gr{\"a}fener}, {Vink}, {de Koter}, \&
  {Langer}}]{grafener2011}
{Gr{\"a}fener}, G., {Vink}, J.~S., {de Koter}, A., \& {Langer}, N. 2011, \aap,
  535, A56

\bibitem[{Gray(1976)}]{Gray}
Gray, D. 1976, The Observation and Analysis of Stellar Photospheres, third
  edition edn. (Cambridge University Press)

\bibitem[{{Grin} {et~al.}(2016){Grin}, {Ram{\'{\i}}rez-Agudelo}, {de Koter},
  {Sana}, {Puls}, {Brott}, {Crowther}, {Dufton}, {Evans}, {Graefener},
  {Herrero}, {Langer}, {Lennon}, {van Loon}, {Markova}, {de Mink}, {Najarro},
  {Schneider}, {Taylor}, {Tramper}, {Vink}, \& {Walborn}}]{grin_et_al_2016}
{Grin}, N.~J., {Ram{\'{\i}}rez-Agudelo}, O.~H., {de Koter}, A., {et~al.} 2016,
  ArXiv e-prints

\bibitem[{{Groenewegen} \& {Lamers}(1989)}]{groenewegen_and_lamers1989}
{Groenewegen}, M.~A.~T. \& {Lamers}, H.~J.~G.~L.~M. 1989, \aaps, 79, 359

\bibitem[{{Groh} {et~al.}(2014){Groh}, {Meynet}, {Ekstr{\"o}m}, \&
  {Georgy}}]{groh2014}
{Groh}, J.~H., {Meynet}, G., {Ekstr{\"o}m}, S., \& {Georgy}, C. 2014, \aap,
  564, A30

\bibitem[{{Herrero} {et~al.}(1992){Herrero}, {Kudritzki}, {Vilchez}, {Kunze},
  {Butler}, \& {Haser}}]{herrero}
{Herrero}, A., {Kudritzki}, R.~P., {Vilchez}, J.~M., {et~al.} 1992, \aap, 261,
  209

\bibitem[{{Hillier} \& {Miller}(1998)}]{hillier_miller_1998}
{Hillier}, D.~J. \& {Miller}, D.~L. 1998, \apj, 496, 407

\bibitem[{{Hirano} {et~al.}(2015){Hirano}, {Hosokawa}, {Yoshida}, {Omukai}, \&
  {Yorke}}]{hirano2015}
{Hirano}, S., {Hosokawa}, T., {Yoshida}, N., {Omukai}, K., \& {Yorke}, H.~W.
  2015, \mnras, 448, 568

\bibitem[{{Hirano} {et~al.}(2014){Hirano}, {Hosokawa}, {Yoshida}, {Umeda},
  {Omukai}, {Chiaki}, \& {Yorke}}]{hirano2014}
{Hirano}, S., {Hosokawa}, T., {Yoshida}, N., {et~al.} 2014, \apj, 781, 60

\bibitem[{{Hubeny}(1988)}]{hubeny1988}
{Hubeny}, I. 1988, Computer Physics Communications, 52, 103

\bibitem[{{Hubeny} \& {Lanz}(1995)}]{hubeny1995}
{Hubeny}, I. \& {Lanz}, T. 1995, \apj, 439, 875

\bibitem[{{K{\"o}hler} {et~al.}(2015){K{\"o}hler}, {Langer}, {de Koter}, {de
  Mink}, {Crowther}, {Evans}, {Gr{\"a}fener}, {Sana}, {Sanyal}, {Schneider}, \&
  {Vink}}]{kohler2015}
{K{\"o}hler}, K., {Langer}, N., {de Koter}, A., {et~al.} 2015, \aap, 573, A71

\bibitem[{{Kudritzki} \& {Puls}(2000)}]{Kudritzki_puls2000}
{Kudritzki}, R.-P. \& {Puls}, J. 2000, \araa, 38, 613

\bibitem[{{Langer}(1992)}]{langer_1992}
{Langer}, N. 1992, \aap, 265, L17

\bibitem[{{Langer}(2012)}]{langer2012}
{Langer}, N. 2012, \araa, 50, 107

\bibitem[{{Langer} \& {Kudritzki}(2014)}]{langer_kudritzki2014}
{Langer}, N. \& {Kudritzki}, R.~P. 2014, \aap, 564, A52

\bibitem[{{Lanz} \& {Hubeny}(2007)}]{lanz2007}
{Lanz}, T. \& {Hubeny}, I. 2007, \apjs, 169, 83

\bibitem[{{Leitherer} {et~al.}(1992){Leitherer}, {Robert}, \&
  {Drissen}}]{LRD92}
{Leitherer}, C., {Robert}, C., \& {Drissen}, L. 1992, \apj, 401, 596

\bibitem[{{L{\'e}pine} \& {Moffat}(2008)}]{lepine2008}
{L{\'e}pine}, S. \& {Moffat}, A.~F.~J. 2008, \aj, 136, 548

\bibitem[{{Ma{\'{\i}}z Apell{\'a}niz} {et~al.}(2016){Ma{\'{\i}}z
  Apell{\'a}niz}, {Sota}, {Arias}, {Barb{\'a}}, {Walborn},
  {Sim{\'o}n-D{\'{\i}}az}, {Negueruela}, {Marco}, {Le{\~a}o}, {Herrero},
  {Gamen}, \& {Alfaro}}]{jesus_apellaniz_et_al_2016}
{Ma{\'{\i}}z Apell{\'a}niz}, J., {Sota}, A., {Arias}, J.~I., {et~al.} 2016,
  ArXiv e-prints

\bibitem[{{Markova} {et~al.}(2004){Markova}, {Puls}, {Repolust}, \&
  {Markov}}]{markova2004}
{Markova}, N., {Puls}, J., {Repolust}, T., \& {Markov}, H. 2004, \aap, 413, 693

\bibitem[{{Martins} {et~al.}(2005){Martins}, {Schaerer}, \&
  {Hillier}}]{martins}
{Martins}, F., {Schaerer}, D., \& {Hillier}, D.~J. 2005, \aap, 436, 1049

\bibitem[{{Massa} {et~al.}(2003){Massa}, {Fullerton}, {Sonneborn}, \&
  {Hutchings}}]{massa2003}
{Massa}, D., {Fullerton}, A.~W., {Sonneborn}, G., \& {Hutchings}, J.~B. 2003,
  \apj, 586, 996

\bibitem[{{Massey} {et~al.}(2004){Massey}, {Bresolin}, {Kudritzki}, {Puls}, \&
  {Pauldrach}}]{massey2004}
{Massey}, P., {Bresolin}, F., {Kudritzki}, R.~P., {Puls}, J., \& {Pauldrach},
  A.~W.~A. 2004, \apj, 608, 1001

\bibitem[{{Massey} {et~al.}(2013){Massey}, {Neugent}, {Hillier}, \&
  {Puls}}]{massey2013}
{Massey}, P., {Neugent}, K.~F., {Hillier}, D.~J., \& {Puls}, J. 2013, \apj,
  768, 6

\bibitem[{{Massey} {et~al.}(2005){Massey}, {Puls}, {Pauldrach}, {Bresolin},
  {Kudritzki}, \& {Simon}}]{massey2005}
{Massey}, P., {Puls}, J., {Pauldrach}, A.~W.~A., {et~al.} 2005, \apj, 627, 477

\bibitem[{{Massey} {et~al.}(2009){Massey}, {Zangari}, {Morrell}, {Puls},
  {DeGioia-Eastwood}, {Bresolin}, \& {Kudritzki}}]{massey2009}
{Massey}, P., {Zangari}, A.~M., {Morrell}, N.~I., {et~al.} 2009, \apj, 692, 618

\bibitem[{{Matteucci}(2008)}]{matteucci2008}
{Matteucci}, F. 2008, in IAU Symposium, Vol. 250, Massive Stars as Cosmic
  Engines, ed. F.~{Bresolin}, P.~A. {Crowther}, \& J.~{Puls}, 391--400

\bibitem[{{McEvoy} {et~al.}(2015){McEvoy}, {Dufton}, {Evans}, {Kalari},
  {Markova}, {Sim{\'o}n-D{\'{\i}}az}, {Vink}, {Walborn}, {Crowther}, {de
  Koter}, {de Mink}, {Dunstall}, {H{\'e}nault-Brunet}, {Herrero}, {Langer},
  {Lennon}, {Ma{\'{\i}}z Apell{\'a}niz}, {Najarro}, {Puls}, {Sana},
  {Schneider}, \& {Taylor}}]{mcevoy2015}
{McEvoy}, C.~M., {Dufton}, P.~L., {Evans}, C.~J., {et~al.} 2015, \aap, 575, A70

\bibitem[{{Meynet} \& {Maeder}(2005)}]{meynet2005}
{Meynet}, G. \& {Maeder}, A. 2005, \aap, 429, 581

\bibitem[{{Mokiem} {et~al.}(2007{\natexlab{a}}){Mokiem}, {de Koter}, {Evans},
  {Puls}, {Smartt}, {Crowther}, {Herrero}, {Langer}, {Lennon}, {Najarro},
  {Villamariz}, \& {Vink}}]{mokiem2007}
{Mokiem}, M.~R., {de Koter}, A., {Evans}, C.~J., {et~al.} 2007{\natexlab{a}},
  \aap, 465, 1003

\bibitem[{{Mokiem} {et~al.}(2006){Mokiem}, {de Koter}, {Evans}, {Puls},
  {Smartt}, {Crowther}, {Herrero}, {Langer}, {Lennon}, {Najarro}, {Villamariz},
  \& {Yoon}}]{mokiem2006}
{Mokiem}, M.~R., {de Koter}, A., {Evans}, C.~J., {et~al.} 2006, \aap, 456, 1131

\bibitem[{{Mokiem} {et~al.}(2005){Mokiem}, {de Koter}, {Puls}, {Herrero},
  {Najarro}, \& {Villamariz}}]{mokiem2005}
{Mokiem}, M.~R., {de Koter}, A., {Puls}, J., {et~al.} 2005, \aap, 441, 711

\bibitem[{{Mokiem} {et~al.}(2007{\natexlab{b}}){Mokiem}, {de Koter}, {Vink},
  {Puls}, {Evans}, {Smartt}, {Crowther}, {Herrero}, {Langer}, {Lennon},
  {Najarro}, \& {Villamariz}}]{mokiem2007b}
{Mokiem}, M.~R., {de Koter}, A., {Vink}, J.~S., {et~al.} 2007{\natexlab{b}},
  \aap, 473, 603

\bibitem[{{Muijres} {et~al.}(2011){Muijres}, {de Koter}, {Vink}, {Krti{\v
  c}ka}, {Kub{\'a}t}, \& {Langer}}]{muijres2011}
{Muijres}, L.~E., {de Koter}, A., {Vink}, J.~S., {et~al.} 2011, \aap, 526, A32

\bibitem[{{Muijres} {et~al.}(2012){Muijres}, {Vink}, {de Koter}, {M{\"u}ller},
  \& {Langer}}]{muijres2012}
{Muijres}, L.~E., {Vink}, J.~S., {de Koter}, A., {M{\"u}ller}, P.~E., \&
  {Langer}, N. 2012, \aap, 537, A37

\bibitem[{{Najarro} {et~al.}(2011){Najarro}, {Hanson}, \& {Puls}}]{najarro2011}
{Najarro}, F., {Hanson}, M.~M., \& {Puls}, J. 2011, \aap, 535, A32

\bibitem[{{Oskinova} {et~al.}(2007){Oskinova}, {Hamann}, \&
  {Feldmeier}}]{oskinova2007}
{Oskinova}, L.~M., {Hamann}, W.-R., \& {Feldmeier}, A. 2007, \aap, 476, 1331

\bibitem[{{Peimbert} {et~al.}(2007){Peimbert}, {Luridiana}, \&
  {Peimbert}}]{peimbert2007}
{Peimbert}, M., {Luridiana}, V., \& {Peimbert}, A. 2007, \apj, 666, 636

\bibitem[{{Pietrzy{\'n}ski} {et~al.}(2013){Pietrzy{\'n}ski}, {Graczyk},
  {Gieren}, {Thompson}, {Pilecki}, {Udalski}, {Soszy{\'n}ski}, {Koz{\l}owski},
  {Konorski}, {Suchomska}, {Bono}, {Moroni}, {Villanova}, {Nardetto},
  {Bresolin}, {Kudritzki}, {Storm}, {Gallenne}, {Smolec}, {Minniti}, {Kubiak},
  {Szyma{\'n}ski}, {Poleski}, {Wyrzykowski}, {Ulaczyk}, {Pietrukowicz},
  {G{\'o}rski}, \& {Karczmarek}}]{pietrzyski2013}
{Pietrzy{\'n}ski}, G., {Graczyk}, D., {Gieren}, W., {et~al.} 2013, \nat, 495,
  76

\bibitem[{{Podsiadlowski} {et~al.}(1992){Podsiadlowski}, {Joss}, \&
  {Hsu}}]{PJH1992}
{Podsiadlowski}, P., {Joss}, P.~C., \& {Hsu}, J.~J.~L. 1992, \apj, 391, 246

\bibitem[{{Prantzos}(2000)}]{prantzos2000}
{Prantzos}, N. 2000, \nar, 44, 303

\bibitem[{{Prinja} \& {Massa}(2010)}]{prinja2010}
{Prinja}, R.~K. \& {Massa}, D.~L. 2010, \aap, 521, L55

\bibitem[{{Puls} {et~al.}(1996){Puls}, {Kudritzki}, {Herrero}, {Pauldrach},
  {Haser}, {Lennon}, {Gabler}, {Voels}, {Vilchez}, {Wachter}, \&
  {Feldmeier}}]{puls1996}
{Puls}, J., {Kudritzki}, R.-P., {Herrero}, A., {et~al.} 1996, \aap, 305, 171

\bibitem[{{Puls} {et~al.}(2000){Puls}, {Springmann}, \& {Lennon}}]{puls2000}
{Puls}, J., {Springmann}, U., \& {Lennon}, M. 2000, \aaps, 141, 23

\bibitem[{{Puls} {et~al.}(2005){Puls}, {Urbaneja}, {Venero}, {Repolust},
  {Springmann}, {Jokuthy}, \& {Mokiem}}]{puls2005}
{Puls}, J., {Urbaneja}, M.~A., {Venero}, R., {et~al.} 2005, \aap, 435, 669

\bibitem[{{Puls} {et~al.}(2008){Puls}, {Vink}, \& {Najarro}}]{puls2008}
{Puls}, J., {Vink}, J.~S., \& {Najarro}, F. 2008, \aapr, 16, 209

\bibitem[{{Ram{\'{\i}}rez-Agudelo} {et~al.}(2015){Ram{\'{\i}}rez-Agudelo},
  {Sana}, {de Mink}, {H{\'e}nault-Brunet}, {de Koter}, {Langer}, {Tramper},
  {Gr{\"a}fener}, {Evans}, {Vink}, {Dufton}, \& {Taylor}}]{ramirezagudelo2015}
{Ram{\'{\i}}rez-Agudelo}, O.~H., {Sana}, H., {de Mink}, S.~E., {et~al.} 2015,
  \aap, 580, A92 (\citetalias{ramirezagudelo2015})

\bibitem[{{Ram{\'{\i}}rez-Agudelo} {et~al.}(2013){Ram{\'{\i}}rez-Agudelo},
  {Sim{\'o}n-D{\'{\i}}az}, {Sana}, {de Koter}, {Sab{\'{\i}}n-Sanjul{\'{\i}}an},
  {de Mink}, {Dufton}, {Gr{\"a}fener}, {Evans}, {Herrero}, {Langer}, {Lennon},
  {Ma{\'{\i}}z Apell{\'a}niz}, {Markova}, {Najarro}, {Puls}, {Taylor}, \&
  {Vink}}]{ramirezagudelo}
{Ram{\'{\i}}rez-Agudelo}, O.~H., {Sim{\'o}n-D{\'{\i}}az}, S., {Sana}, H.,
  {et~al.} 2013, \aap, 560, A29 (\citetalias{ramirezagudelo})

\bibitem[{{Repolust} {et~al.}(2004){Repolust}, {Puls}, \&
  {Herrero}}]{repolust2004}
{Repolust}, T., {Puls}, J., \& {Herrero}, A. 2004, \aap, 415, 349

\bibitem[{{Rivero Gonz{\'a}lez} {et~al.}(2012{\natexlab{a}}){Rivero
  Gonz{\'a}lez}, {Puls}, {Massey}, \& {Najarro}}]{rivero_gonzalez2012b}
{Rivero Gonz{\'a}lez}, J.~G., {Puls}, J., {Massey}, P., \& {Najarro}, F.
  2012{\natexlab{a}}, \aap, 543, A95

\bibitem[{{Rivero Gonz{\'a}lez} {et~al.}(2011){Rivero Gonz{\'a}lez}, {Puls}, \&
  {Najarro}}]{rivero_gonzalez2011}
{Rivero Gonz{\'a}lez}, J.~G., {Puls}, J., \& {Najarro}, F. 2011, \aap, 536, A58

\bibitem[{{Rivero Gonz{\'a}lez} {et~al.}(2012{\natexlab{b}}){Rivero
  Gonz{\'a}lez}, {Puls}, {Najarro}, \& {Brott}}]{rivero_gonzalez2012a}
{Rivero Gonz{\'a}lez}, J.~G., {Puls}, J., {Najarro}, F., \& {Brott}, I.
  2012{\natexlab{b}}, \aap, 537, A79

\bibitem[{{Rolleston} {et~al.}(2002){Rolleston}, {Trundle}, \&
  {Dufton}}]{rolleston2002}
{Rolleston}, W.~R.~J., {Trundle}, C., \& {Dufton}, P.~L. 2002, \aap, 396, 53

\bibitem[{{Rubele} {et~al.}(2012){Rubele}, {Kerber}, {Girardi}, {Cioni},
  {Marigo}, {Zaggia}, {Bekki}, {de Grijs}, {Emerson}, {Groenewegen},
  {Gullieuszik}, {Ivanov}, {Miszalski}, {Oliveira}, {Tatton}, \& {van
  Loon}}]{rubele2012}
{Rubele}, S., {Kerber}, L., {Girardi}, L., {et~al.} 2012, \aap, 537, A106

\bibitem[{{Sabbi} {et~al.}(2015){Sabbi}, {Lennon}, {Anderson}, {Cignoni}, {van
  der Marel}, {Zaritsky}, {de Marchi}, {Panagia}, {Gouliermis}, {Grebel},
  {Gallager}, {Smith}, {Sana}, {Aloisi}, {Tosi}, {Evans}, {Arab}, {Boyer}, {de
  Mink}, {Gordon}, {Koekemoer}, {Larsen}, {Ryon}, \& {Zeidler}}]{sabbi2015b}
{Sabbi}, E., {Lennon}, D.~J., {Anderson}, J., {et~al.} 2015, ArXiv e-prints

\bibitem[{{Sab{\'{\i}}n-Sanjuli{\'a}n}
  {et~al.}(2014){Sab{\'{\i}}n-Sanjuli{\'a}n}, {Sim{\'o}n-D{\'{\i}}az},
  {Herrero}, {Walborn}, {Puls}, {Ma{\'{\i}}z Apell{\'a}niz}, {Evans}, {Brott},
  {de Koter}, {Garcia}, {Markova}, {Najarro}, {Ram{\'{\i}}rez-Agudelo}, {Sana},
  {Taylor}, \& {Vink}}]{sabin_sanjulian2014}
{Sab{\'{\i}}n-Sanjuli{\'a}n}, C., {Sim{\'o}n-D{\'{\i}}az}, S., {Herrero}, A.,
  {et~al.} 2014, \aap, 564, A39 (\citetalias{sabin_sanjulian2014})

\bibitem[{{Salpeter}(1955)}]{salpeter1955}
{Salpeter}, E.~E. 1955, \apj, 121, 161

\bibitem[{{Sana} {et~al.}(2013){Sana}, {de Koter}, {de Mink}, {Dunstall},
  {Evans}, {H{\'e}nault-Brunet}, {Ma{\'{\i}}z Apell{\'a}niz},
  {Ram{\'{\i}}rez-Agudelo}, {Taylor}, {Walborn}, {Clark}, {Crowther},
  {Herrero}, {Gieles}, {Langer}, {Lennon}, \& {Vink}}]{sana}
{Sana}, H., {de Koter}, A., {de Mink}, S.~E., {et~al.} 2013, \aap, 550, A107
  (\citetalias{sana})

\bibitem[{{Sana} {et~al.}(2001){Sana}, {Rauw}, \& {Gosset}}]{sana2001}
{Sana}, H., {Rauw}, G., \& {Gosset}, E. 2001, \aap, 370, 121

\bibitem[{{Schneider} {et~al.}(2014){Schneider}, {Langer}, {de Koter}, {Brott},
  {Izzard}, \& {Lau}}]{schneider2014}
{Schneider}, F.~R.~N., {Langer}, N., {de Koter}, A., {et~al.} 2014, \aap, 570,
  A66

\bibitem[{{Seaton}(1958)}]{seaton_et_al_1958}
{Seaton}, M.~J. 1958, \mnras, 118, 504

\bibitem[{{Sim{\'o}n-D{\'{\i}}az} {et~al.}(2011){Sim{\'o}n-D{\'{\i}}az},
  {Caballero}, \& {Lorenzo}}]{simon-diaz2011}
{Sim{\'o}n-D{\'{\i}}az}, S., {Caballero}, J.~A., \& {Lorenzo}, J. 2011, \apj,
  742, 55

\bibitem[{{Sim{\'o}n-D{\'{\i}}az} {et~al.}(2015){Sim{\'o}n-D{\'{\i}}az},
  {Caballero}, {Lorenzo}, {Ma{\'{\i}}z Apell{\'a}niz}, {Schneider},
  {Negueruela}, {Barb{\'a}}, {Dorda}, {Marco}, {Montes}, {Pellerin},
  {Sanchez-Bermudez}, {S{\'o}dor}, \& {Sota}}]{simon-diaz2015}
{Sim{\'o}n-D{\'{\i}}az}, S., {Caballero}, J.~A., {Lorenzo}, J., {et~al.} 2015,
  \apj, 799, 169

\bibitem[{{Sim{\'o}n-D{\'{\i}}az} {et~al.}(2014){Sim{\'o}n-D{\'{\i}}az},
  {Herrero}, {Sab{\'{\i}}n-Sanjuli{\'a}n}, {Najarro}, {Garcia}, {Puls},
  {Castro}, \& {Evans}}]{simon2014}
{Sim{\'o}n-D{\'{\i}}az}, S., {Herrero}, A., {Sab{\'{\i}}n-Sanjuli{\'a}n}, C.,
  {et~al.} 2014, \aap, 570, L6

\bibitem[{{Sota} {et~al.}(2014){Sota}, {Ma{\'{\i}}z Apell{\'a}niz}, {Morrell},
  {Barb{\'a}}, {Walborn}, {Gamen}, {Arias}, \& {Alfaro}}]{sota2014}
{Sota}, A., {Ma{\'{\i}}z Apell{\'a}niz}, J., {Morrell}, N.~I., {et~al.} 2014,
  \apjs, 211, 10

\bibitem[{{Sota} {et~al.}(2011){Sota}, {Ma{\'{\i}}z Apell{\'a}niz}, {Walborn},
  {Alfaro}, {Barb{\'a}}, {Morrell}, {Gamen}, \& {Arias}}]{sota2011}
{Sota}, A., {Ma{\'{\i}}z Apell{\'a}niz}, J., {Walborn}, N.~R., {et~al.} 2011,
  \apjs, 193, 24

\bibitem[{{Sundqvist} {et~al.}(2010){Sundqvist}, {Puls}, \&
  {Feldmeier}}]{sundqvist2010}
{Sundqvist}, J.~O., {Puls}, J., \& {Feldmeier}, A. 2010, \aap, 510, A11

\bibitem[{{Sundqvist} {et~al.}(2011){Sundqvist}, {Puls}, {Feldmeier}, \&
  {Owocki}}]{sundqvist2011}
{Sundqvist}, J.~O., {Puls}, J., {Feldmeier}, A., \& {Owocki}, S.~P. 2011, \aap,
  528, A64

\bibitem[{{Sundqvist} {et~al.}(2014){Sundqvist}, {Puls}, \&
  {Owocki}}]{sundqvist_et_al_2014}
{Sundqvist}, J.~O., {Puls}, J., \& {Owocki}, S.~P. 2014, \aap, 568, A59

\bibitem[{{Tramper} {et~al.}(2011){Tramper}, {Sana}, {de Koter}, \&
  {Kaper}}]{tramper2011}
{Tramper}, F., {Sana}, H., {de Koter}, A., \& {Kaper}, L. 2011, \apjl, 741, L8

\bibitem[{{Tramper} {et~al.}(2014){Tramper}, {Sana}, {de Koter}, {Kaper}, \&
  {Ram{\'{\i}}rez-Agudelo}}]{tramper2014}
{Tramper}, F., {Sana}, H., {de Koter}, A., {Kaper}, L., \&
  {Ram{\'{\i}}rez-Agudelo}, O.~H. 2014, \aap, 572, A36

\bibitem[{{Trundle} \& {Lennon}(2005)}]{trundle2005}
{Trundle}, C. \& {Lennon}, D.~J. 2005, \aap, 434, 677

\bibitem[{{Trundle} {et~al.}(2004){Trundle}, {Lennon}, {Puls}, \&
  {Dufton}}]{trundle_et_al_2004}
{Trundle}, C., {Lennon}, D.~J., {Puls}, J., \& {Dufton}, P.~L. 2004, \aap, 417,
  217

\bibitem[{{{\v S}urlan} {et~al.}(2013){{\v S}urlan}, {Hamann}, {Aret},
  {Kub{\'a}t}, {Oskinova}, \& {Torres}}]{surlan2013}
{{\v S}urlan}, B., {Hamann}, W.-R., {Aret}, A., {et~al.} 2013, \aap, 559, A130

\bibitem[{{Vink} {et~al.}(2010){Vink}, {Brott}, {Gr{\"a}fener}, {Langer}, {de
  Koter}, \& {Lennon}}]{vink2010}
{Vink}, J.~S., {Brott}, I., {Gr{\"a}fener}, G., {et~al.} 2010, \aap, 512, L7

\bibitem[{{Vink} {et~al.}(1999){Vink}, {de Koter}, \& {Lamers}}]{vink1999}
{Vink}, J.~S., {de Koter}, A., \& {Lamers}, H.~J.~G.~L.~M. 1999, \aap, 350, 181

\bibitem[{{Vink} {et~al.}(2001){Vink}, {de Koter}, \& {Lamers}}]{vink2001}
{Vink}, J.~S., {de Koter}, A., \& {Lamers}, H.~J.~G.~L.~M. 2001, \aap, 369, 574

\bibitem[{{Vink} \& {Gr{\"a}fener}(2012)}]{vink2012}
{Vink}, J.~S. \& {Gr{\"a}fener}, G. 2012, \apjl, 751, L34

\bibitem[{{Walborn}(1973)}]{walborn1973}
{Walborn}, N.~R. 1973, \aj, 78, 1067

\bibitem[{{Walborn} \& {Blades}(1997)}]{walborn}
{Walborn}, N.~R. \& {Blades}, J.~C. 1997, \apjs, 112, 457

\bibitem[{{Walborn} {et~al.}(2014){Walborn}, {Sana}, {Sim{\'o}n-D{\'{\i}}az},
  {Ma{\'{\i}}z Apell{\'a}niz}, {Taylor}, {Evans}, {Markova}, {Lennon}, \& {de
  Koter}}]{walborn2014}
{Walborn}, N.~R., {Sana}, H., {Sim{\'o}n-D{\'{\i}}az}, S., {et~al.} 2014, \aap,
  564, A40 (\citetalias{walborn2014})

\bibitem[{{Zhang} {et~al.}(2009){Zhang}, {Zhang}, {Virgili}, {Liang}, {Kann},
  {Wu}, {Proga}, {Lv}, {Toma}, {M{\'e}sz{\'a}ros}, {Burrows}, {Roming}, \&
  {Gehrels}}]{zhang2009}
{Zhang}, B., {Zhang}, B.-B., {Virgili}, F.~J., {et~al.} 2009, \apj, 703, 1696

\end{thebibliography}

\begin{appendix} 

\section{Projected rotational velocity \vrot }
\label{subsec:vsini_paperXII_GA}

\citetalias{ramirezagudelo} presents the rotational properties of the spectroscopic
single O-type stars by taking into account line broadening 
due to  macro-turbulent motions. The stars analyzed 
in this paper are a subset of those analysed in \citetalias{ramirezagudelo}.
Here we determine \vrot\ neglecting macro-turbulent motions. Our results may, therefore,  
differ from \citetalias{ramirezagudelo}. 

\begin{figure}
\centering
\includegraphics[scale=0.45]{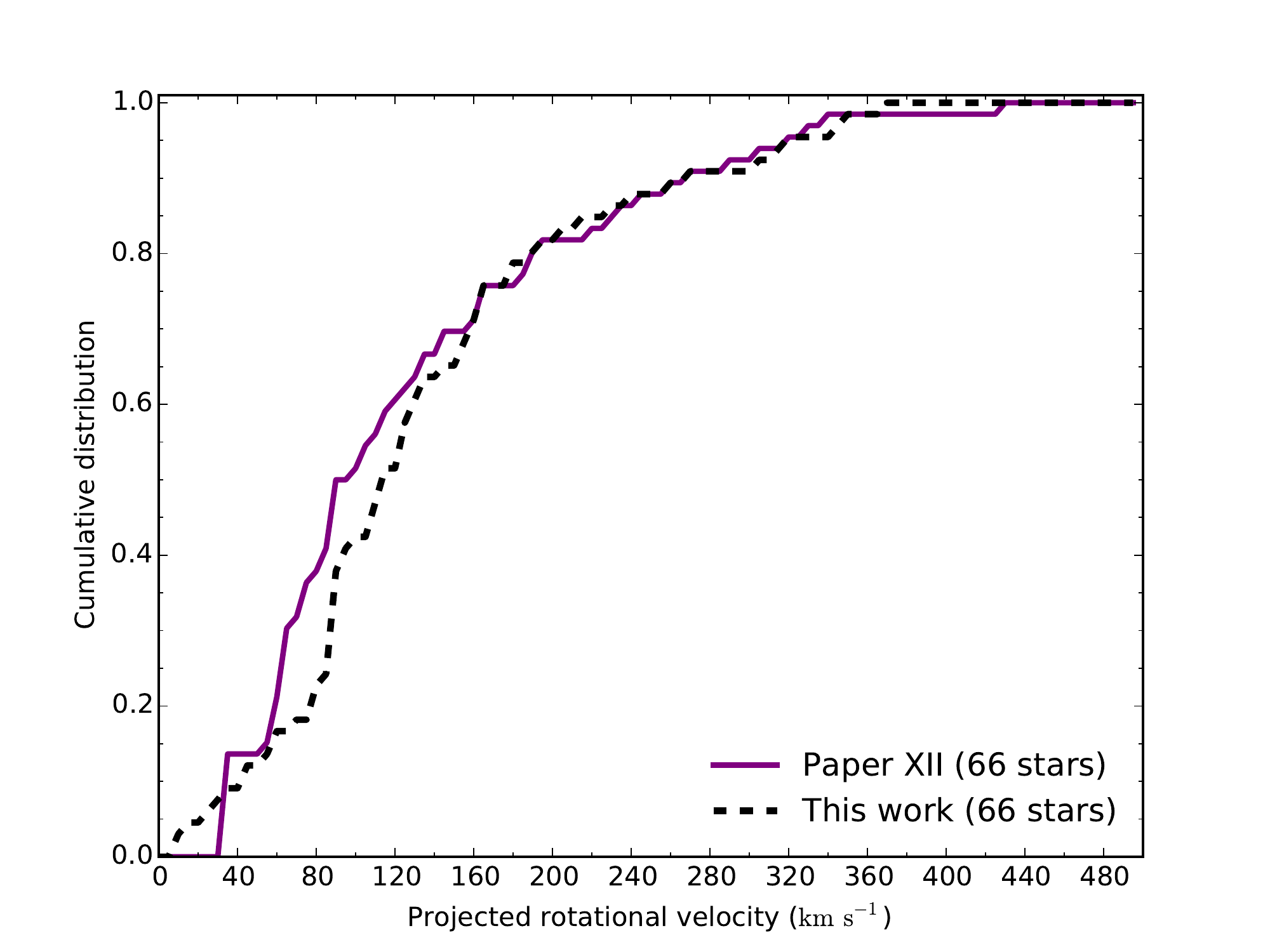}
\includegraphics[scale=0.45]{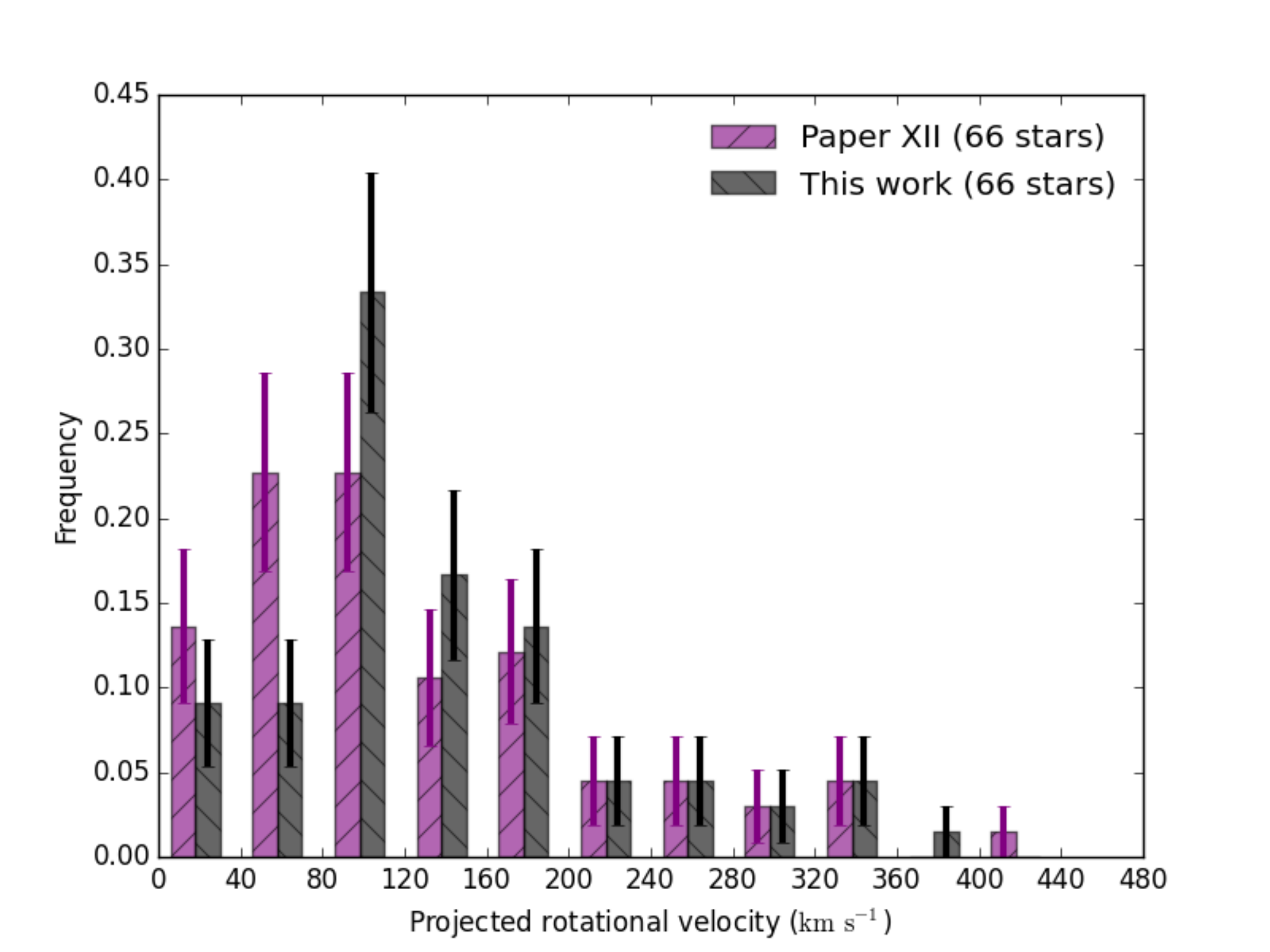}
\includegraphics[scale=0.45]{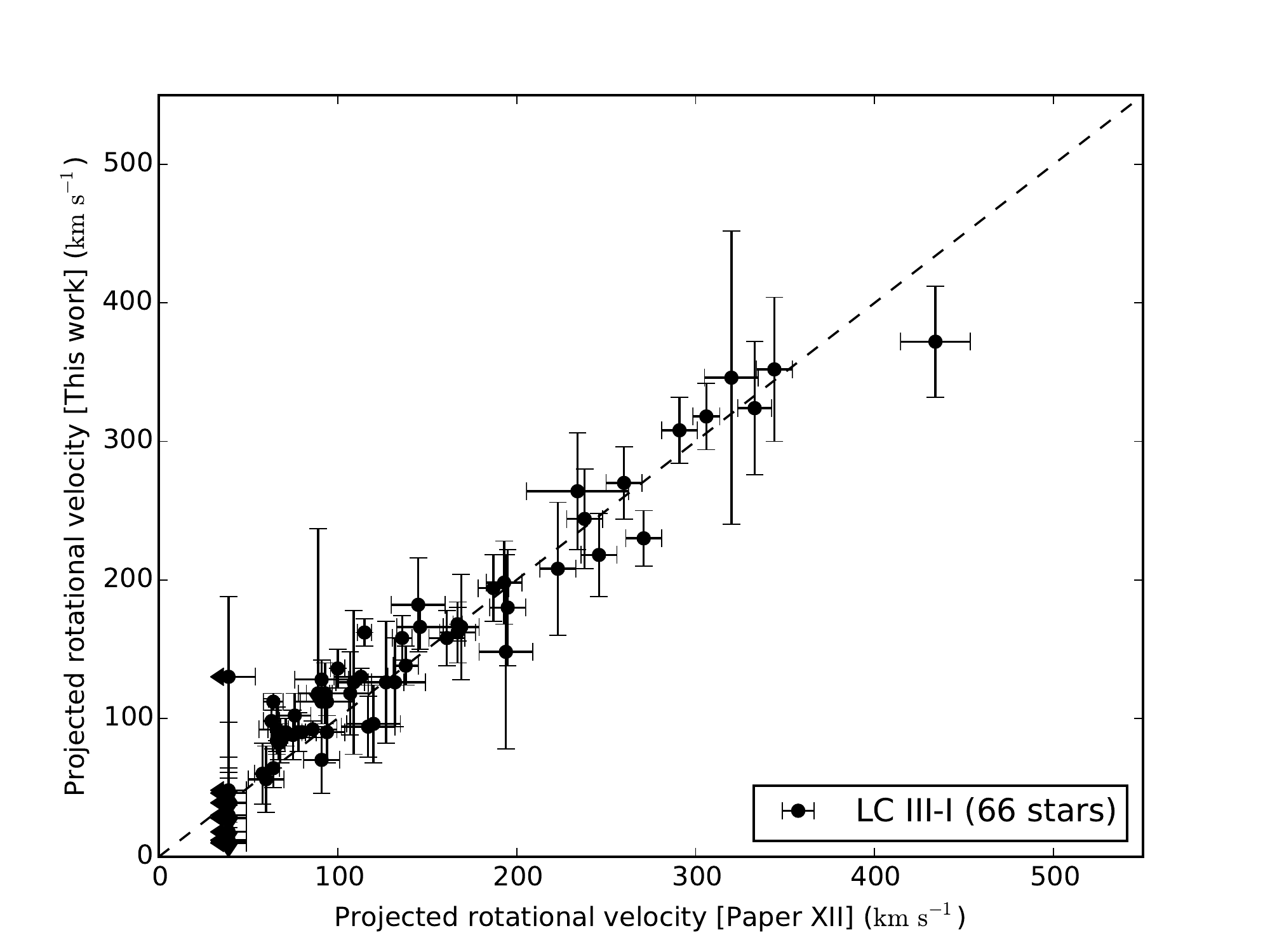}
\caption{Cumulative (upper panel) and frequency (middle panel) distributions of the 
projected rotational velocities of the O-type LC\,III to I 
as derived here from the automated {\sc fastwind} analysis and \citetalias{ramirezagudelo}. 
The middle panel shows Poissonian error bars. The
lower panel compares the actual \vrot\ of these two samples.}
\label{fig:vsini}
\end{figure}

Figure~\ref{fig:vsini} shows a comparison of both \vrot\ estimates using different
representations of the difference. The systematic difference of all stars in common 
is approximately 7\,\kms\ with a standard deviation 21\,\kms. This is in
agreement with the uncertainties discussed in \citetalias{ramirezagudelo}. 
Qualitatively, the shape of the (cumulative) \vrot\ distribution is similar in both
methodologies (see upper and middle panel). Below 160\,\kms,
the values presented here tend to somewhat overestimate \vrot. 
This is a consequence of not distinguishing between broadening from 
rotation and macro turbulence in the regime where rotation does not dominate 
the line width. Hence, the \vrot\ values up to 160\,\kms\ derived here may be 
overestimated by up to several tens of \kms. Such overestimates
do not, however, impact the determination of other stellar properties in any significant way.

\section{HRD as a function of the spatial location in 30\,Dor}\label{sec:hrd_clusters}

\begin{figure}
\centering
\includegraphics[width=\columnwidth]{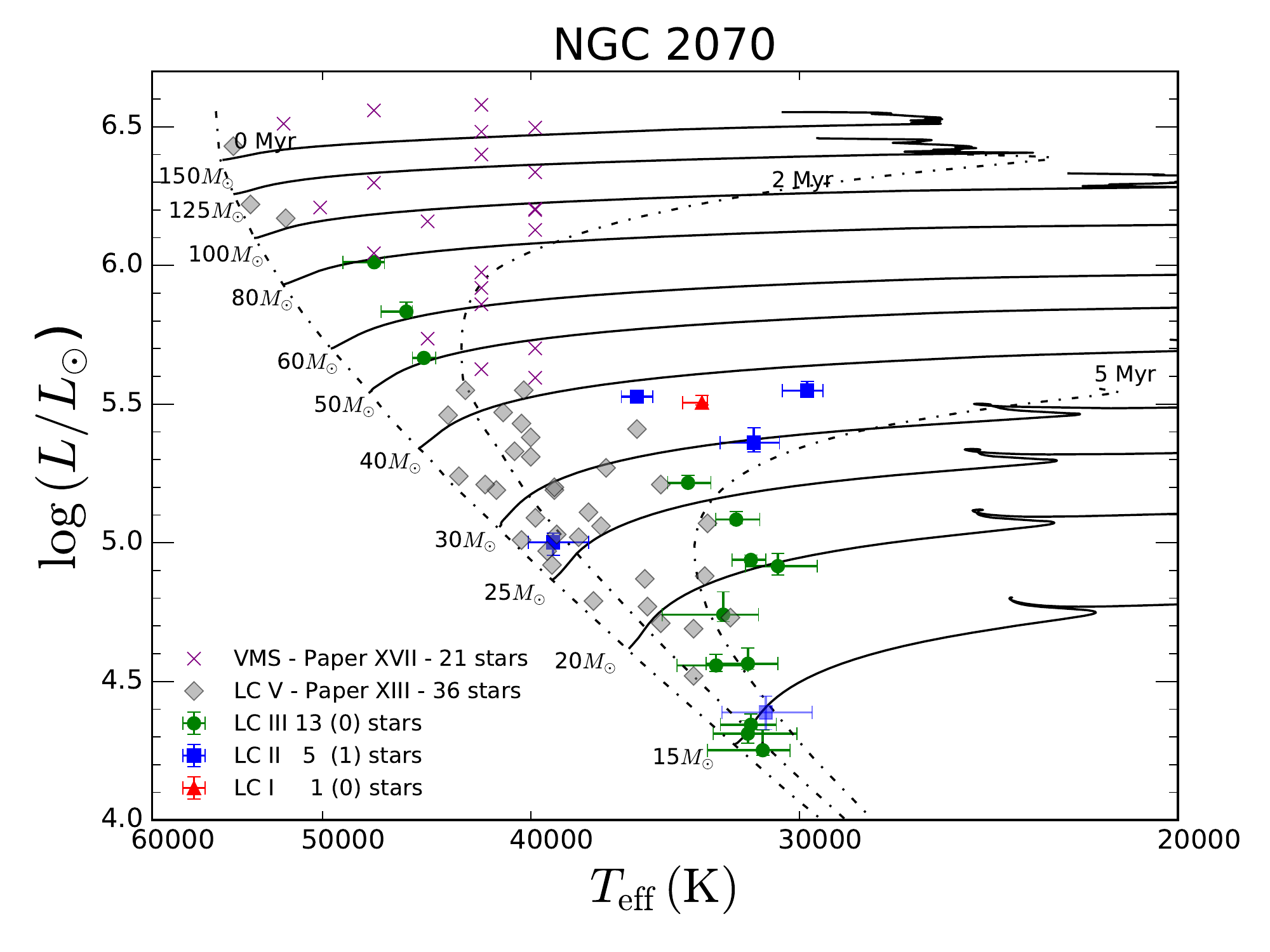}
\includegraphics[width=\columnwidth]{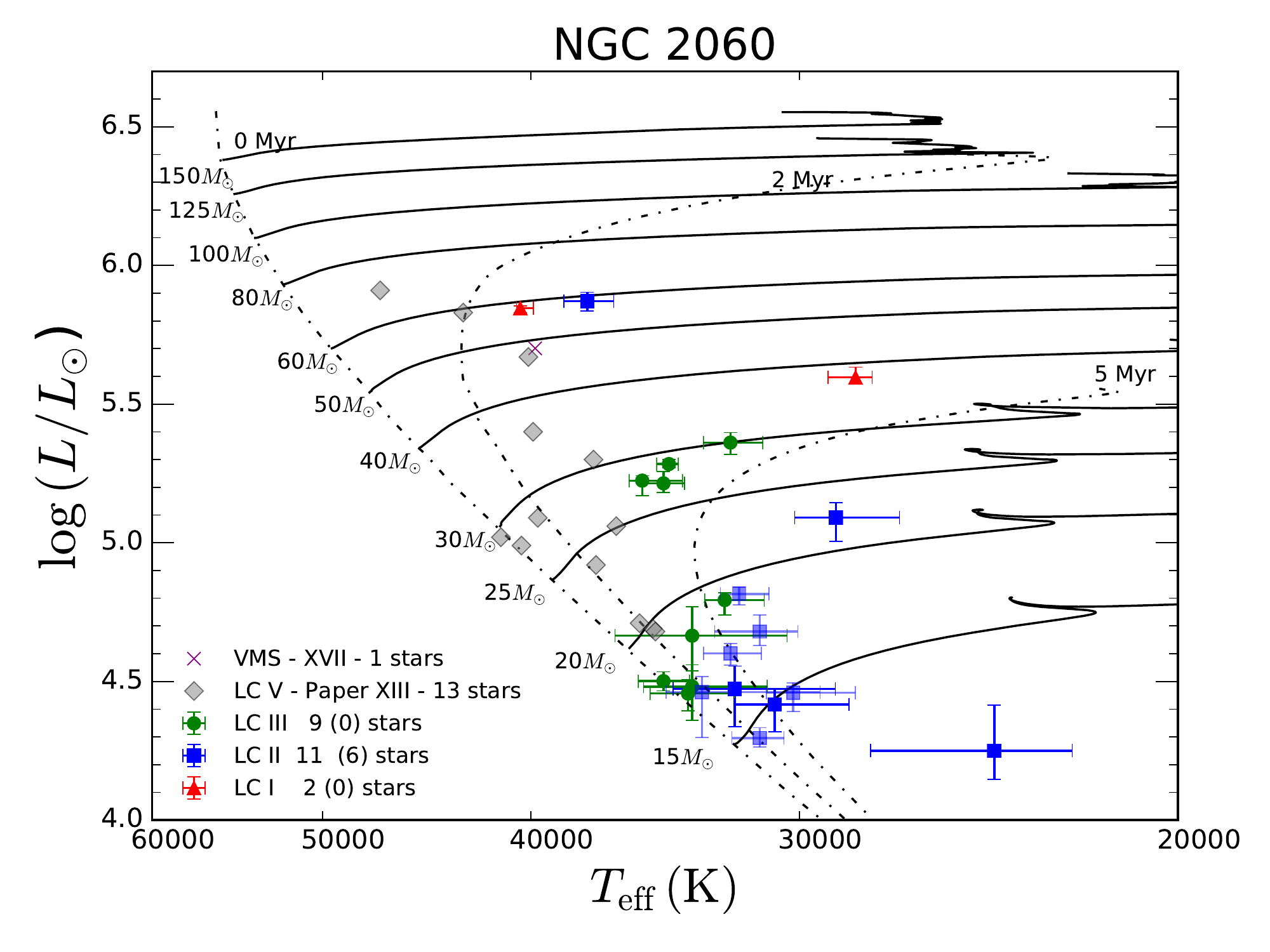}
\includegraphics[width=\columnwidth]{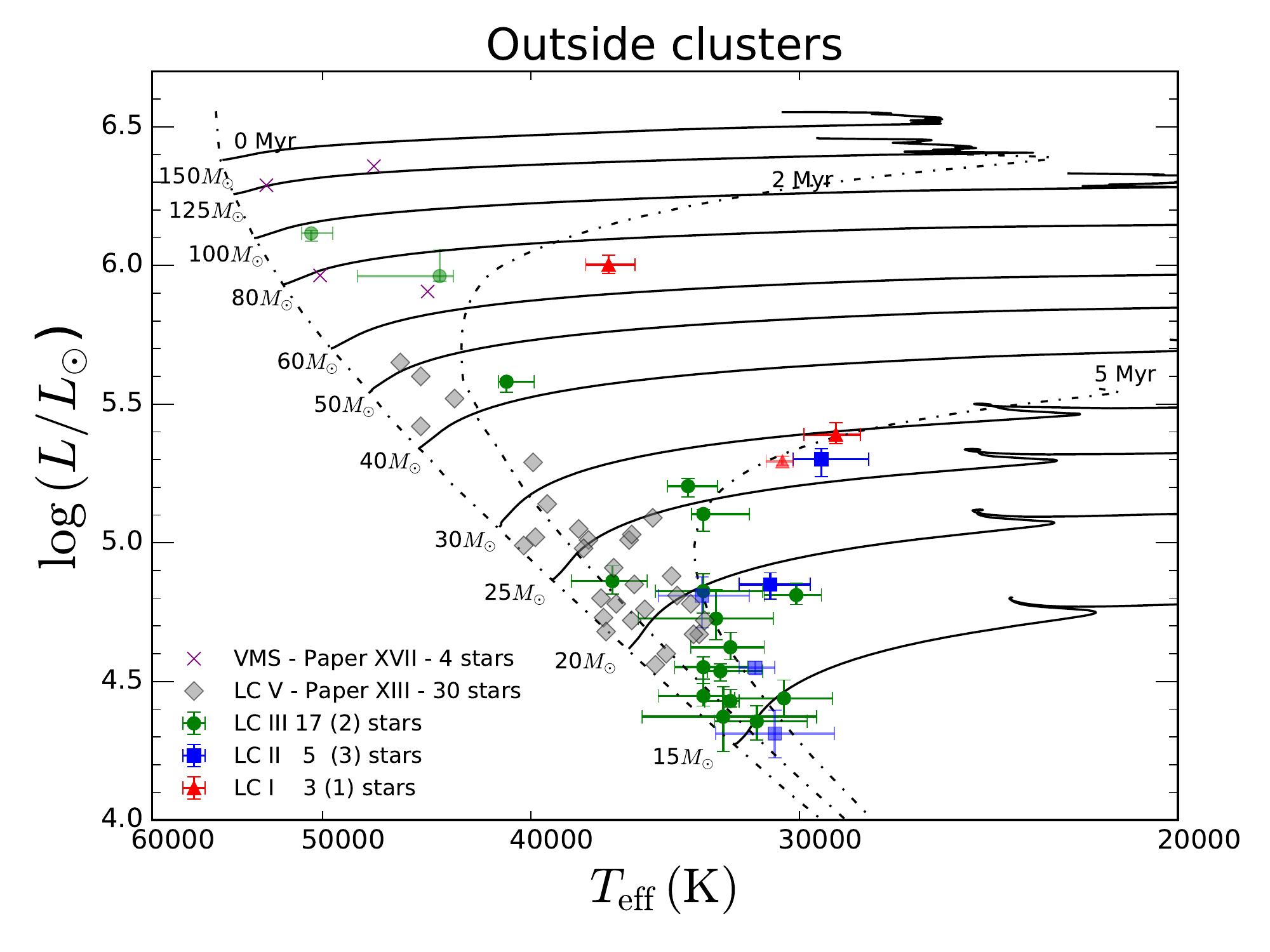}
\caption{Hertzsprung-Russell diagram of the sample as a function of spatial location: NGC\,2070 (upper panel), 
NGC\,2060 (middle panel) and stars \textit{outside star-forming complexes} (lower panel).
Evolutionary tracks (solid lines) and isochrones (dot-dashed lines) are from 
stellar models initially rotating with approximately  200\,\kms\  \citep[][]{brott,kohler2015}. 
}
\label{fig:HR_clusters}
\end{figure}

In Sect.~\ref{ref:hrd}, we investigated the evolutionary status of
our sample stars by placing them in the Hertzsprung-Russell diagram. Fig.~\ref{fig:HR_clusters} shows HRDs of sub-populations selected
with respect to their spatial location: NGC\,2070 (upper panel),  NGC\,2060 (middle panel) and
stars outside the two star-forming complexes (lower panel), all complemented with the VFTS dwarfs and VMS 
stars. As pointed out, stars of LC\,III to I are, on average, more evolved than the LC\,V. 
Stars in R136 are not resolved spatially with VLT-FLAMES and are therefore omitted from our sample.  
Though it is likely that the stars in R136 are younger, no apparent age differences are present in the three populations
specified here: they all show an age spread between approximately 2 and 5 Myrs.

\section{Tables}
\label{online_table}
\onecolumn
\tiny{
\begin{landscape}
\setlength\LTcapwidth{\textwidth} 
\setlength\LTleft{0pt}            
\setlength\LTright{0pt}           
\begin{longtable}[H]{cccccccccccccccccccccccc}     
\caption{List of the diagnostic lines for the O giants, bright giants, and supergiants for which
hydrogen, helium, and nitrogen have been included and used simultaneously in the determination of the stellar and wind parameters.  
 The `X' and `0' indicate if the line-profile has been used or not, respectively.
}             
\label{table:stats_lines_N}\\     
\hline\hline\\[-4pt]                 
	& Ion & $\rm{H}$  & $\rm{H}$ & $\rm{H}$ & $\rm{H}$ & {He\,{\sc i+ii}}&{He\,{\sc i}} & {He\,{\sc i}}& {He\,{\sc i}} &   {He\,{\sc i}}   & {He\,{\sc ii}} & {He\,{\sc ii}}&  {He\,{\sc ii}} & N\,{\sc ii} & N\,{\sc iii} &  N\,{\sc iii}&  N\,{\sc iii}&  N\,{\sc iii}&  N\,{\sc iii}& N\,{\sc iii} &   N\,{\sc iv} &  N\,{\sc v}&  N\,{\sc v}\\
VFTS & Lines & $\delta$ & $\gamma$ & $\beta$  & $\alpha$  	& $\lambda$4026 & $\lambda$4387 & $\lambda$4471  & $\lambda$4713& $\lambda$4922 		& $\lambda$4200$^{a}$& $\lambda$4541 & $\lambda$4686 &
  $\lambda$3995 &  $\lambda$4097 &  $\lambda$4195 &  $\lambda$4379 &  qua$^{b}$ &  $\lambda$4523&  trip$^{c}$ &  $\lambda$4058 &   $\lambda$4603 &  $\lambda$4619\\    
\hline\\[-4pt]                        
\endfirsthead
\hline\hline\\[-4pt]                 
	& Ion & $\rm{H}$  & & &            & {He\,{\sc i+ii}}&{He\,{\sc i}} & & &       & {He\,{\sc ii}} & &   & N\,II & N\,III & & & & & &  N\,IV & N\,V& \\
VFTS & Lines & $\delta$ & $\gamma$ & $\beta$  & $\alpha$  	& $\lambda$4026 & $\lambda$4387 & $\lambda$4471  & $\lambda$4713& $\lambda$4922 		& $\lambda$4200$^{a}$& $\lambda$4541 & $\lambda$4686  &
  $\lambda$3995 &  $\lambda$4097 &  $\lambda$4195 &  $\lambda$4379 &  qua$^{a}$ &  $\lambda$4523&  trip$^{b}$ &  $\lambda$4058 &   $\lambda$4603 &  $\lambda$4619\\    
\hline\\[-4pt]                        
\endhead
\hline
\endfoot
\hline
\multicolumn{24}{l}{$^{(a)}$\hel{ii}{4200} also includes \Nl{iii}{4200};  $^{(b)}$ \Nlmod{III}{qua} refers to \Nl{III}{4511}, 4515, 4518; $^{(c)}$ \Nlmod{III}{trip} refers to \Nl{III}{4634}, 4640.}
\endlastfoot  
016  & 14 & X	 & X	 & X	 & X	 & X	 & X	 & X	 & 0	 & X	 & X	 &X	 & X	 & 0	 &0	 & 0	 & 0	 & 0	 &0	 & 0	 & X	 &X	 & X \\
087  & 17 & X	 & X	 & X	 & 0	 & X	 & X	 & X	 & X	 & X	 & X	 &X	 & X	 & X	 &X	 & X	 & X	 & X	 & X	 & 0	 & 0	 & 0	 & 0 \\
178	 & 17 & X	 & X	 & X	 & 0	 & X	 & X	 & X	 & X	 & X	 & X	 &X	 & X	 & X	 &X	 & X	 & X	 & X	 &X	 & 0  	 & 0	 &0	 & 0\\
180  & 12 & X	 & X	 & X	 & 0	 & X	 & X	 & 0	 & 0	 & X	 & X	 &X	 & X	 & 0	 &0	 & 0	 & 0	 & X	 &0	 & X	 & X	 &0	 & 0 \\
259  & 13 & X	 & X	 & X	 & 0	 & X	 & X	 & 0	 & 0	 & X	 & X	 &X	 & X	 & 0	 &X	 & 0	 & 0	 & 0	 &0	 & X	 & 0	 &0	 & 0 \\
267	 & 14 & X	 & X	 & X	 & X	 & X	 & X	 & X	 & 0	 & X	 & X	 &X	 & X	 & 0	 &0	 & 0	 & 0	 & 0	 &0	 & 0	 & X	 &X	 & X \\
518	 & 15 & X	 & X	 & X	 & X	 & X	 & X	 & X	 & X	 & 0	 & X	 &X	 & X	 & 0	 &0	 & 0	 & 0	 & 0	 &0	 & X	 & X	 &X	 & X \\
566	 & 15 & X	 & X	 & X	 & X	 & X	 & X	 & X	 & 0	 & X	 & X	 &X	 & X	 & 0	 &0	 & 0 & 0	 & X	 &0	 & 0	 & X	 &X	 & X \\
\end{longtable}
\end{landscape}
}

\setlength\LTcapwidth{\textwidth} 
\begin{longtable}[H]{cccccccccccccc}        
\caption{List of the H and He diagnostic lines included in the analysis of the O giants, bright giants, and supergiants. 
The `X' and `0'  indicate if the line-profile has been used or not, as in Table~\ref{table:stats_lines_N}.
Newly identified binaries (see Sect.~\ref{sec:poor_fits}) are marked with 
an asterisk (*) in the first column.
}              
\label{table:stats_lines}\\     
\hline\hline\\[-4pt]                 
	& Ion & $\rm{H}$  & $\rm{H}$ & $\rm{H}$ & $\rm{H}$ & {He\,{\sc i+ii}} &{He\,{\sc i}} &{He\,{\sc i}} &{He\,{\sc i}} & {He\,{\sc i}}  & {He\,{\sc ii}} & {He\,{\sc ii}} & {He\,{\sc ii}} \\
VFTS & Lines & $\delta$ & $\gamma$ & $\beta$  & $\alpha$  	& $\lambda$4026 & $\lambda$4387 & $\lambda$4471  & $\lambda$4713& $\lambda$4922 		& $\lambda$4200& $\lambda$4541 & $\lambda$4686  \\    
\hline\\[-4pt]                        
\endfirsthead
\hline\hline\\[-4pt]                 
\hline\\[-4pt]                        
\endhead
\hline
\endfoot
035$\phantom{*}$ 	 & 12	 & X	 & X	 & X	 & X	 & X	 & X	 & X	 & X	 & X	 & X	 &X	 & X \\
046$\phantom{*}$ 	 & 12	 & X	 & X	 & X	 & X	 & X	 & X	 & X	 & X	 & X	 & X	 &X	 & X \\
070$\phantom{*}$ 	 & 12	 & X	 & X	 & X	 & X	 & X	 & X	 & X	 & X	 & X	 & X	 &X	 & X \\
076$\phantom{*}$ 	 & 12	 & X	 & X	 & X	 & X	 & X	 & X	 & X	 & X	 & X	 & X	 &X	 & X \\
077$\phantom{*}$ 	 & 12	 & X	 & X	 & X	 & X	 & X	 & X	 & X	 & X	 & X	 & X	 &X	 & X \\
080$\phantom{*}$ 	 & 12	 & X	 & X	 & X	 & X	 & X	 & X	 & X	 & X	 & X	 & X	 &X	 & X \\
091$\phantom{*}$ 	 & 12	 & X	 & X	 & X	 & X	 & X	 & X	 & X	 & X	 & X	 & X	 &X	 & X \\
103$\phantom{*}$ 	 & 12	 & X	 & X	 & X	 & X	 & X	 & X	 & X	 & X	 & X	 & X	 &X	 & X \\
\end{longtable}

\begin{centering}
\setlength\LTcapwidth{\textwidth} 
\begin{longtable}[H]{cccccccccccccc}        
\caption{List of the H and He diagnostic lines included in the analysis of O stars without luminosity class. 
The `X' and `0' indicate if the line-profile has been used or not, as in Table~\ref{table:stats_lines_N}.
Stars rated as \textit{poor quality} fits in Sect.~\ref{sec:poor_fits} are marked with 
an asterisk (*) in the first column.
}              
\label{table:stats_lines_pqf}\\     
\hline\hline\\[-4pt]                 
	& Ion & $\rm{H}$  & H & H &  H          & {He\,{\sc i+ii}} &{He\,{\sc i}} &{He\,{\sc i}} &{He\,{\sc i}} &   {He\,{\sc i}}                   & {He\,{\sc ii}}& {He\,{\sc ii}}& {He\,{\sc ii}}  \\
VFTS & Lines &$\delta$ & $\gamma$ & $\beta$  & $\alpha$  	& $\lambda$4026 & $\lambda$4387 & $\lambda$4471  & $\lambda$4713& $\lambda$4922 		& $\lambda$4200& $\lambda$4541 & $\lambda$4686  \\    
\hline\\[-4pt]                        
\endfirsthead
\endhead
\hline
\multicolumn{14}{l}{$^{(a)}$ For VFTS\,125, the following Nitrogen lines have also been used: \Nl{III}{4634}, 4640, \Nl{iv}{4058}, \Nl{v}{4603}, 4619.}
\endlastfoot 
051$\phantom{*}$ 	 & 11	 & X	 & X	 & X	 & 0	 & X	 & X	 & X	 & X	 & X	 & X	 &X	 & X \\
125$^a$ 	 & 15	 & X	 & X	 & X	 & 0	 & X	 & X	 & X	 & X	 & X	 & X	 &X	 & X \\
131$\phantom{*}$ 	 & 12	 & X	 & X	 & X	 & X	 & X	 & X	 & X	 & X	 & X	 & X	 &X	 & X \\
142$\phantom{*}$ 	 & 11	 & X	 & X	 & X	 & 0	 & X	 & X	 & X	 & X	 & X	 & X	 &X	 & X \\
177$\phantom{*}$ 	 & 10	 & X	 & X	 & X	 & 0	 & X	 & X	 & X	 & X	 & X	 & X	 &X	 & 0 \\
208$\phantom{*}$ 	 & 10	 & X	 & X	 & X	 & 0	 & X	 & X	 & X	 & X	 & X	 & X	 &X	 & 0 \\
373$\phantom{*}$ 	 & 12	 & X	 & X	 & X	 & X	 & X	 & X	 & X	 & X	 & X	 & X	 &X	 & X \\
393$\phantom{*}$ 	 & 11	 & X	 & X	 & X	 & 0	 & X	 & X	 & X	 & X	 & X	 & X	 &X	 & X \\
\end{longtable}
\end{centering}

\begin{landscape}
\setlength\LTcapwidth{\columnwidth} 
\setlength\LTleft{0pt}            
\setlength\LTright{0pt}           
\begin{longtable}[H]{clcccccccccccccc}       
\caption{Best-fitting atmospheric and wind parameters of our sample stars.
Parameters of stars that include nitrogen in the spectral analysis are  marked with $\dag$
in the first column (see also list in Table~\ref{table:stats_lines_N}). 
Newly detected binaries fits are listed separately. 
The quoted uncertainties identify the locations, with respect to the best-fit values, of the upper and lower boundaries of the 95\%\ confidence intervals. They reflect statistical uncertainties of the fitting process and hence do not account for systematic uncertainties (see Sects.~\ref{subsec:error_calculation} and~\ref{subsec:systematics}). 
}              
\label{table:new}\\      
\hline\hline\\[-9pt]
VFTS & SpT   & $M_\mathrm{K}$ & $T_{\rm eff}$ & $\log\,g$                   & $\log\,g_c^{a}$                   & $\log\,\dot{M}$                     & $Y^{b}$ & $\varv_{\mathrm{turb}}$       & \vrot\  & $\varv_{\infty}$ & $\log\,L$                           &             $R$                      &        $\log\,D_{\rm{mom}}$                                             &  $M_{\mathrm{spec}}^{c}$  &  $M_{\mathrm{evol}}$ \\    
           &          &  (mag) &      (kK)            & ($\rm{cm\,s^{-2}}$) & ($\rm{cm\,s^{-2}}$) & ($\rm{M_{\odot}\, yr^{-1}}$)&                   & (\kms)                     & (\kms) &      (\kms)           & ($\mathrm{L_{\odot}}$)  & ($\mathrm{R_{\odot}}$)&  ($\mathrm{g\,cm\,s^{-2}}$)  & ($M_{\odot}$)        & ($M_{\odot}$) \\    
\hline\\[-8pt]                        
\endfirsthead
\hline\hline\\[-9pt]
VFTS & SpT   & $M_\mathrm{K}$ & $T_{\rm eff}$ & $\log\,g$                   & $\log\,g_c^{a}$                   & $\log\,\dot{M}$                     & $Y^{b}$ & $\varv_{\mathrm{turb}}$       & \vrot\  & $\varv_{\infty}$ & $\log\,L$                           &             $R$                      &        $\log\,D_{\rm{mom}}$                                             &  $M_{\rm{spec}}^{c}$  &  $M_{\mathrm{evol}}$ \\    
           &          &  (mag) &      (kK)            & ($\rm{cm\,s^{-2}}$) & ($\rm{cm\,s^{-2}}$) & ($\rm{M_{\odot}\, yr^{-1}}$)&                   & (\kms)                     & (\kms) &      (\kms)           & ($\mathrm{L_{\odot}}$)  & ($\mathrm{R_{\odot}}$)&  ($\mathrm{g\,cm\,s^{-2}}$)  & ($M_{\odot}$)        & ($M_{\odot}$) \\    
\hline\\[-8pt]                        
\endhead
\hline
\endfoot
\hline
\multicolumn{15}{l}{$^{(a)}\log\,g_{\rm c} = \log\,[g+(\vrot)^2/R]$;  $^{(b)} Y$ is the helium mass fraction; 
$^{c}$ Spectroscopic mass computed with Newtonian gravity ($g_c$).}
\endlastfoot 
 016$^{\dag}$ & O2III-If*  & -5.3 & $50.60^{+0.51}_{-1.15}$ & $4.03^{+0.02}_{-0.03}$ & $4.03^{+0.02}_{-0.03}$ & $-5.00^{+0.05}_{-0.05}$ & $0.28^{+0.06}_{-0.03}$ & $27^{\uparrow}_{-4}$ & $112^{+20}_{-10}$ & $3631^{+85}_{-122}$ & $6.12^{+0.01}_{-0.03}$ & $15.09^{+0.19}_{-0.08}$ & $29.95^{+0.01}_{-0.05}$ & $89.9^{+4.2}_{-6.0}$ & $94.0^{+3.4}_{-3.8}$\\[2pt]	
 035$\phantom{*}$ & O9.5IIIn  & -2.4 & $32.55^{+2.95}_{-3.10}$ & $4.18^{+0.32}_{-0.47}$ & $4.27^{+0.31}_{-0.43}$ & $-6.60^{+0.30}_{-0.35}$ & $0.24^{+0.20}_{-0.08}$ & $24^{\uparrow}_{\downarrow}$ & $346^{+126}_{-106}$ & $2460^{+1039}_{-987}$ & $4.37^{+0.11}_{-0.12}$ & $4.90^{+0.28}_{-0.23}$ & $27.94^{+0.44}_{\downarrow}$ & $16.3^{+11.2}_{-7.4}$ & $16.2^{+1.0}_{-0.9}$\\[2pt]	
 046$\phantom{*}$ & O9.7II((n))  & -4.5 & $28.85^{+1.30}_{-1.90}$ & $3.23^{+0.10}_{-0.18}$ & $3.30^{+0.10}_{-0.16}$ & $-6.40^{+0.20}_{\downarrow}$ & $0.39^{+0.08}_{-0.08}$ & $5^{+10}_{\downarrow}$ & $168^{+24}_{-12}$ & $1405^{+157}_{-188}$ & $5.09^{+0.05}_{-0.08}$ & $14.25^{+0.56}_{-0.35}$ & $28.12^{+0.21}_{\downarrow}$ & $14.7^{+2.6}_{-2.9}$ & $20.8^{+2.0}_{-1.6}$\\[2pt]	
 070$\phantom{*}$ & O9.7II  & -2.7 & $32.15^{+2.20}_{-3.28}$ & $4.22^{+0.30}_{-0.26}$ & $4.23^{+0.30}_{-0.26}$ & $-7.35^{+0.60}_{\downarrow}$ & $0.25^{+0.10}_{\downarrow}$ & $9^{+18}_{\downarrow}$ & $126^{+44}_{-44}$ & $2762^{+904}_{-575}$ & $4.47^{+0.08}_{-0.14}$ & $5.64^{+0.35}_{-0.20}$ & $27.27^{+0.60}_{\downarrow}$ & $19.7^{+14.3}_{-6.9}$ & $15.2^{+1.2}_{-1.1}$\\[2pt]	
 076$\phantom{*}$ & O9.2III  & -4.1 & $33.25^{+0.42}_{-1.60}$ & $3.55^{+0.09}_{-0.14}$ & $3.56^{+0.09}_{-0.14}$ & $-6.25^{+0.10}_{-0.59}$ & $0.28^{+0.13}_{-0.03}$ & $14^{+2}_{\downarrow}$ & $90^{+14}_{-4}$ & $1775^{+122}_{-251}$ & $5.10^{+0.02}_{-0.06}$ & $10.89^{+0.30}_{-0.09}$ & $28.32^{+0.03}_{-0.52}$ & $15.8^{+2.1}_{-3.7}$ & $22.6^{+1.8}_{-1.7}$\\
\end{longtable}
\end{landscape}

\begin{landscape}
\setlength\LTcapwidth{\columnwidth} 
\setlength\LTleft{0pt}            
\setlength\LTright{0pt}           
\begin{longtable}[H]{clcccccccccccccc}       
\caption{Best-fitting atmospheric and wind parameters of the stars with no luminosity class.
Parameters of the star VFTS\,125 include nitrogen in the spectroscopic analysis and hence this star is marked with $\dag$
in the first column (see also list in Table~\ref{table:stats_lines_N}). 
Stars with \textit{poor-quality} fits are listed separately. 
The quoted uncertainties identify the locations, with respect to the best-fit values, of the upper and lower boundaries of the 95\%\ confidence intervals. They reflect statistical uncertainties of the fitting process and hence do not account for systematic uncertainties (see Sects.~\ref{subsec:error_calculation} and~\ref{subsec:systematics}).
}              
\label{table:new_noLC}\\      
\hline\hline\\[-9pt]
VFTS & SpT   & $M_\mathrm{K}$ & $T_{\rm eff}$ & $\log\,g$                   & $\log\,g_c^{a}$                   & $\log\,\dot{M}$                     & $Y^{b}$ & $\varv_{\mathrm{turb}}$       & \vrot\  & $\varv_{\infty}$ & $\log\,L$                           &             $R$                      &        $\log\,D_{\rm{mom}}$                                             &  $M_{\rm{spec}}^{c}$  &  $M_{\mathrm{evol}}$ \\    
           &          &  (mag) &      (kK)            & ($\rm{cm\,s^{-2}}$) & ($\rm{cm\,s^{-2}}$) & ($\rm{M_{\odot}\, yr^{-1}}$)&                   & (\kms)                     & (\kms) &      (\kms)           & ($\mathrm{L_{\odot}}$)  & ($\mathrm{R_{\odot}}$)&  ($\mathrm{g\,cm\,s^{-2}}$)  & ($M_{\odot}$)        & ($M_{\odot}$) \\    
\hline\\[-8pt]                        
\endfirsthead
\hline\hline\\[-9pt]
VFTS & SpT   & $M_\mathrm{K}$ & $T_{\rm eff}$ & $\log\,g$                   & $\log\,g_c^{a}$                   & $\log\,\dot{M}$                     & $Y^{b}$ & $\varv_{\mathrm{turb}}$       & \vrot\  & $\varv_{\infty}$ & $\log\,L$                           &             $R$                      &        $\log\,D_{\rm{mom}}$                                             &  $M_{\rm{spec}}^{c}$  &  $M_{\mathrm{evol}}$ \\    
           &          &  (mag) &      (kK)            & ($\rm{cm\,s^{-2}}$) & ($\rm{cm\,s^{-2}}$) & ($\rm{M_{\odot}\, yr^{-1}}$)&                   & (\kms)                     & (\kms) &      (\kms)           & ($\mathrm{L_{\odot}}$)  & ($\mathrm{R_{\odot}}$)&  ($\mathrm{g\,cm\,s^{-2}}$)  & ($M_{\odot}$)        & ($M_{\odot}$) \\    
\hline\\[-8pt]                        
\endhead
\hline
\endfoot
\hline
\multicolumn{15}{l}{$^{(a)}\log\,g_{\rm c} = \log\,[g+(\vrot)^2/R]$;  $^{(b)} Y$ is the helium mass fraction; 
$^{c}$ Spectroscopic mass computed with Newtonian gravity ($g_c$).}
\endlastfoot  
 051$\phantom{*}$ & OBpe  & -4.5 & $28.25^{+2.05}_{-3.55}$ & $3.01^{+0.22}_{-0.29}$ & $3.44^{+0.08}_{-0.07}$ & $-6.00^{+0.25}_{\downarrow}$ & $0.17^{+0.04}_{\downarrow}$ & $5^{+6}_{\downarrow}$ & $412$ & $1077^{+308}_{-285}$ & $5.03^{+0.09}_{-0.17}$ & $13.91^{+1.11}_{-0.54}$ & $28.40^{+0.36}_{-1.63}$ & $19.6^{+4.6}_{-2.8}$ & $20.4^{+0.7}_{-1.8}$\\[2pt]
 125$^{\dag}$ & Ope  & -4.5 & $55.15^{\uparrow}_{\downarrow}$ & $4.00^{+0.21}_{-0.37}$ & $4.04^{+0.21}_{\downarrow}$ & $-6.05^{+0.51}_{\downarrow}$ & $0.19^{+0.16}_{\downarrow}$ & $16^{\uparrow}_{\downarrow}$ & $274^{+100}_{-88}$ & $2847^{+551}_{-850}$ & $5.90^{\uparrow}_{-0.38}$ & $9.94^{+1.70}_{\downarrow}$ & $28.70^{+0.73}_{-1.60}$ & $39.9^{+20.2}_{\downarrow}$ & $60.8^{\uparrow}_{\downarrow}$\\[2pt]
 131$\phantom{*}$ & O9.7  & -2.2 & $33.55^{+4.12}_{-3.10}$ & $4.58^{+0.43}_{-0.30}$ & $4.59^{+0.44}_{-0.30}$ & $-7.50^{+1.10}_{\downarrow}$ & $0.18^{+0.13}_{\downarrow}$ & $16^{\uparrow}_{\downarrow}$ & $124^{+192}_{-78}$ & $3730^{+2223}_{-1048}$ & $4.35^{+0.16}_{-0.13}$ & $4.49^{+0.30}_{-0.32}$ & $27.20^{+1.06}_{\downarrow}$ & $28.2^{+34.7}_{-12.8}$ & $15.4^{+1.1}_{-0.7}$\\[2pt]	
 142$\phantom{*}$ & Op  & -3.2 & $37.80^{+1.05}_{-1.50}$ & $4.22^{+0.15}_{-0.14}$ & $4.22^{+0.15}_{-0.14}$ & $-6.50^{+0.30}_{-0.90}$ & $0.24^{+0.08}_{-0.05}$ & $9^{+15}_{\downarrow}$ & $72^{+34}_{-36}$ & $3014^{+557}_{-396}$ & $4.91^{+0.03}_{-0.05}$ & $6.71^{+0.15}_{-0.10}$ & $28.19^{+0.32}_{\downarrow}$ & $27.4^{+9.6}_{-6.7}$ & $23.6^{+0.6}_{-0.8}$\\[2pt]	
 177$\phantom{*}$ & O7n(f)p  & -4.7 & $34.60^{+1.30}_{-0.35}$ & $3.55^{+0.09}_{-0.17}$ & $3.66^{+0.09}_{-0.15}$ & $-5.45^{+0.10}_{-0.20}$ & $0.26^{+0.10}_{-0.06}$ & $30^{\uparrow}_{-6}$ & $310^{+38}_{-28}$ & $2018^{+220}_{-358}$ & $5.40^{+0.05}_{-0.01}$ & $14.08^{+0.08}_{-0.29}$ & $29.23^{+0.09}_{-0.26}$ & $32.7^{+4.9}_{-8.8}$ & $31.2^{+1.0}_{-1.1}$\\
\end{longtable}
\end{landscape}

\section{Special remarks on the fitted stars} \label{app:remarks}
This section presents notes on the fitting results of some of the sample stars as well as additional remarks on peculiar properties of the objects that are presented in Appendix~\ref{app:example}. Some stars from our presumably single-star sample were part of the Tarantula Massive Binary Monitoring (TMBM), a 32-epoch radial velocity program that followed up on VFTS RV variable stars \citep[][]{almeida_et_al_2016}. In those cases, preliminary results are also included. Details on the spectral classification are included in Tables~A.1 and A.2 of \citet{walborn2014} and hence are not repeated here. 
\begin{itemize}
    \item \textbf{VFTS\,046} is one of the helium-enriched stars. The cores of the Balmer lines are clipped because of remaining residuals of nebular contamination. The best models overpredict \hel{i}{4387} by 5\%\ with respect to the continuum. The rest of the line profiles are very well reproduced (rms better than 1\%\ of the continuum). 
    \item \textbf{VFTS\,064 (SB1)}. The cores of the Balmer lines are clipped  because of remaining residuals of nebular contamination. Except for \hel{ii}{4686}, the lines are well fitted (rms better than 1\%\ of the continuum). The data of \hel{ii}{4686} are red-shifted by $\sim$2~\AA\ compared to the models. 
    TMBM has identified the star as a single-lined (SB1) spectroscopic binary with a tentative period of approximately 900~d. Its binary nature is likely the cause of the \hel{ii}{4686} discrepancy. This star is excluded from the analysis in Sect.~\ref{sec:results}.
    
    \item \textbf{VFTS\,087}'s H$\alpha$ observed profile is not used because it presents strong nebular emission. The cores of  H$\beta-\gamma$ are clipped because of remaining residuals of nebular contamination. Its spectrum also shows strong Nitrogen lines (i.e., \Nl{ii}{3995}, \Nl{iii}{4097}, 4195, 4379, 4511, 4515, 4518, 4523) and hence they are taken into account for the fit. Most of the lines are well fitted (rms better than 1\%\ of the continuum) except for \hel{ii}{4686}. The data of \hel{ii}{4686} are slightly red-shifted by $\sim$0.2~\AA\ compared to the models. Further, the star shows  a significant but moderate RV variability with a peak-to-peak amplitude of approximately 12~\kms. It was followed up by TMBM but no periodicity was found. 

    \item \textbf{VFTS\,103'}s Balmer line cores and \hel{i}{4713,4922} cores are clipped
    because of remaining residuals of nebular contamination. The models overpredict 
     \hel{ii}{4200,4541} by 3\%\  and 2\%\  with respect to the continuum, respectively. For the remaining lines the differences are less than 1\% (with respect to the continuum). \hel{ii}{4686} shows a tentative structure, possibly reminiscent of a blended profile.

    \item \textbf{VFTS\,104}'s VISTA  $K$-band measurement is not available. We have adopted the $K$ magnitude from the InfraRed Survey Facility \citepalias[IRSF; see also Table~6 of][]{evans}. The cores of the Balmer lines are clipped because of residuals of nebular contamination. The models fairly reproduce the line depths and widths of the spectral lines.
    \item \textbf{VFTS\,113}'s Balmer line cores are clipped because of significant nebular correction residuals. The models reproduce the spectral lines well. The star was followed up by TMBM but no periodicity was found despite a rather large peak-to-peak RV variability of 28~\kms.
    \item \textbf{VFTS\,141}'s Balmer line cores are clipped. The models underpredict \hel{ii}{4200,4541} by $\sim$2\%\  with respect to the continuum, possibly indicating that the temperature has been underestimated. The remaining  lines are all well reproduced by the model (rms better than 1\%). 

    \item \textbf{VFTS\,151}'s H$\alpha$ observed profile is not used because it presents strong emission. The cores of  H$\beta-\gamma$ are clipped because of remaining residuals of nebular contamination.  The best models overpredict \hel{i}{4387} by 3\%\ with respect to the continuum and fail to reproduce the wings of \hel{i}{4471} and the red wing of \hel{ii}{4686}. The remaining lines are well reproduced by the models and hence this star is included in the analysis of Sect.~\ref{sec:results} as an acceptable fit.
    
    \item \textbf{VFTS\,153}'s Balmer line cores are clipped. The best models overpredict \hel{ii}{4200} by 2\%\ with respect to the continuum and fail to reproduce the left wing of \hel{ii}{4686}. For the remaining lines, the models fairly reproduce the spectral lines (rms better than 1\%\ of the continuum) and hence this star is included in the analysis of Sect.~\ref{sec:results} as an acceptable fit.

    \item \textbf{VFTS\,160}'s H$\alpha-\beta-\gamma$ and \hel{i}{4387,4713,4922} cores are clipped. The models fail to reproduce the line depth and width of \hel{i}{4387}. For the remaining lines, the models (within the 95\%\ confidence interval) acceptably reproduce the spectral lines and hence this star is included in the analysis of Sect.~\ref{sec:results}. 
    \item \textbf{VFTS\,171 (SB1)}. The cores of the Balmer lines are clipped.
    The observed profile of \hel{i}{4026} and the wings of the Balmer lines are well fitted by the model. 
    The data of \hel{ii}{4686} are red-shifted by $\sim$0.5\,\AA\ compared to the best model. For the remaining lines, the model presents incoherent line depths and widths.
    The differences are approximately 3\%\ with respect to the continuum. VFTS\,171 has been dentified by TMBM as a 670-d SB1 object. Hence this star is excluded from 
    the analysis in Sect.~\ref{sec:results}.
    
    \item \textbf{VFTS\,178}'s H$\alpha$ observed profile is not used because it presents strong nebular emission. The core of  H$\beta$ is clipped because of remaining residuals of nebular contamination. Its spectrum also shows strong Nitrogen lines (i.e., \Nl{ii}{3995}, \Nl{iii}{4097}, 4195, 4379, 4511, 4515, 4518, 4523) and hence they are taken into account for the fit. Most of the lines are well fitted (rms better than 1\%\ of the continuum). VFTS\,178 was also followed up by TMBM but no periodicity was found ($\Delta RV=15$~\kms).

    \item \textbf{VFTS\,180} is one of the helium enriched stars. The  \hel{i}{4471,4713} and H$\alpha$ observed profiles are not used due to strong residuals of nebular contamination. The cores of the remaining Balmer lines are also clipped. Its spectrum shows strong Nitrogen lines (i.e, \Nl{iii}{4511}, 4515, 4518, 4630, 4640, \Nl{iv}{4058}) and hence they are taken into account for the fit. The data of \Nl{iv}{4058} are red-shifted by $\sim$1~\AA\ compared to the models. The remaining lines are fairly well reproduced by the models (rms better than 1\%\ of the continuum), including  the rather strong \hel{ii}{4686} emission line, which is typical for strong stellar winds. 
    
    \item \textbf{VFTS\,192}'s  Balmer line cores are clipped. The lines are well fitted (rms of the order of 1\% of the continuum), except for \hel{i}{4471}, although the latter differences probably result from a poor nebular correction in the observed data of that line. Finally, the \hel{ii}{4686} spectral line seems slightly blue-shifted with respect to the best fit-model ($\sim$0.2\,\AA) but it is within the 95\%\ probability models.  While this may indicate a binary nature, the best fit model is still representative of the observed data.
    
    \item \textbf{VFTS\,244}'s  Balmer line cores are clipped. The fit quality of the models
    is acceptable, with the exception of the \hel{ii}{4686} line. The latter displays a small P-Cygni shape, 
    with an absorption component that is significantly red-shifted (by $\sim$\,1\,\AA) with respect to the model.   
    
    \item \textbf{VFTS\,253}'s H$\alpha-\beta-\gamma$ cores are clipped. The fit quality of the models is acceptable, though the \hhel{i} line core seems narrower than the best-fit models but their wings are broader. This may indicate a composite nature or a significant macro-turbulent component (not included in our fitting approach).

    \item \textbf{VFTS\,259}'s H$\alpha$ observed profile is not used in the fit because of strong residuals of nebular contamination. The cores of H$\beta-\gamma$ are also clipped. For the fit, we have also used \Nl{iii}{4634}, 4640. The observed profile of \hel{ii}{4686} is blue-shifted by 1\,\AA\ with respect to the best fit model. The remaining lines are well reproduced by the 95\%\ models. The star displays RV variation amplitude of $\Delta RV = 23$~\kms\ according to TMBM. While a periodicity of 3.7~d was identified, the folded RV-curve  does not resemble that of a spectroscopic binary system, suggesting another origin for the line variability. Hence VFTS\,259 is included in the analysis of Sect.~\ref{sec:results}. 
    
    \item \textbf{VFTS\,267}'s  Balmer line and \hel{i}{4471} cores  are clipped because of remaining residuals of nebular contamination. The 
    observed profiles of the \hhel{i} lines are relatively weak suggesting a very hot star. The \Nl{IV}{4058}, and \Nl{V}{4603}, 4619 lines are thus included in the fitting. The \hel{i}{4026}, \hel{ii}{4200,4541} and the wings of H$\delta-\gamma-\beta$ are fairly well fitted by the best models. \hel{ii}{4686} and H$\alpha$ show P-cygni profiles, though  \hel{ii}{4686}  is peculiar with an apparent excess absorption on top of the P-cygni profile. The model reproduces the red wings of these two lines and \Nl{IV}{4058}, and \Nl{V}{4603,4619} within 2\%\ with respect to the continuum. The data of the Nitrogen lines are slightly red-shifted (0.5\,\AA) with respect to the best model. TMBM identified a significant $\Delta RV$ of 22~\kms\ but no periodicity.

    \item \textbf{VFTS\,306}'s Balmer line cores are are clipped. 
    The  \hel{ii}{4686} line profile  is red-shifted by $\sim$1~\AA\ compared to the best model.  For the remaining lines, the fit quality of the 95\%\ probability models is acceptable (rms of 1\%\ with respect to the continuum).
    
   \item \textbf{VFTS\,332  (SB1)}. The cores of the 
   Balmer lines are clipped because of remaining residuals of nebular contamination. Only \hel{i}{4026} and the wings of H$\delta-\gamma-\beta$ are well fitted. 
   The data of \hel{ii}{4686} is red-shifted by $\sim$0.5\,\AA\ compared to the model. For the remaining lines, models within our 95\%\ confidence interval manage to reproduce most of the profile properties. TMBM results however revealed it is a long orbital period ($P\sim 3$~yr) binary.
   
   \item \textbf{VFTS\,333  (SB1)}. The cores of the Balmer lines are clipped because of remaining residuals of nebular contamination. The data of \hel{ii}{4686} is red-shifted by $\sim$1.0\,\AA\ compared to the model. For the remaining lines, models within our 95\%\ confidence interval manage to reproduce most of the profile properties.    TMBM results however revealed a binary nature with a period of $P\sim 3$~yr.
    
    \item \textbf{VFTS\,370}'s Balmer line cores and the red wing of \hel{i}{4471} are clipped because of remaining residuals of nebular contamination. The \hel{i}{4026}, left wing of \hel{i}{4471} and the wings of the Balmer lines are well fitted (rms better than 1\%\ of the continuum). The model underpredicts \hel{ii}{4200}, 4541 by 1\%\ with respect to the continuum. For the remaining lines,  the line depths (widths) are slightly underestimated (overestimated). 
    The differences are, however,  small ($<$2\%\ with respect to the continuum) and seemingly within the range of models that meet our 95\%-confidence criteria. 
    
    \item \textbf{VFTS\,399}'s  H$\alpha$ observed profile is not used because it presents strong emission. Cores of the remaining Balmer and \hel{i}{4471} lines are clipped  because of residuals of nebular contamination.  
    The model underpredicts \hel{i}{4200}, 4541, and 4686 by 2\%\ with respect to the continuum. The remaining lines are well fitted by the model (rms better that 1\%\ with respect to the continuum). VFTS\,399 was identified as an X-ray binary by \citet{clark2015}.

    \item \textbf{VFTS\,440 (SB1).} The cores of the Balmer lines are clipped because of remaining residuals of nebular contamination. The observed profile of \hel{ii}{4686} presents an inverse P-cygni profile, of which the absorption is red-shifted by 2\,\AA\ compared to the best-fit model. The model fails to reproduce the line depths and widths of all lines. Hence this star is excluded from further analysis in Sect.~\ref{sec:results}. TMBM result indicate that it is a $P\sim100$~d binary, providing an explanation for the shift in the \hel{ii}{4686} spectral line. In such a scenario, the peculiar  \hel{ii}{4686} profile may resuls from an isothermal wind-wind collision zone, as in the case of, HD~152248 \citep{sana2001}, for example.
    
    \item \textbf{VFTS\,466}'s Balmer line cores are clipped. The \hel{i}{4026} and the wings of the Balmer lines are well fitted by the model (rms better than 1\%\ of the continuum).
    The data of \hel{i}{4471} present an asymmetry in the core and hence it is not well reproduced by the model; such core-infilling may result from limited nebular correction quality. The lines \hel{i}{4713} and \hel{ii}{4200} are underpredicted by 2\%\ with respect to the continuum by the best-fit models, but some models within the 95\%-confidence interval better reproduce these lines. The remaining lines are all well fitted by the model (rms better than 1\%\ of the continuum).

    \item \textbf{VFTS\,502}'s Balmer line and \hel{i}{4471}, and 4922 cores are clipped because of remaining residuals of nebular contamination. The fit quality of the models is acceptable, with the exception of the right wing of  H$\gamma$ and  \hel{ii}{4686}. The former presents a blend that is not reproduced by the models. The latter is slightly red-shifted with respect to the best-fit model ($\sim$0.5\,\AA) but it is within the 95\%\ probability models.    

    \item \textbf{VFTS\,503}'s $K$-band magnitude from VISTA is not available, thus we have adopted the $K$ magnitude from VLT-MAD observations of \citet{campbell2010}.  The cores of the Balmer and \hel{i}{4026}, 4471, and 4922 lines are clipped because of remaining residuals of nebular contamination. The fit quality of the models is acceptable. 
    The data of \hel{ii}{4686} are slightly red-shifted with respect to the best-fit model ($\sim$0.5\,\AA) but still it is within the 95\%\ probability models.    
    
    \item \textbf{VFTS\,513}'s \hel{i}{4713} and H$\alpha$ observed profiles are not used  because of contamination from strong emission. Large part of the cores of the remaining Balmer and \hhel{i} lines are clipped. The H$\delta$ and \hel{ii}{4200} lines are well fitted (rms better than 1\%\ of the continuum). The fit quality of the models for the remaining lines is acceptable, with the exception of the \hel{ii}{4686} line. The latter is red-shifted with respect to the best-fit model by $\sim$1.0\,\AA\ but it is within the 95\%\ probability models.

    \item \textbf{VFTS\,518} is one of the helium enriched stars. The cores of the Balmer and \hel{i}{4026}, 4471 are clipped. The data of the \hhel{i} lines are relatively weak, thus the data of  \Nl{iii}{4634}, 4640, 4641, \Nl{IV}{4058}, and \Nl{V}{4603}, 4619 are included. Despite that the data of \hel{ii}{4686} and \Nl{IV}{4058} are slightly red-shifted with respect to the best-fit model ($<$0.3\,\AA), the model reproduces the depths and widths of the spectral lines (rms approximately 1\% of the continuum). 

    \item \textbf{VFTS\,546}'s Balmer line and \hhel{i} line cores are widely clipped because of residuals of nebular contamination. Except for the left wing of \hel{i}{4922}, the model well reproduces the data of \hel{ii}{4200}, 4541, and 4686  and the wings of the remaining lines  (rms of 1\% of the continuum). 

    \item \textbf{VFTS\,566}'s Balmer line and \hhel{i} line cores are clipped because of residuals of nebular contamination.  The data of the \hhel{i} lines are relatively weak, thus the data of \Nl{iii}{4634}, 4640, 4641, \Nl{IV}{4058}, and \Nl{V}{4603}, 4619 are included. Though the data of \hel{ii}{4686} and the nitrogen lines are slightly red-shifted with respect to the model ($<$0.5\,\AA), the model reproduces the depths and widths of the lines relatively well (rms approximately 1\% of the continuum). 

    \item \textbf{VFTS\,569}'s data of H$\alpha$ and \hel{i}{4713} are not used  because of contamination from strong nebular emission. The cores of Balmer and \hhel{i} lines are widely clipped. The models reproduce the profiles of \hel{ii}{4200}, 4541, and 4686 and the wings of the remaining lines  (rms of 1\% of the continuum).
    
    \item \textbf{VFTS\,571}'s $K$-band magnitude from VISTA is not available, thus we have adopted the $K$-band magnitude from VLT-MAD observations of \citet[][]{campbell2010}. The cores of the Balmer and \hhel{i} lines are clipped because of remaining nebular contamination. Best model underpredicts \hel{ii}{4200}, 4541 by 4\% with respect to the continuum. The other diagnostic lines are strong and are relatively well reproduced by the model (rms of 1\%\ of the continuum).
 
    \item \textbf{VFTS\,599'}s Balmer line  and \hel{i}{4387}, 4471 cores are clipped. The data of the \hhel{i} lines are relatively weak, thus the data of \Nl{iii}{4634}, 4640, 4641, \Nl{IV}{4058}, and \Nl{V}{4603,4619} are included. 	The data of \hel{ii}{4686} and the Nitrogen lines are slightly red-shifted with respect to the best-fit model ($<$0.5\,\AA), the model relatively well reproduces the depths and widths of the other lines (rms approximately 1\% of the continuum), despite an incomplete sampling of the parameter space.

    \item \textbf{VFTS\,620}'s $K$-band magnitude from VISTA is not available, thus we have adopted the $K$-band magnitude from VLT-MAD observations of \citet[][]{campbell2010}. The observed profile of H$\alpha$ is not considered because of contamination of strong nebular emission. The cores of the Balmer lines are clipped because of residual nebular contamination. The model fairly reproduces the remaining diagnostic lines (rms better than1\%\ of the continuum).
    
    \item \textbf{VFTS\,664}'s Balmer line cores are clipped because of remaining residuals of nebular contamination. The \hel{i}{4026} and \hel{ii}{4541} lines are well fitted by the best model (rms of 1\%\ of the continuum). For the remaining lines, the fit quality of the models are acceptable, except that they tend to over predict the full width at half maximum (FWHM) (hence probably the rotation rate as well) of the \hhel{i} lines.
    
    \item \textbf{VFTS\,669}'s H$\alpha$ observed profile is not used because it presents strong nebular emission. The cores of  H$\beta-\gamma$ are clipped because of remaining residuals of nebular contamination. To fit its spectrum we also used Nitrogen lines (i.e, \Nl{ii}{3995}, \Nl{iii}{4097}, 4195, 4379, 4511, 4515, 4518, 4523). Most of the lines are well fitted (rms better than 1\%\ of the continuum) except for \hel{ii}{4686}. The data of \hel{ii}{4686} was 2\%\ re-normalized  with respect to the continuum.
    
    \item \textbf{VFTS\,711}'s Balmer line and \hel{i}{4471} cores are clipped. Given the quality of the spectrum, we have fixed the \vrot\ to the value obtained in \citetalias{ramirezagudelo} (\vrot\,=39\,\kms). The \hel{ii}{4200}, 4541 and the Balmer wings are well fitted by the best model (rms of 1\%\ of the continuum). For the remaining lines, the fit quality of the models is acceptable.       

    \item \textbf{VFTS\,764}'s H$\alpha$ observed profile is not used because it presents strong nebular emission.   The data of the \hhel{ii} lines are relatively weak, the following Nitrogen lines: \Nl{ii}{3995}, \Nl{iii}{4097}, 4195, 4379, 4511, 4515, 4518, 4523 are included in the analysis. The fit quality of the 
    models is acceptable, with the exception of \Nl{ii}{3995}, H$\beta$ and \hel{ii}{4686}.
    TMBM revealed a RV variability with a $\Delta RV$ of 27\,\kms. The data are compatible with a periodicity of 1.2~d though 
    the folded RV-curve does not support a binary nature.  
    
   \item \textbf{VFTS\,777}'s Balmer line and \hel{i}{4471} cores are clipped. The fit best-fit model is acceptable, with the exception of the right wing of H$\delta$. The latter presents a blend that is not reproduced by the models.
\end{itemize}

\section*{Stars without luminosity classes}

Here we also provide stellar parameters of VFTS stars for which no LC classification could be assigned 
\citep[see][]{walborn2014}. Most of these stars present low $S/N$, strong residuals of the nebular correction, or both. The Balmer and \hhel{i} lines are often widely clipped. In a number of cases, the \hhel{i} information (width, amplitude) is almost entirely lost. Nevertheless, we comment here on our results in case they are useful for follow up investigations. 

\begin{itemize}
    \item \textbf{VFTS\,051}'s H$\alpha$ observed profile is not used because it presents strong emission. Given the quality of the spectrum, we have fixed the \vrot\ to the value obtained in \citetalias{ramirezagudelo} (\vrot\,= 412\,\kms). For the remaining Balmer lines, the cores are widely clipped because of remaining residuals of nebular contamination. The model well reproduces the wings of the Balmer lines. The shallow and noisy \hhel{i+ii} spectral lines do not add much information and hence the stellar parameters are poorly constrained.

    \item \textbf{VFTS\,125}'s Balmer line cores are widely clipped. The observed profiles of the \hhel{i} lines are relatively weak suggesting a very hot star. The \Nl{iii}{4634}, 4640, 4641 \Nl{IV}{4058}, and \Nl{V}{4603}, 4619 line profiles are thus included in the fitting. The model reproduces \hel{ii}{4200,4541} and the wings of H$\beta-\gamma-\delta$. Given the significant noise in the spectral lines, the stellar parameters present sizeable error bars.
    
    \item \textbf{VFTS\,145 is rated as a poor-quality  fit}. The observed profile of \hel{ii}{4686} is not used because it has a suspicious normalization. The cores of H$\alpha-\beta$ lines were clipped. Except for \hel{i+ii}{4026}, the model fails to reproduce the line depths and widths of the observed lines. The differences vary between 2\%\ (\hel{i}{4713}) and 5\%\ (H$\gamma$) with respect to the continuum.

    \item \textbf{VFTS\,360 is rated as a poor-quality fit}. The observed profiles of H$\alpha-\beta$ are not used because they present strong emission. The parameter space is deficiently explored and hence the model fails to reproduce line depths and widths of the most of the observed lines. 
    \item \textbf{VFTS\,373}'s Balmer line and \hel{i}{4026}, 4387, 4471 cores are clipped.  Except for \hel{i}{4387}, 4471, the model reproduces the observed diagnostic lines. The differences are of 1\%\ with respect to the continuum.

    \item \textbf{VFTS\,400 is rated as a poor-quality fit}. The observed profile of H$\alpha$ is not used because it presents strong emission. The cores of the remaining Balmer lines are clipped. Except for H$\gamma$, the model fails to reproduce the line depths and widths of the observed lines. The differences vary between 2\%\ (\hel{ii}{4541}) and 10\%\ (\hel{i}{4922}).

    \item \textbf{VFTS\,412}'s H$\alpha$ and \hel{i}{4713} observed profiles are not 
    used because they present strong emission. The cores of the remaining Balmer lines and \hel{i}{4471}, 4922 are clipped. The  models overpredict the FWHM (hence likely the rotation) of all He~{\sc i} lines and under-predict their amplitudes.
    
    \item \textbf{VFTS\,444}'s H$\alpha$ observed profile is not used because it presents strong emission.  The cores of the remaining Balmer lines are clipped. The core of \hel{ii}{4541} is also removed because of a strong absorption. Except for \hel{ii}{4200}, the model well reproduces the line depths and widths of the observed profiles (rms of 1\%\ with respect to the continuum).
    
    \item \textbf{VFTS\,446 is rated as a poor-quality fit}. The observed profiles of H$\alpha-\beta$  and \hel{i}{4922} are not used because they present strong emission. The cores of H$\gamma-\delta$ and \hel{i}{4026}, 4387, 4471 are widely clipped. Because the model can only rely on the \hhel{ii} lines, the parameters are poorly constrained.

    \item \textbf{VFTS\,451 is rated as a poor-quality fit.} The cores of the Balmer lines and \hel{i}{4387}, 4471 are widely clipped because of remaining residuals of nebular contamination. The data of \hel{ii}{4686} are red-shifted by 2\,\AA\ compared to the model. Except for \hel{ii}{4200}, 4541 the model has little information from the remaining observed lines and hence parameters are poorly constrained.
    \item \textbf{VFTS\,456}'s H$\alpha$ and \hel{i}{4713} observed profiles are not used because of strong emission. The cores of H$\beta-\gamma-\delta$, \hel{i}{4387}, 4471 are clipped. For the remaining lines, the model reproduces the observed profiles with 1\%\ with respect to the continuum. 

    \item \textbf{VFTS\,477}'s H$\alpha$ and \hel{i}{4713} observed profiles are not used because of strong emission. The cores of H$\beta-\gamma-\delta$ and \hel{i}{4387}, 4471, and 4922 are widely clipped. As a result, almost no information is provided by the \hhel{i} lines. The observed profile  of \hel{ii}{4686} is slightly red-shifted with respect to the model (0.5\,\AA), but in acceptable agrement given the large errors.
    
    \item \textbf{VFTS\,515}'s H$\alpha$ and \hel{i}{4713} observed profiles are not 
    used because they present strong emission. The cores of H$\beta-\gamma-\delta$ and \hel{i}{4387}, 4471, 4922 are widely clipped; most of the information provided by \hhel{i} is lost.  The model slightly underestimates \hel{ii}{4200}, 4541 but failed to reproduce \hel{ii}{4686} (no clear P-Cygni profile seen in the data). The estimated mass-loss rate is probably overestimated.
    
    \item \textbf{VFTS\,519}'s  H$\alpha$, \hel{i}{4713} and \hel{ii}{4686} observed profiles are not used because of strong emission. Except for H$\gamma-\delta$ and \hel{ii}{4200}, the observed profiles present a wrong normalization (i.e., continuum shifted by 5\% with respect to the continuum).     

    \item \textbf{VFTS\,565 is rated as a poor-quality fit}. The observed profile of H$\alpha$ is not used because it presents strong emission. The cores of the Balmer lines and \hel{i}{4387}, 4713, 4922 are clipped. The data of \hel{i}{4387}, 4471 are red-shifted by 1\,\AA\ while \hel{ii}{4686} is blue-shifted by 2\,\AA\ compared to the best model. Therefore, the observed profiles are poorly fitted.
    \item \textbf{VFTS\,594}. The observed profiles of H$\alpha$ and \hel{i}{4713} are not used because of strong emission.  The cores of H$\beta-\gamma-\delta$ and \hhel{i} lines are widely clipped. The models reproduce \hel{ii}{4200}, 4541 and the wings of the Balmer and \hhel{i} lines. The data of \hel{ii}{4686} are slightly red-shifted (0.3\,\AA) compared to the model.  
\end{itemize}

\section{Fitting results} \label{app:example}

In this appendix we show the spectra and the model fits for a selection of stars (20 stars). 
We first show the results for the LC\,III to I stars (14 out of the 72 stars) and then we proceed 
with the stars without LCs (6 out of the 31 stars).  The version of the appendix for the full sample will be available on A\&A.


\begin{figure}
\includegraphics[scale=0.45,trim={5cm 0.5cm 0cm 0cm}]{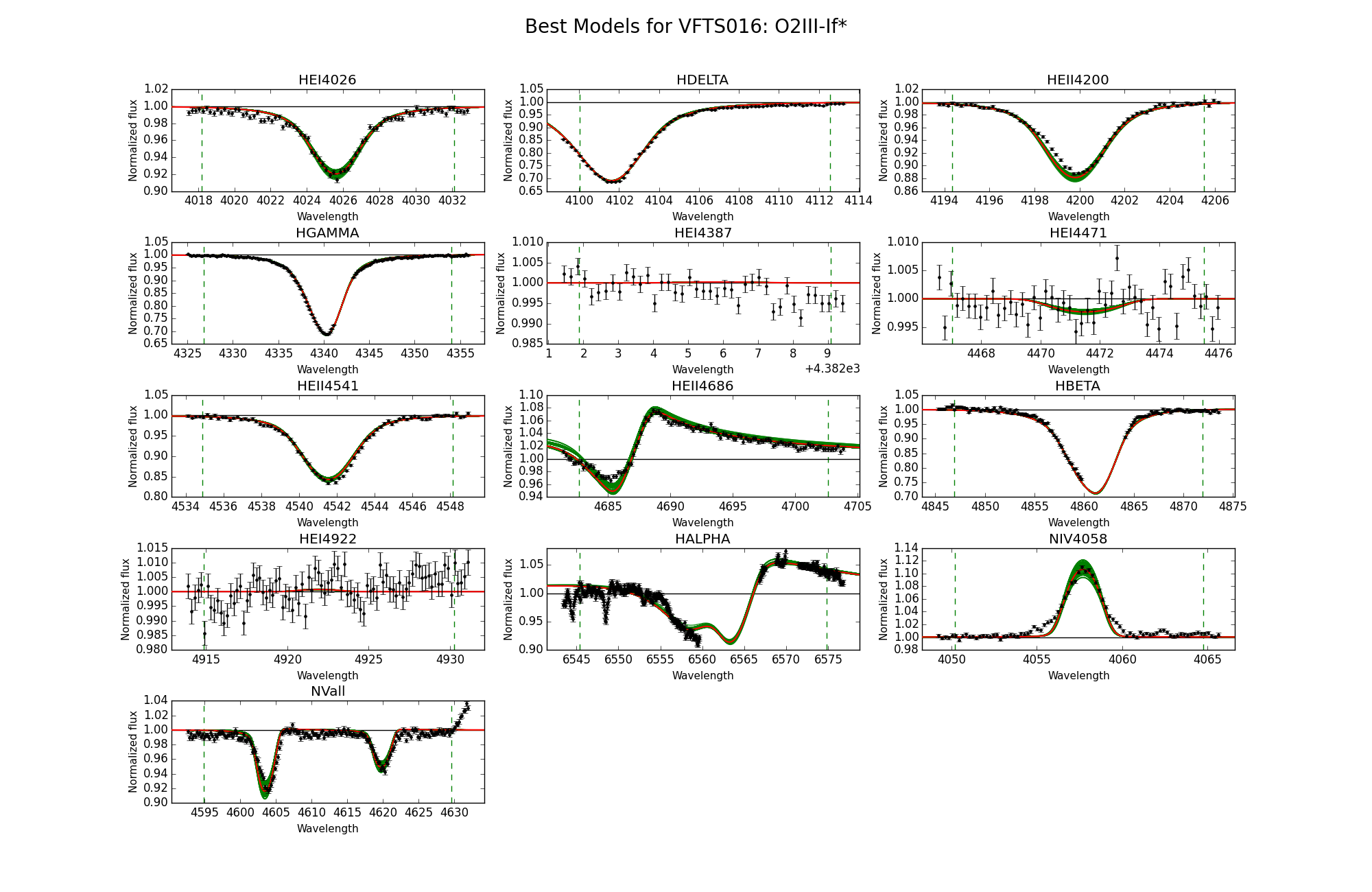}
\includegraphics[scale=0.45,trim={5cm 7.5cm 0cm 0cm}]{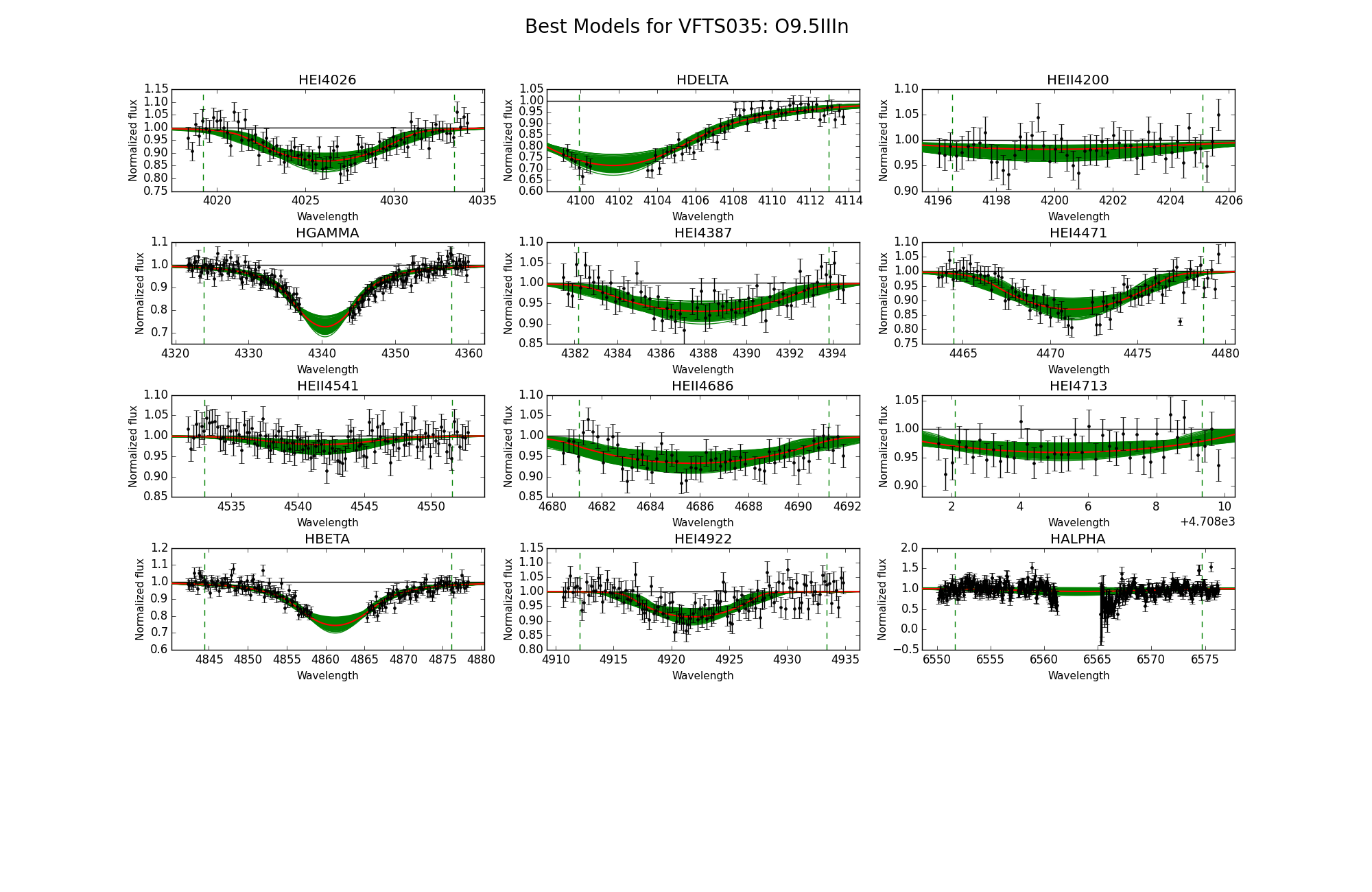}
\caption{Observed spectra, the 95\%\ probability models (green) and the best fit model (red) for VFTS\,016 and 035. The vertical dashed lines indicate 
the wavelength range used to fit the corresponding diagnostic line. [Color version available online.]}
\end{figure}

\FloatBarrier

\begin{figure}
\includegraphics[scale=0.45,trim={5cm 6.5cm 1cm 0cm}]{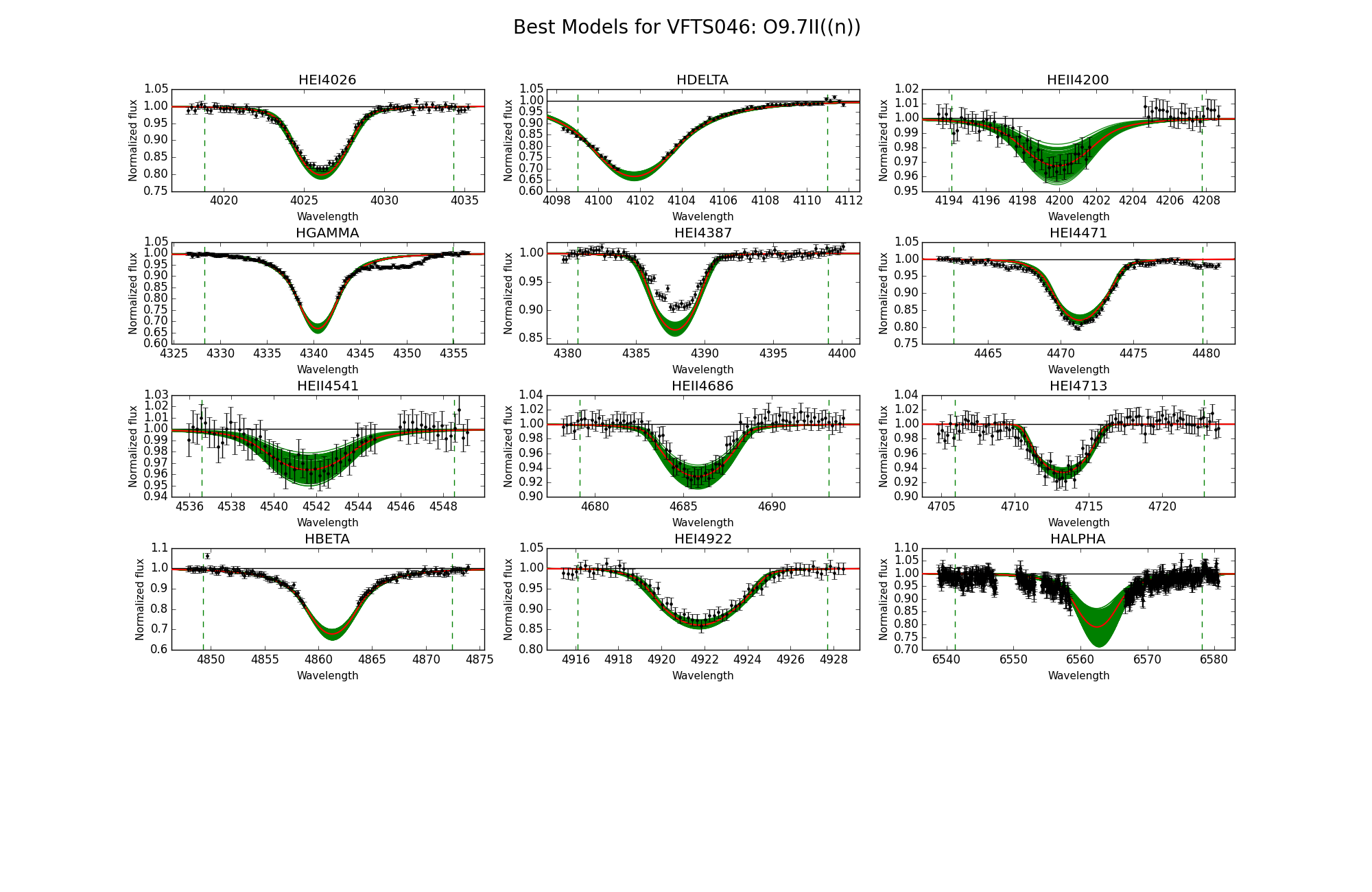}
\includegraphics[scale=0.45,trim={5cm 6.5cm 1cm 0cm}]{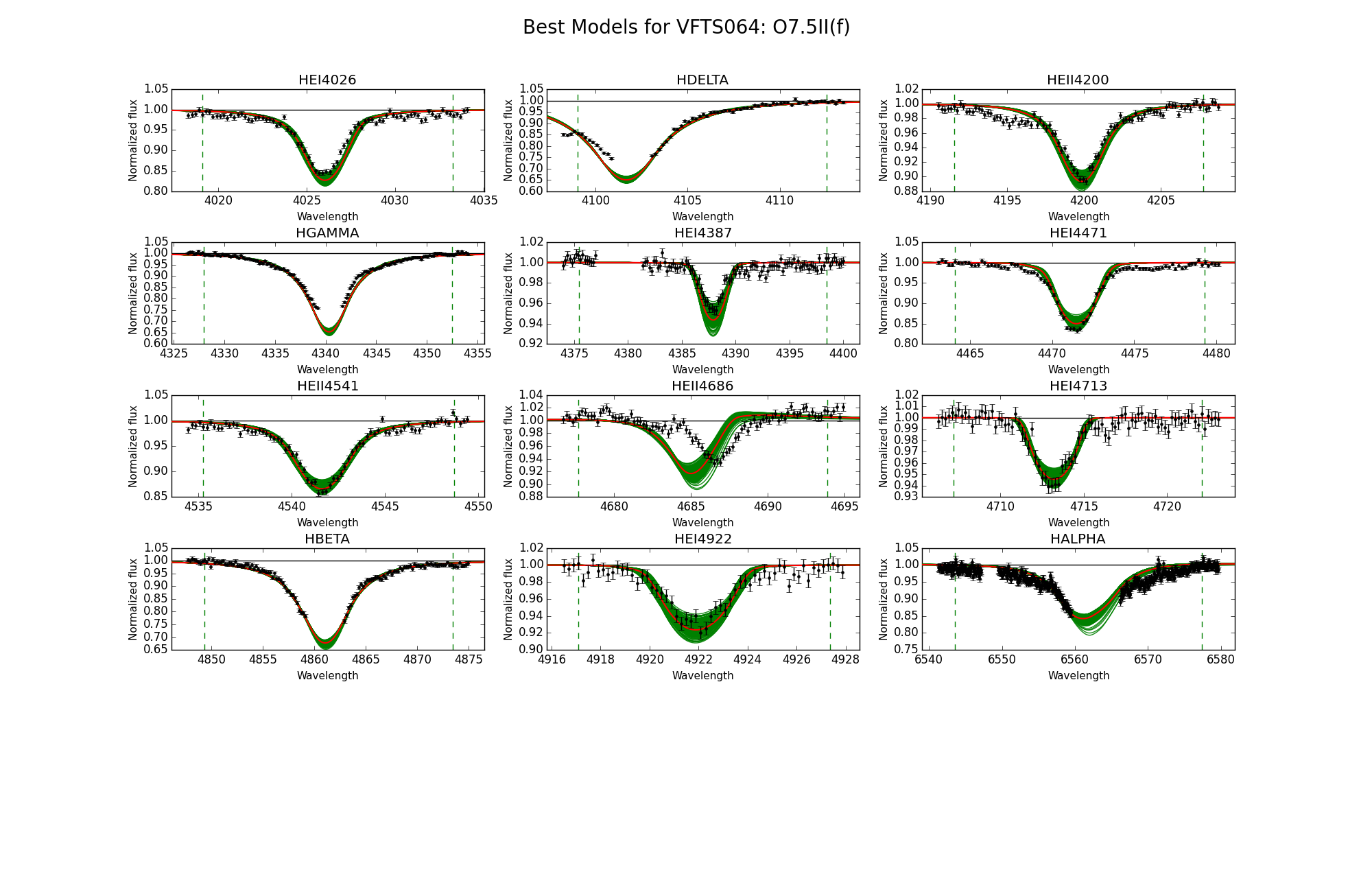}
\caption{Observed spectra, the 95\%\ probability models (green) and the best fit model (red) for VFTS\,046 and 064. The vertical dashed lines indicate 
the wavelength range used to fit the corresponding diagnostic line. [Color version available online.]}
\end{figure}

\begin{figure}
\includegraphics[scale=0.45,trim={5cm 6.5cm 1cm 0cm}]{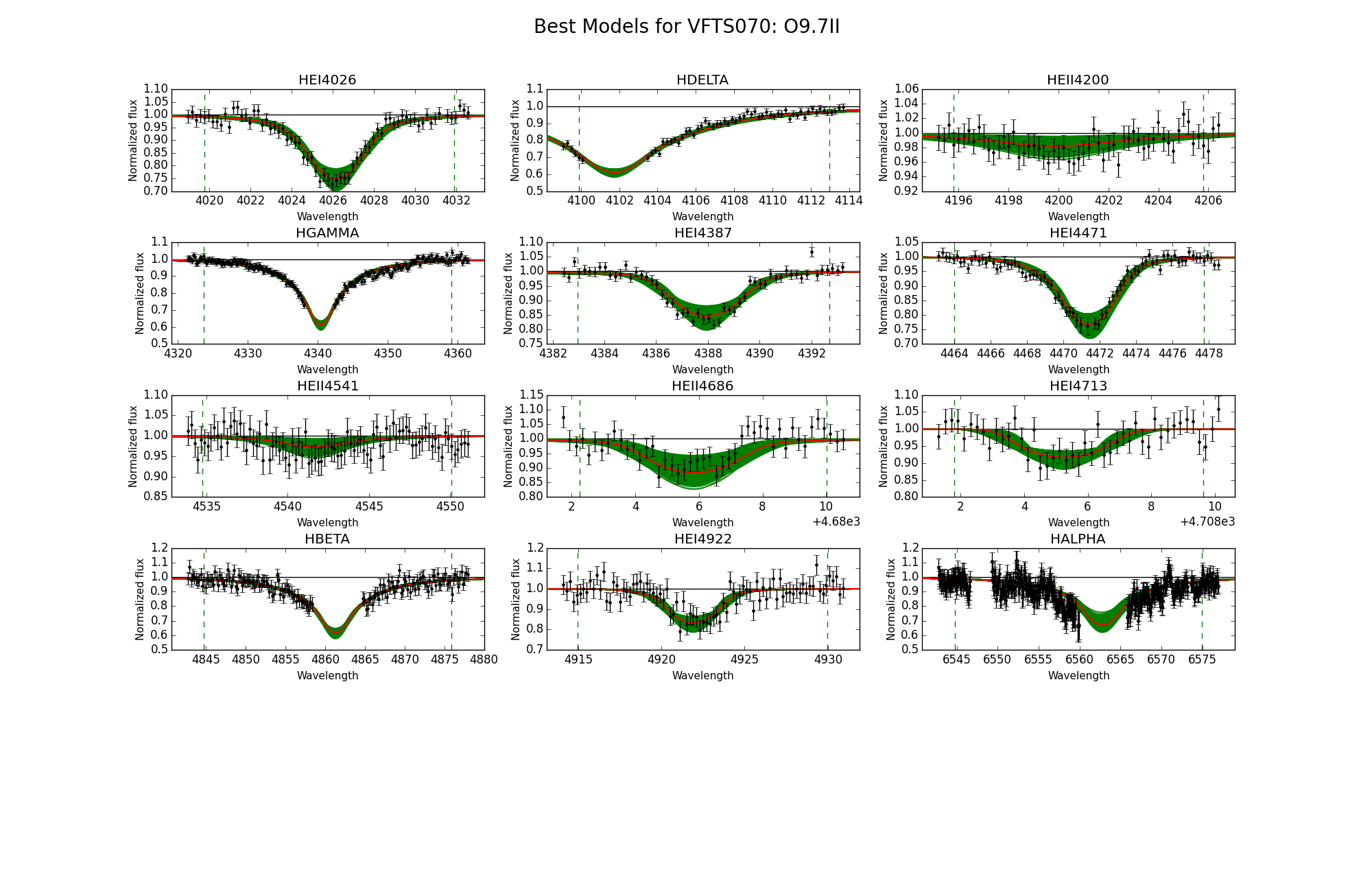}
\includegraphics[scale=0.45,trim={5cm 6.5cm 1cm 0cm}]{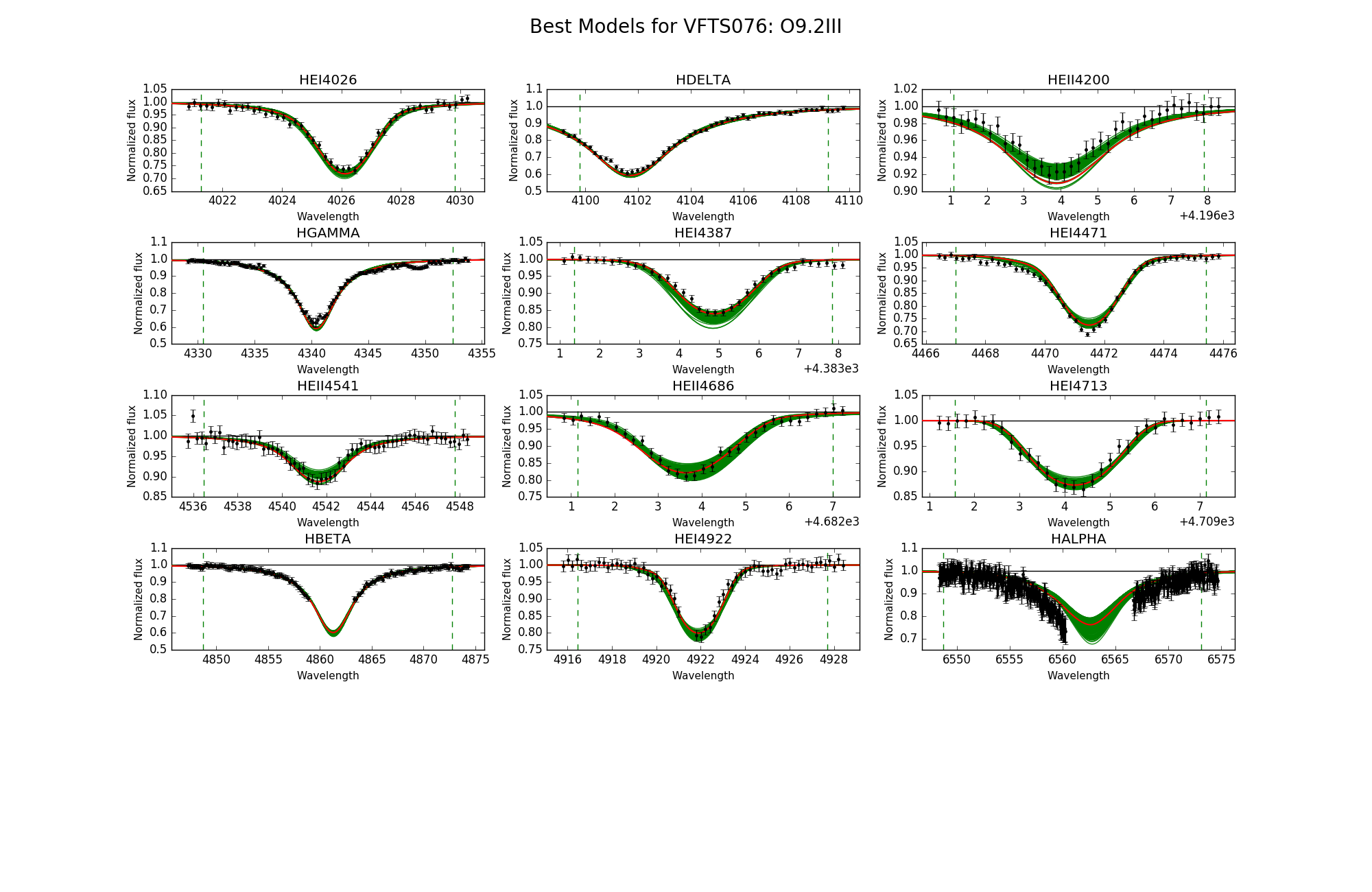}
\caption{Observed spectra, the 95\%\ probability models (green) and the best fit model (red) for VFTS\,070 and 076. The vertical dashed lines indicate 
the wavelength range used to fit the corresponding diagnostic line. [Color version available online.]}
\end{figure}

\begin{figure}
\includegraphics[scale=0.45,trim={5cm 6.5cm 1cm 0cm}]{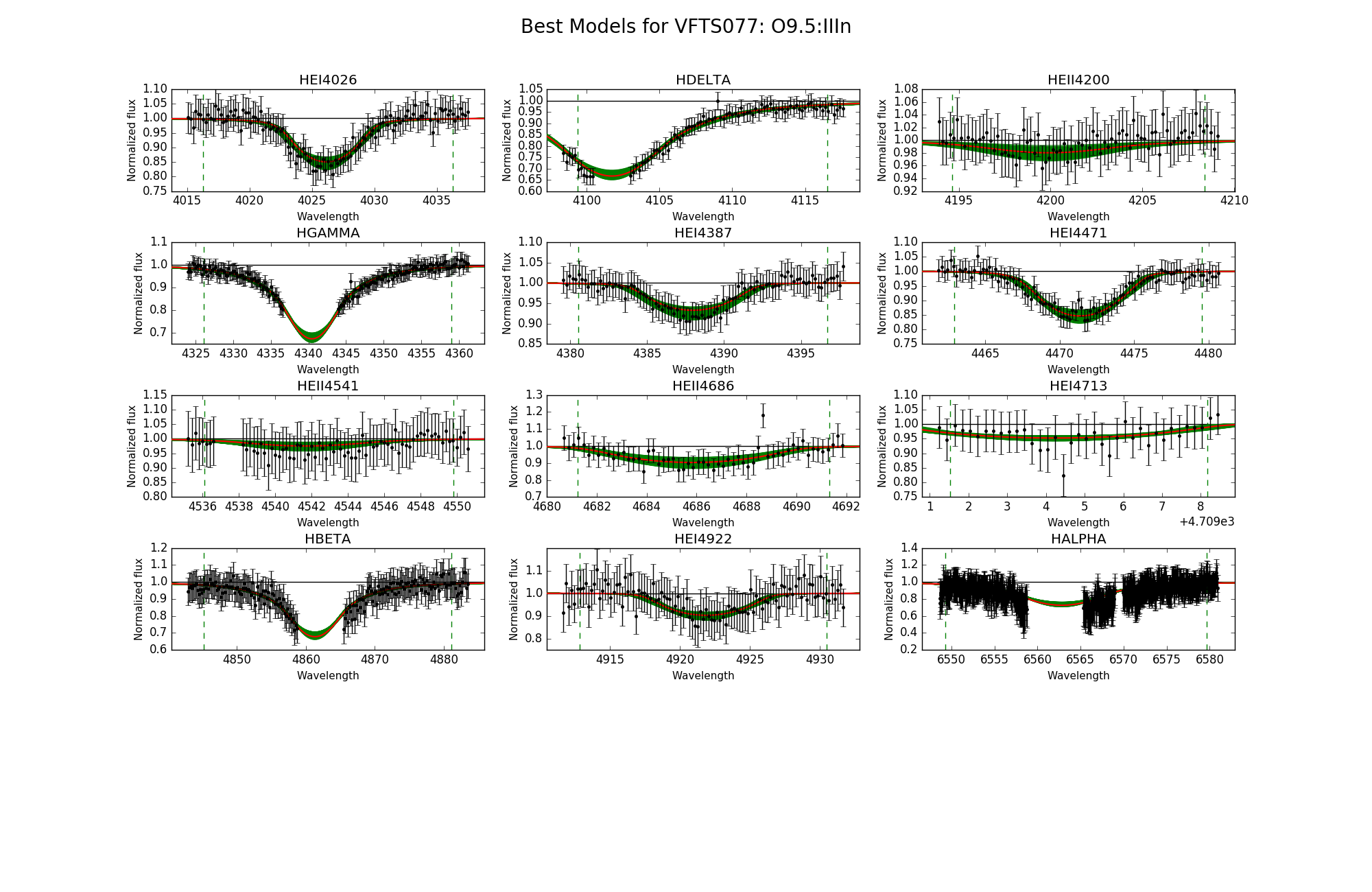}
\includegraphics[scale=0.45,trim={5cm 6.5cm 1cm 0cm}]{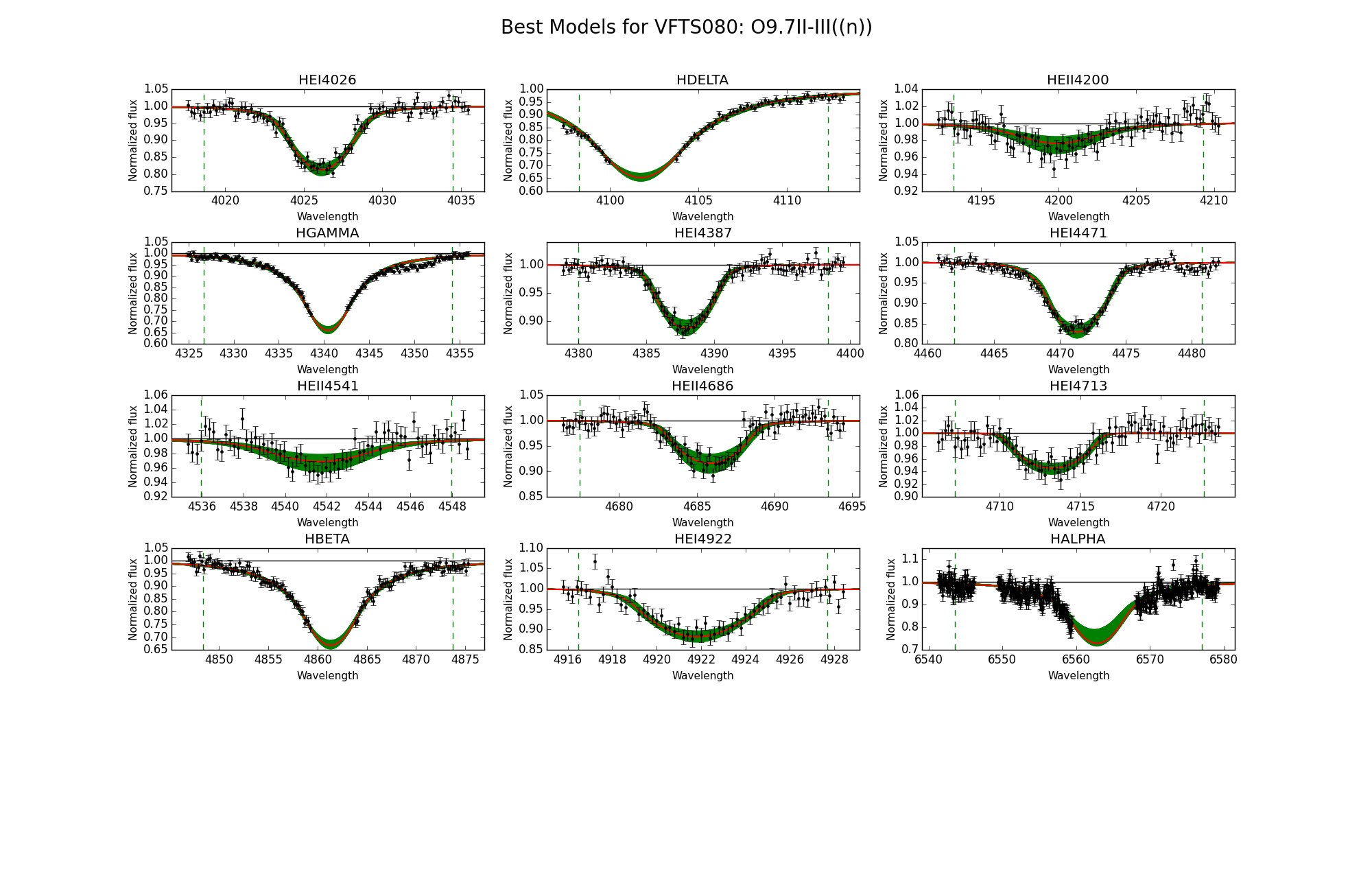}
\caption{Observed spectra, the 95\%\ probability models (green) and the best fit model (red) for VFTS\,077 and 080. The vertical dashed lines indicate 
the wavelength range used to fit the corresponding diagnostic line. [Color version available online.]}
\end{figure}

\begin{figure}
\includegraphics[scale=0.45,trim={5cm 6.5cm 1cm 0cm}]{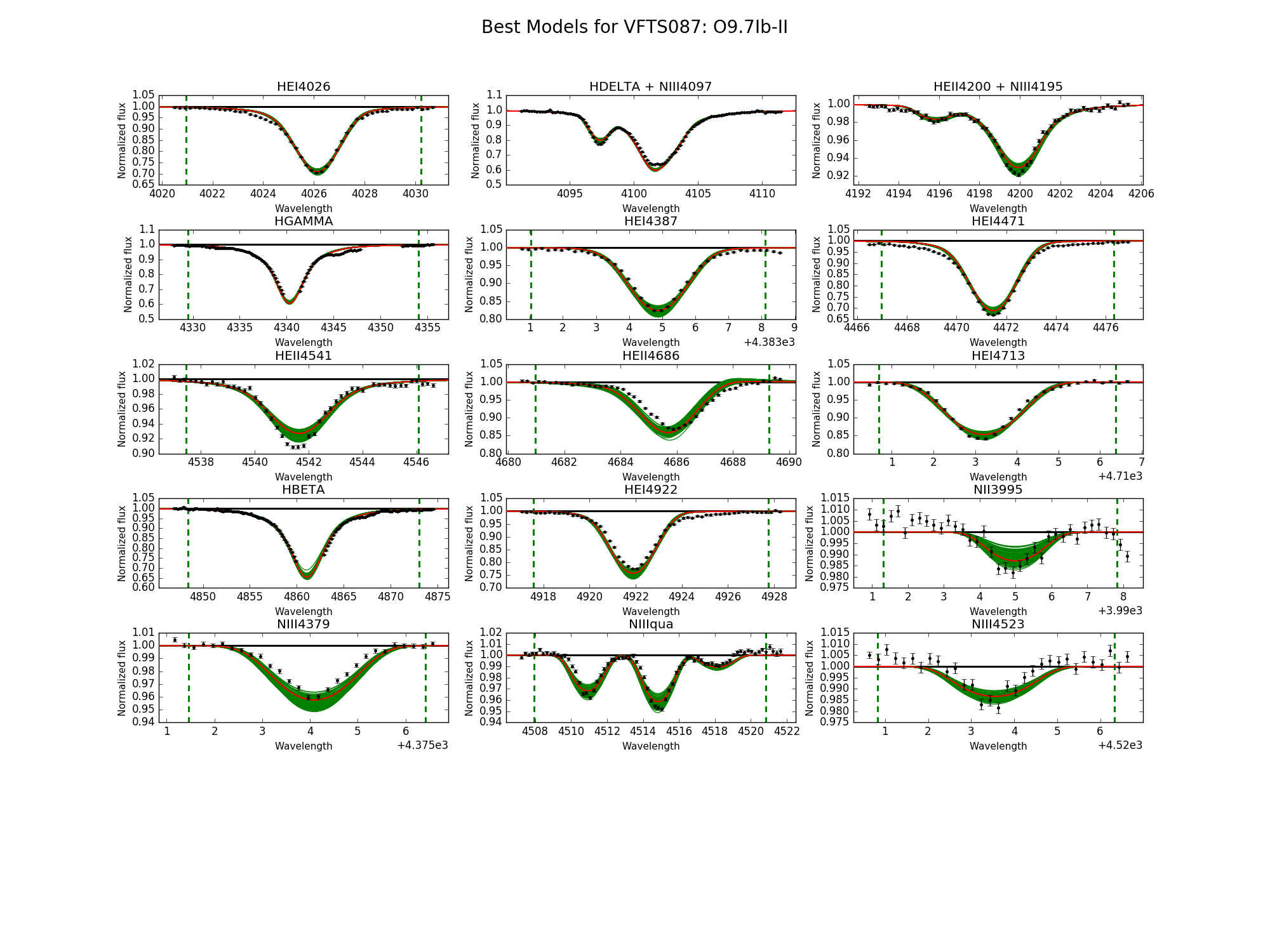}
\includegraphics[scale=0.45,trim={5cm 6.5cm 1cm 0cm}]{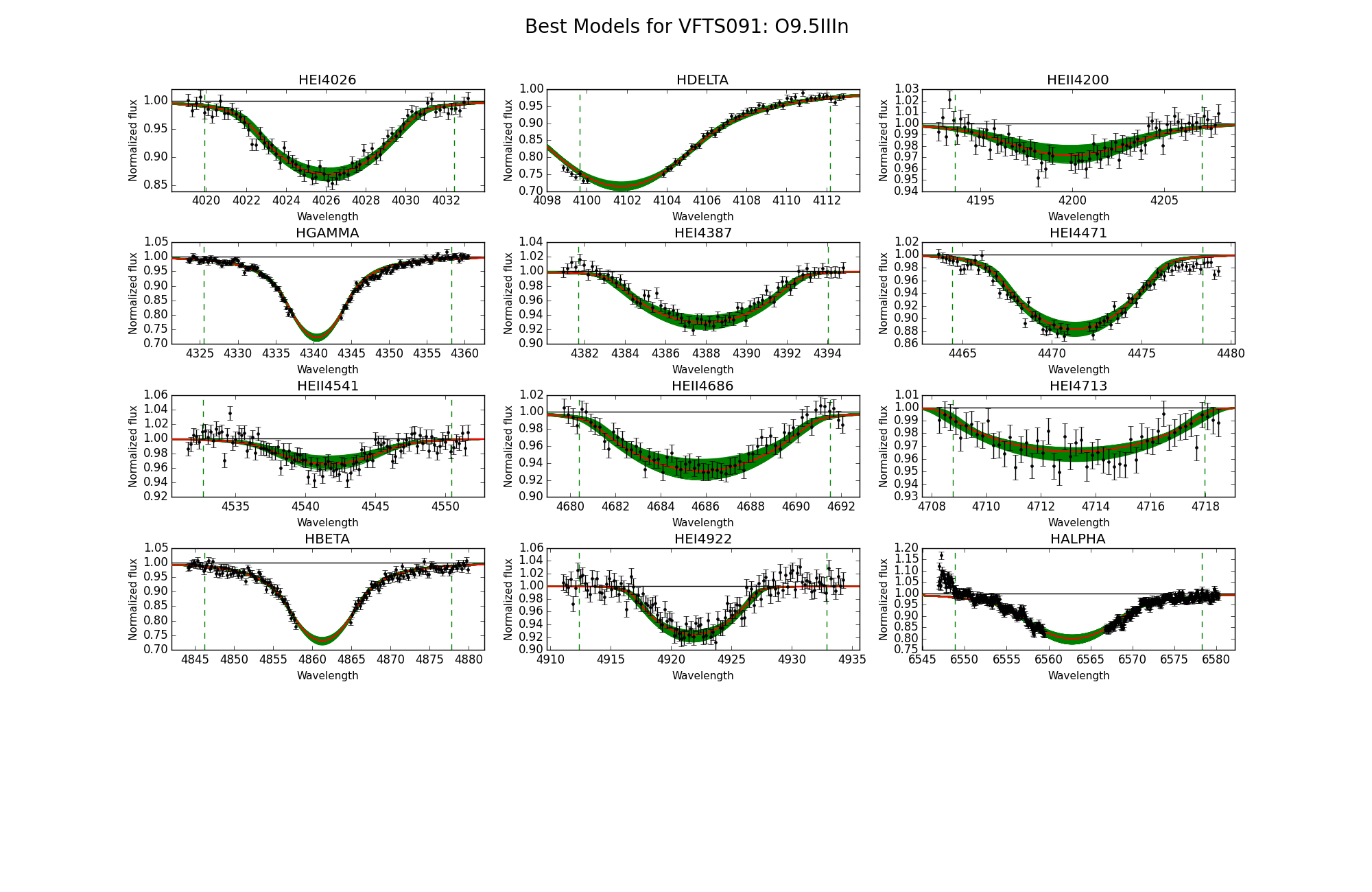}
\caption{Observed spectra, the 95\%\ probability models (green) and the best fit model (red) for VFTS\,087 and 091. The vertical dashed lines indicate 
the wavelength range used to fit the corresponding diagnostic line. [Color version available online.]}
\end{figure}

\begin{figure}
\includegraphics[scale=0.45,trim={5cm 6.5cm 1cm 0cm}]{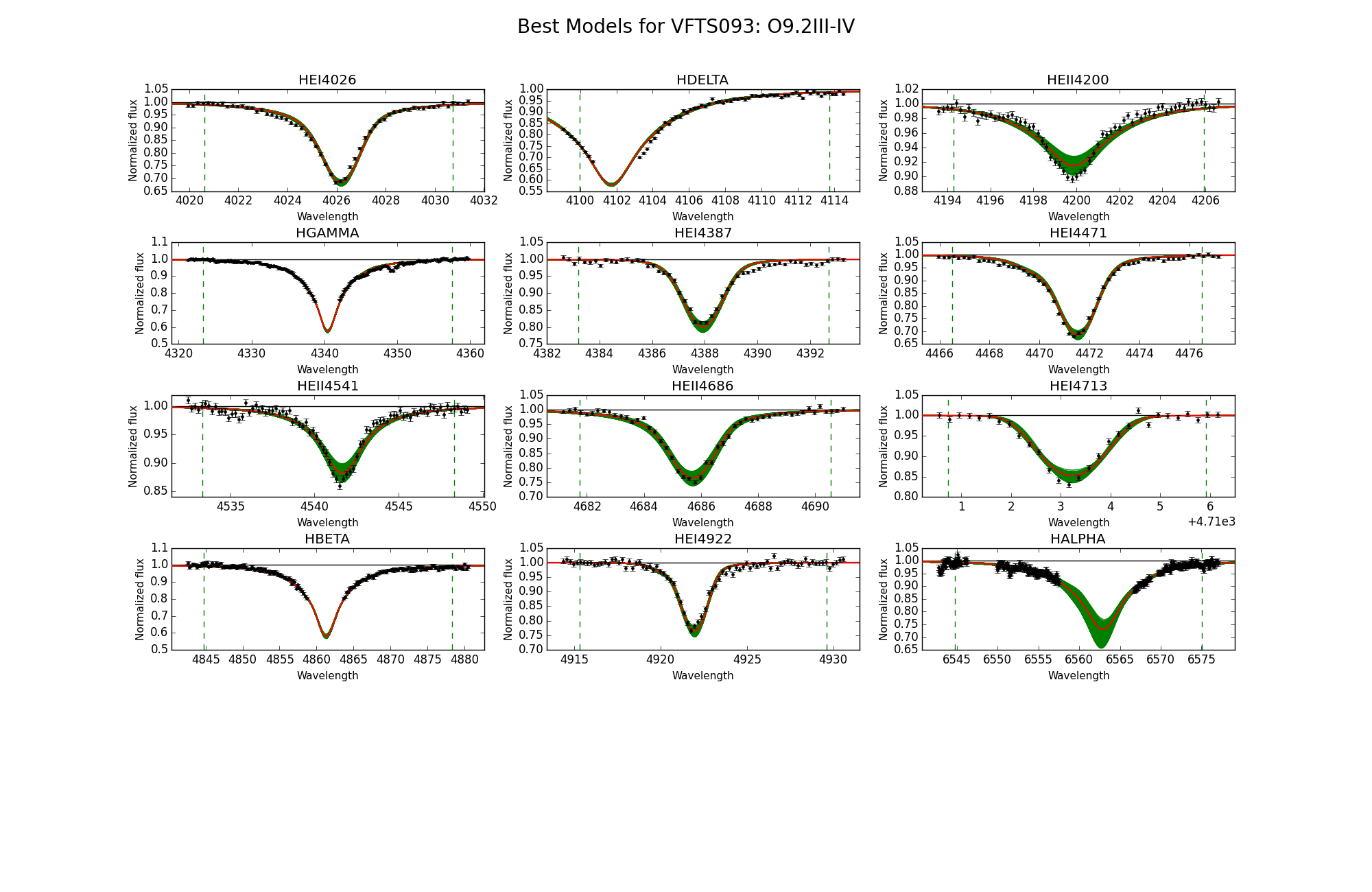}
\includegraphics[scale=0.45,trim={5cm 6.5cm 1cm 0cm}]{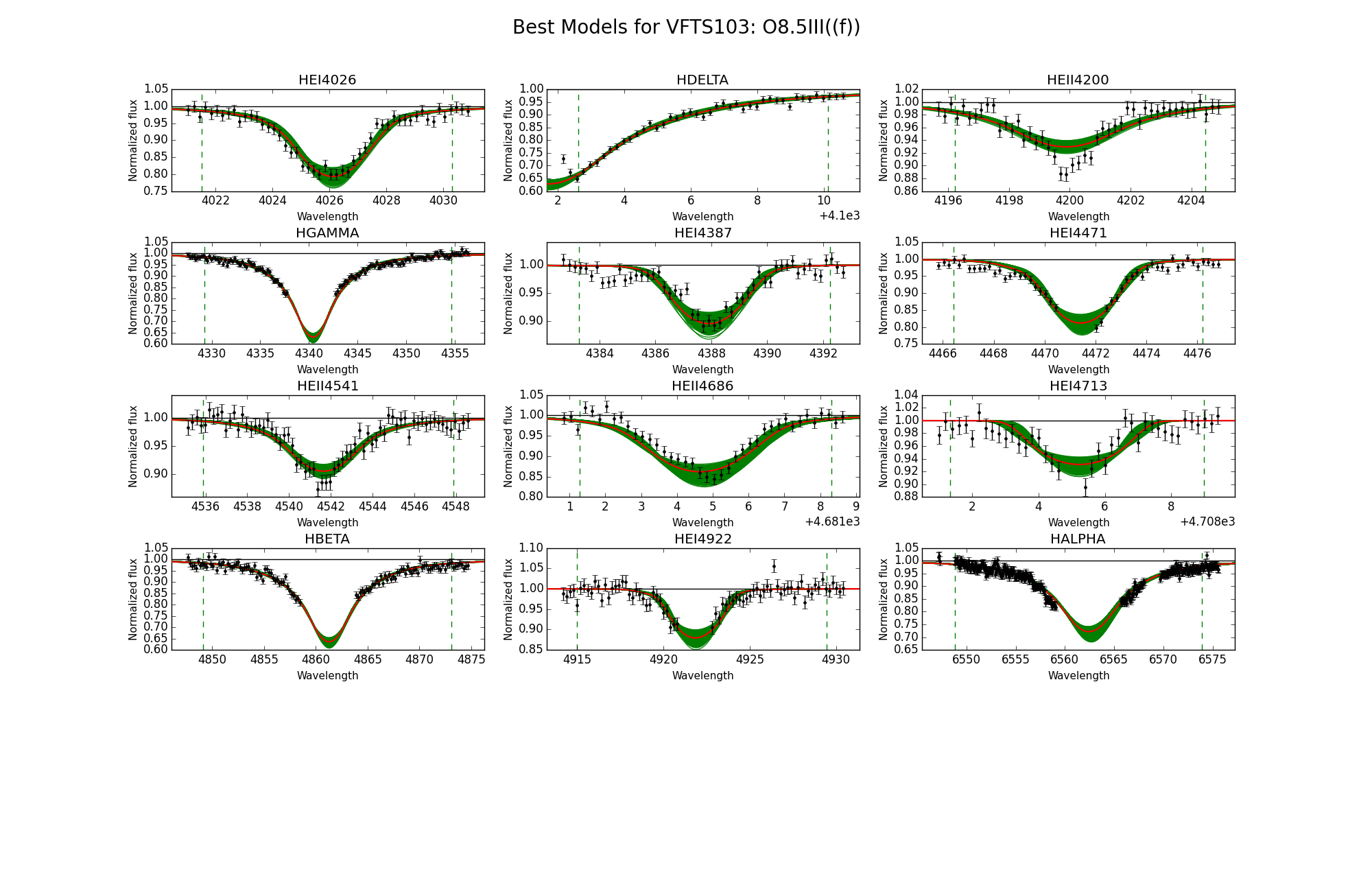}
\caption{Observed spectra, the 95\%\ probability models (green) and the best fit model (red) for VFTS\,093 and 103. The vertical dashed lines indicate 
the wavelength range used to fit the corresponding diagnostic line. [Color version available online.]}
\end{figure}

\begin{figure}
\includegraphics[scale=0.45,trim={5cm 6.5cm 1cm 0cm}]{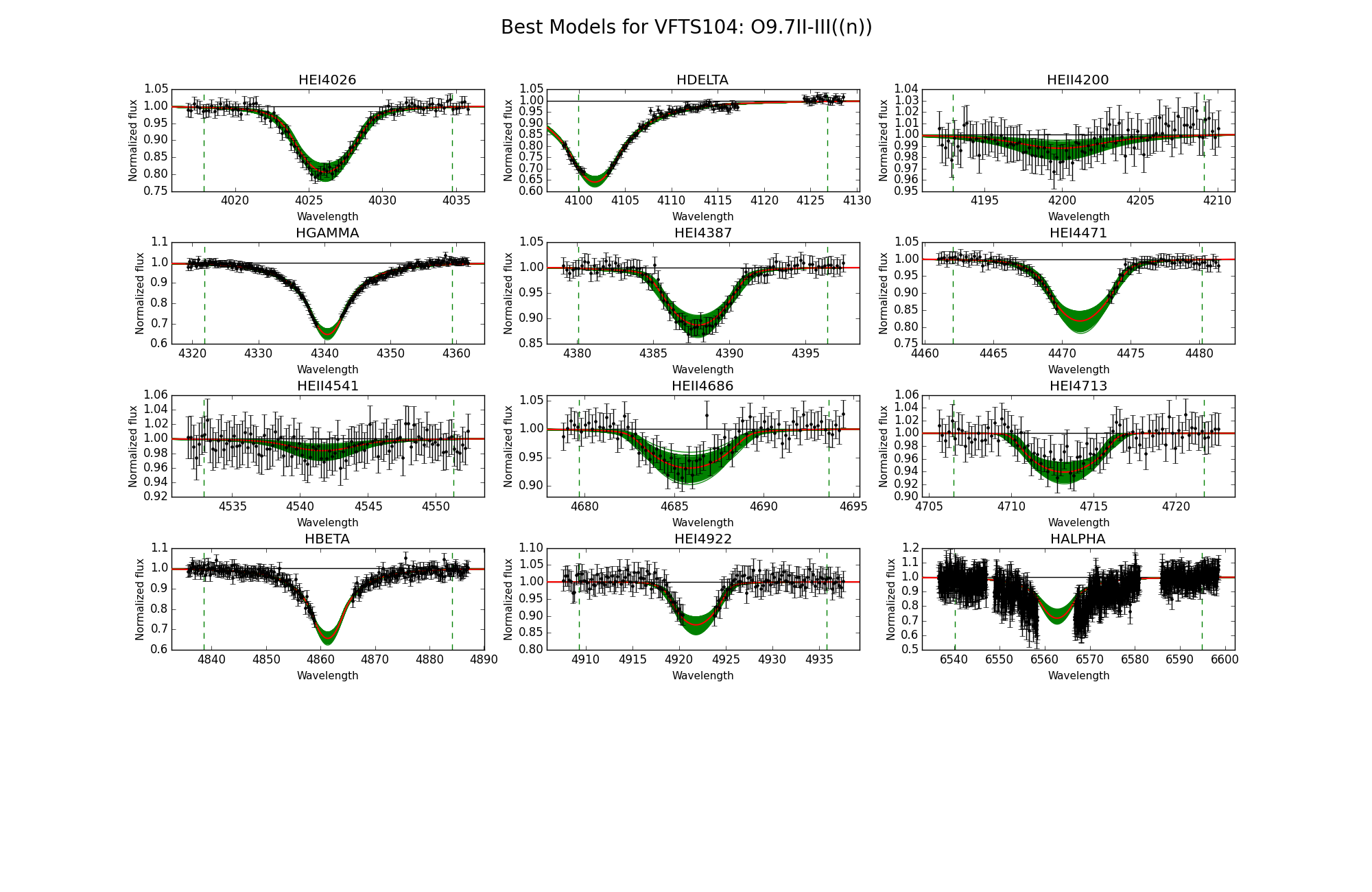}
\includegraphics[scale=0.45,trim={5cm 6.5cm 1cm 0cm}]{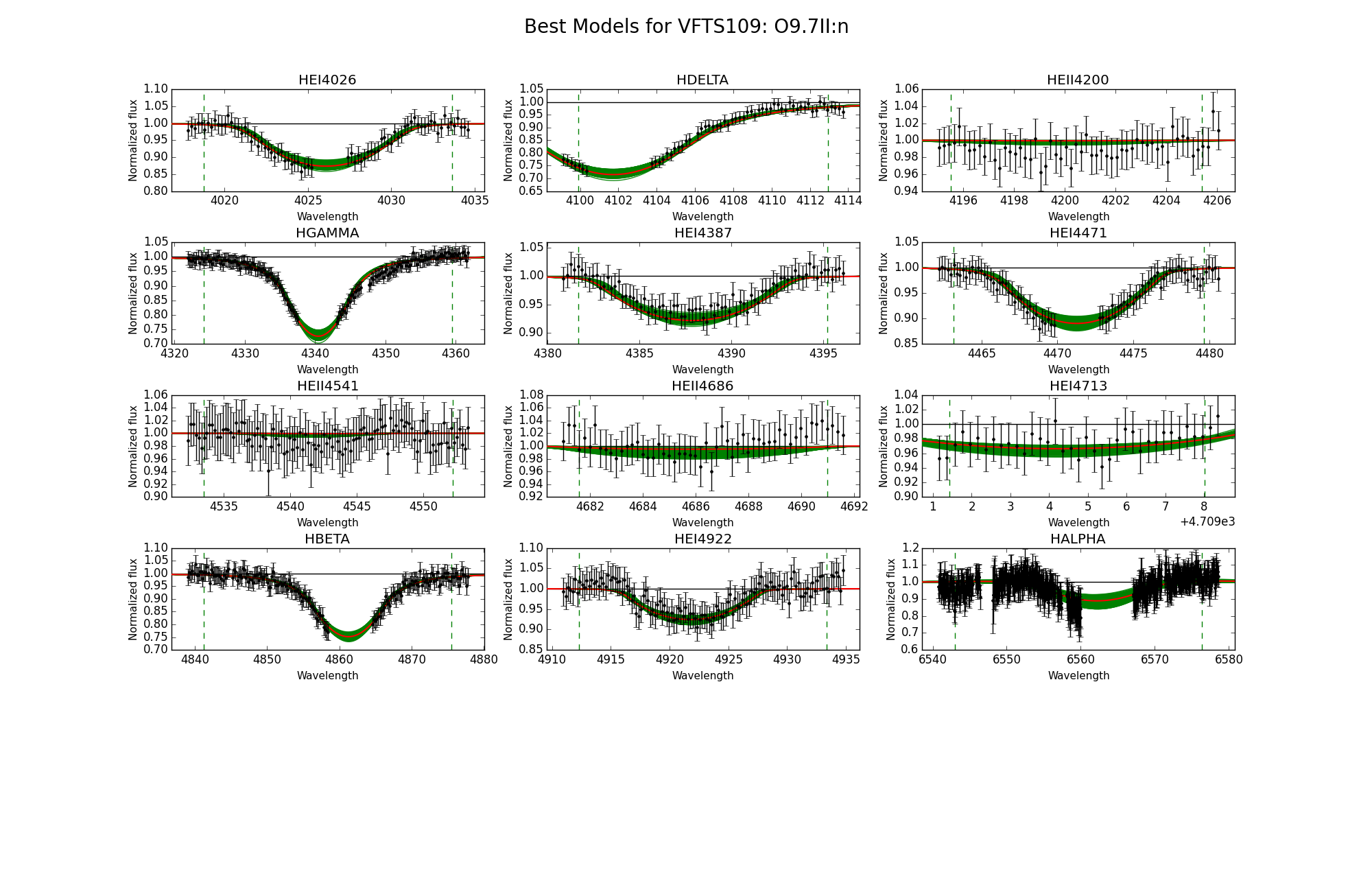}
\caption{Observed spectra, the 95\%\ probability models (green) and the best fit model (red) for VFTS\,104 and 109. The vertical dashed lines indicate 
the wavelength range used to fit the corresponding diagnostic line. [Color version available online.]}
\end{figure}

\begin{figure}
\includegraphics[scale=0.42,trim={5cm 1.5cm 1cm 0cm}]{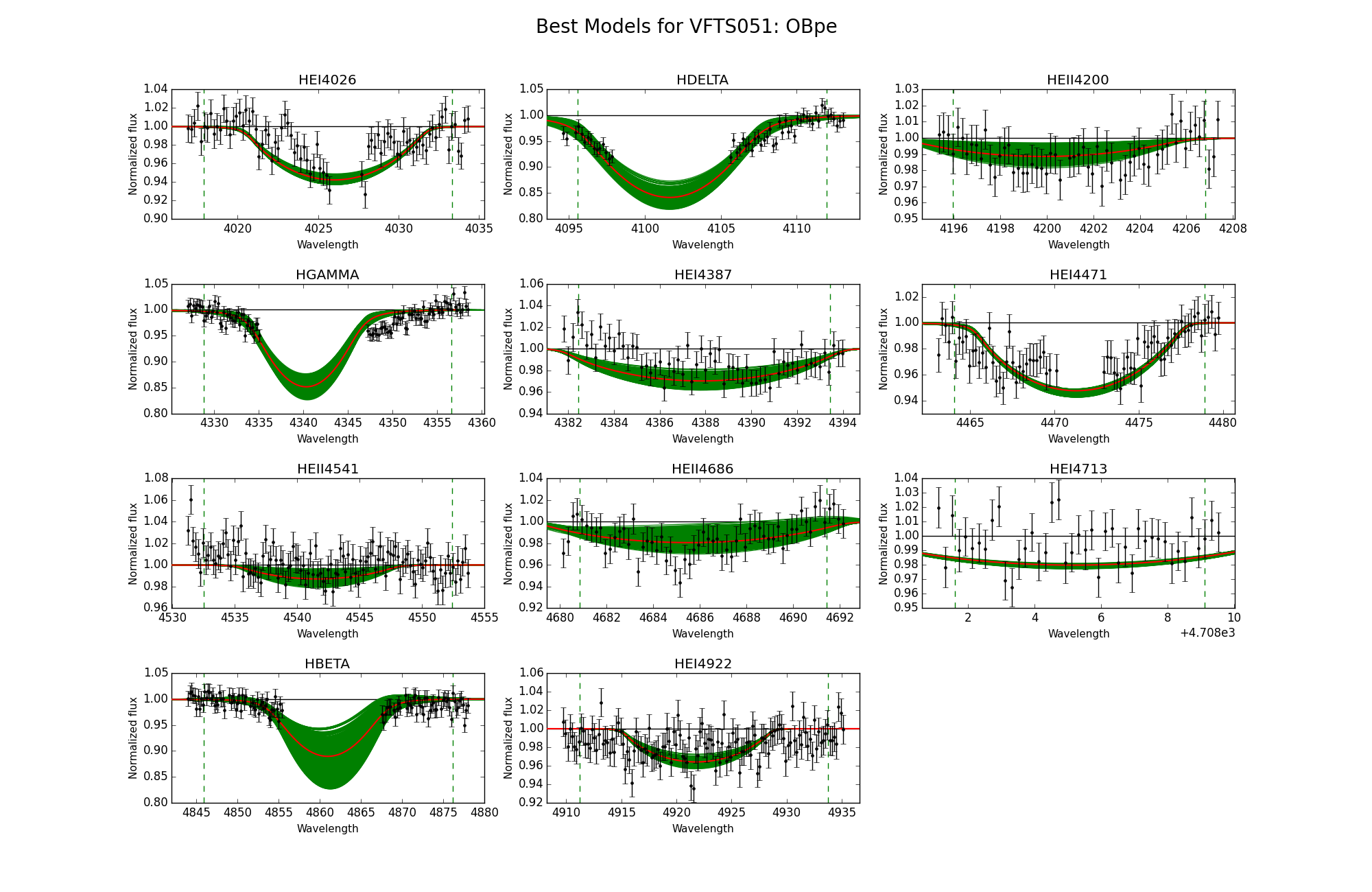}
\includegraphics[scale=0.42,trim={5cm 1.5cm 1cm 0cm}]{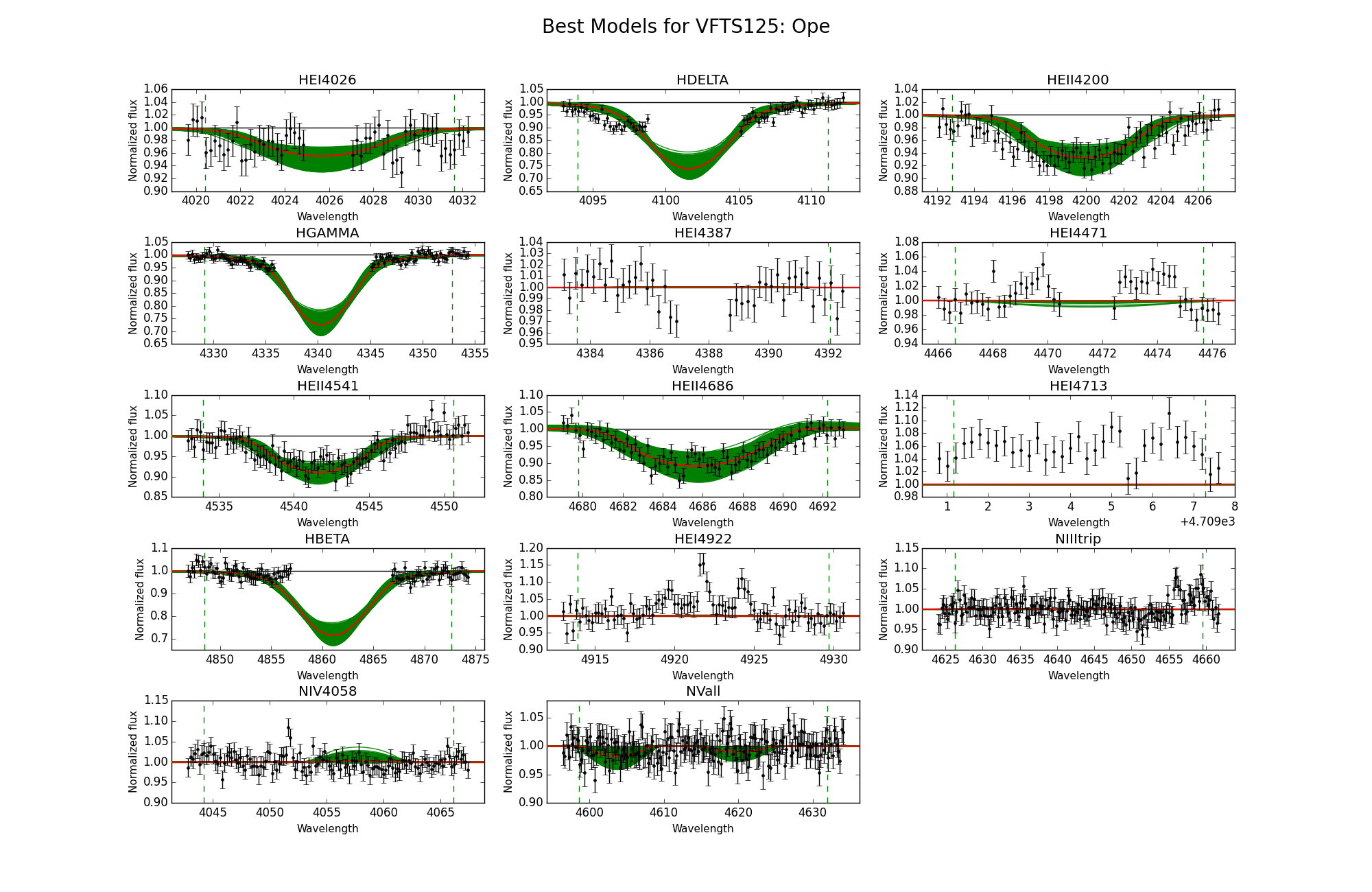}
\caption{Observed spectra, the 95\%\ probability models (green) and the best fit model (red) for VFTS\,051 and 125. [Color version available online.]}
\end{figure}

\begin{figure}
\includegraphics[scale=0.45,trim={5cm 6.5cm 1cm 0cm}]{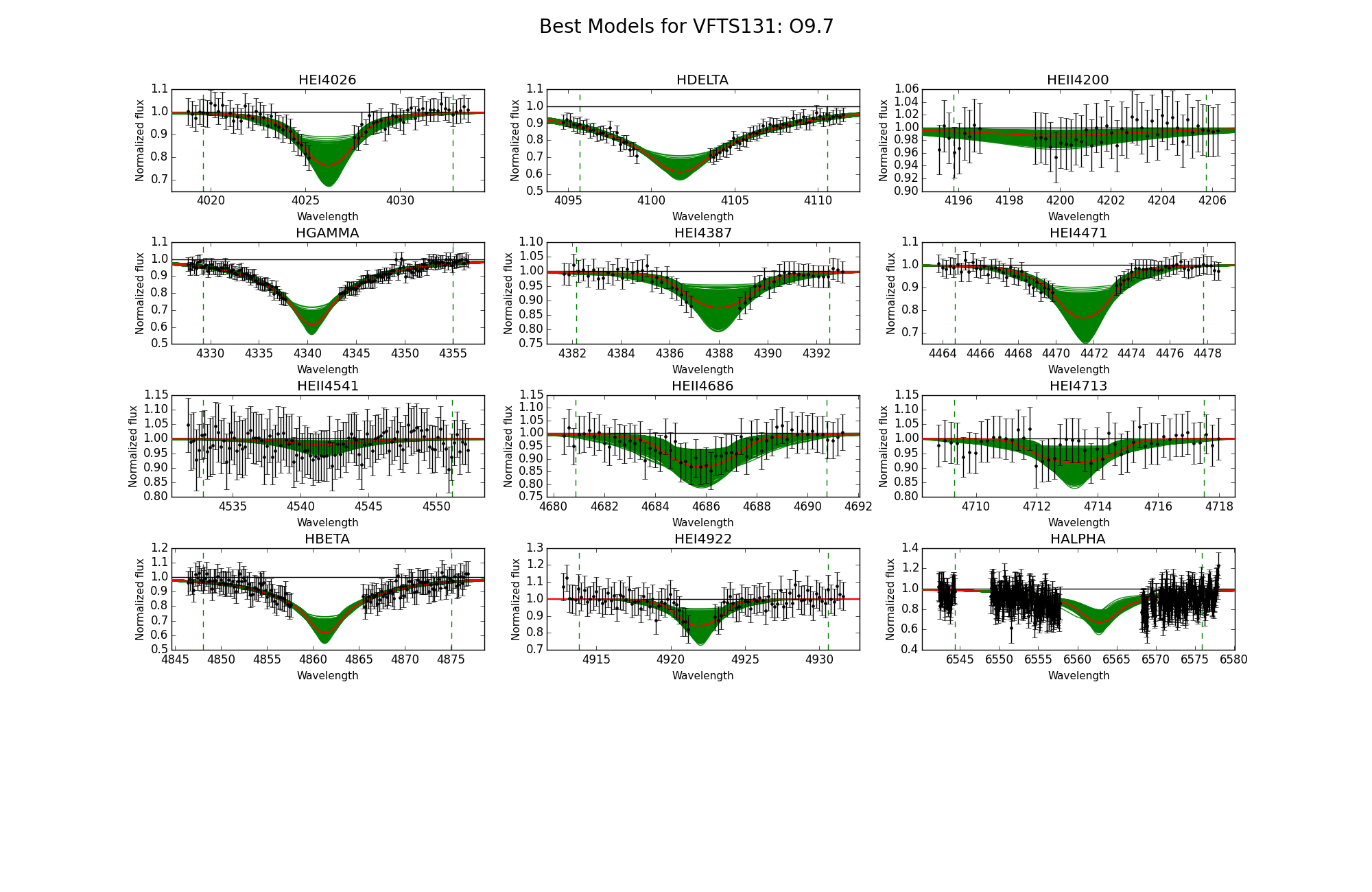}
\includegraphics[scale=0.45,trim={5cm 1.5cm 1cm 0cm}]{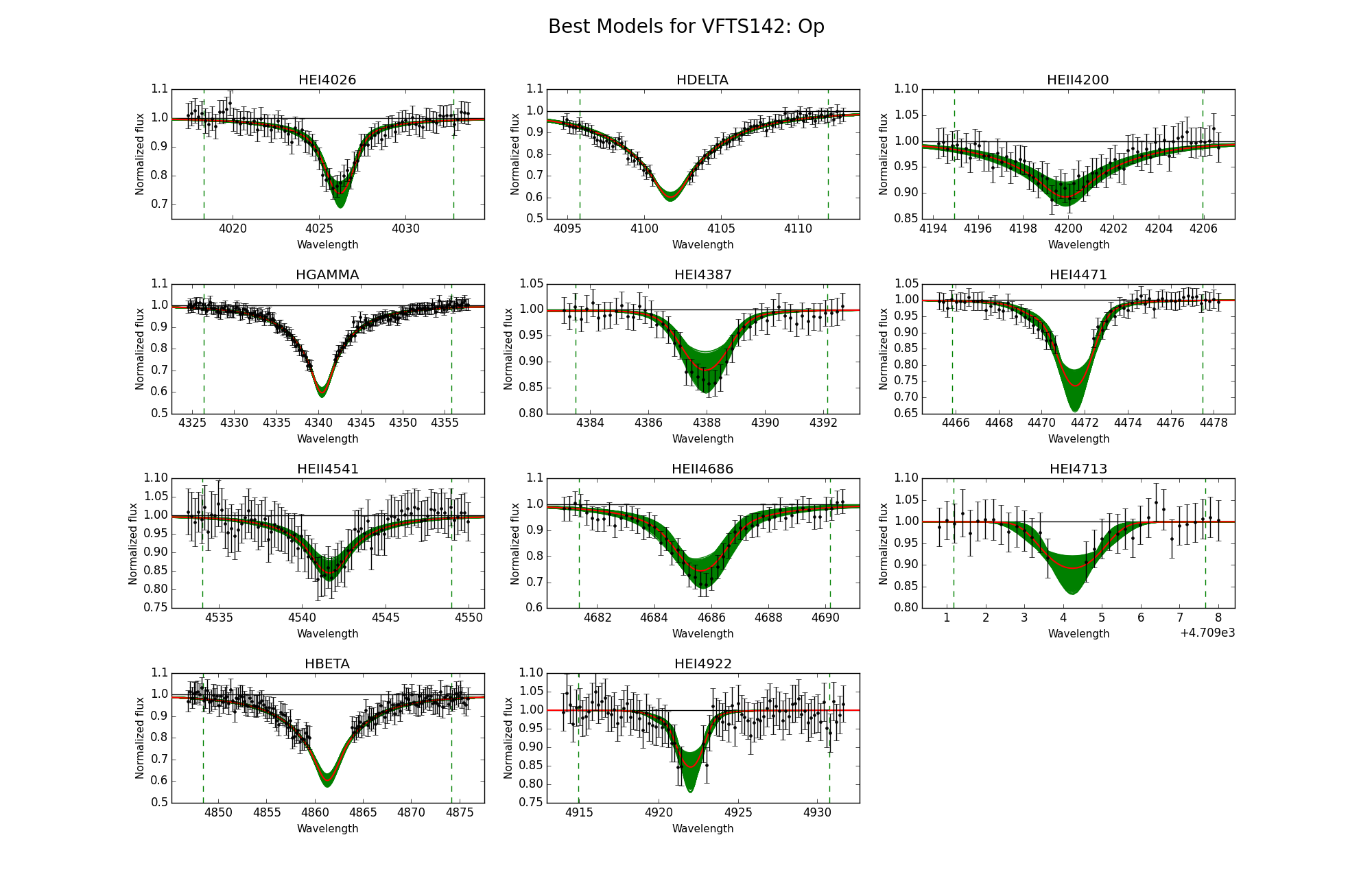}
\caption{Observed spectra, the 95\%\ probability models (green) and the best fit model (red) for VFTS\,131 and 142. The vertical dashed lines indicate 
the wavelength range used to fit the corresponding diagnostic line. [Color version available online.]}
\end{figure}

\begin{figure}
\includegraphics[scale=0.42,trim={5cm 1.5cm 1cm 0cm}]{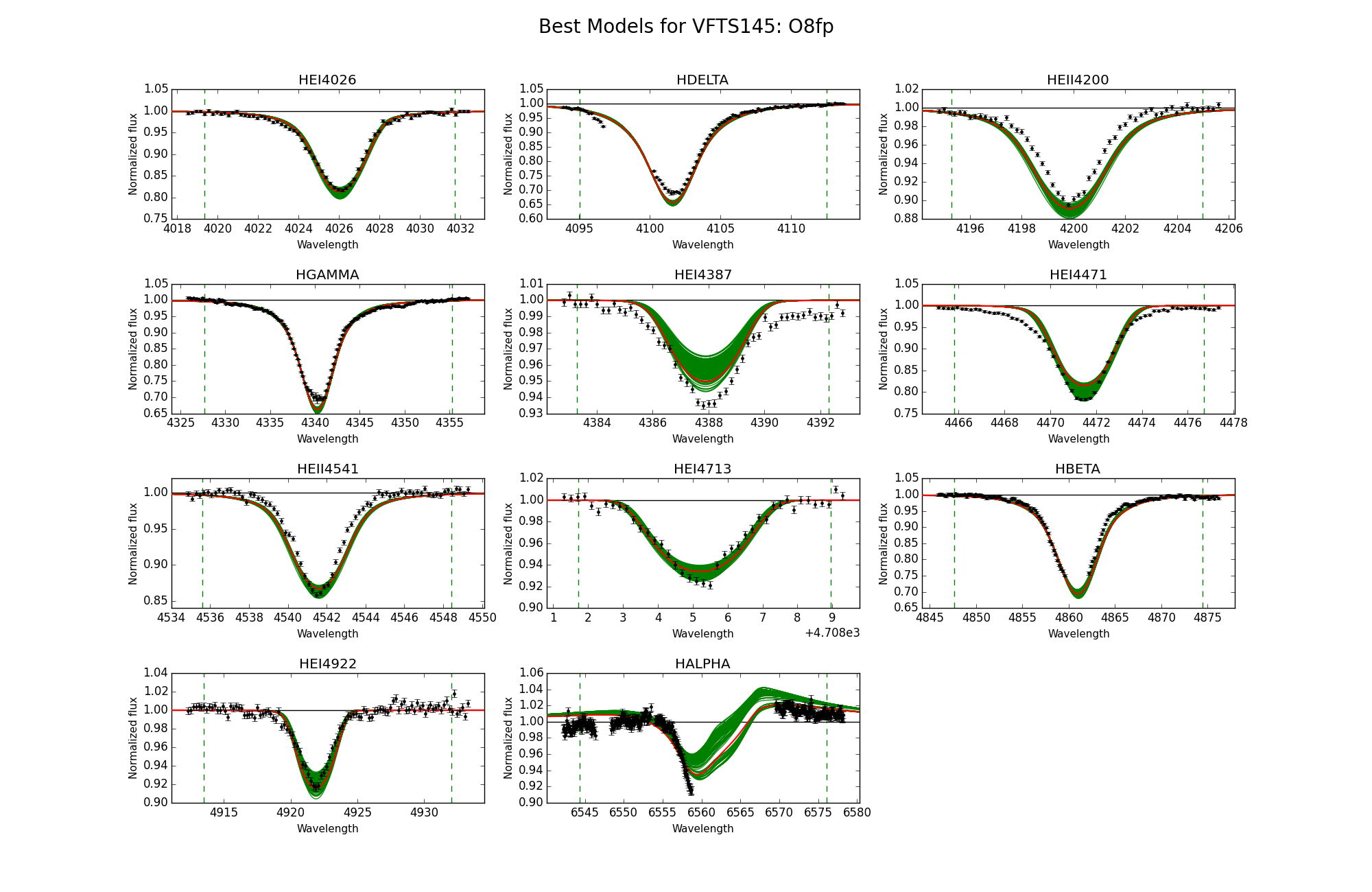}
\includegraphics[scale=0.42,trim={5cm 1.5cm 1cm 0cm}]{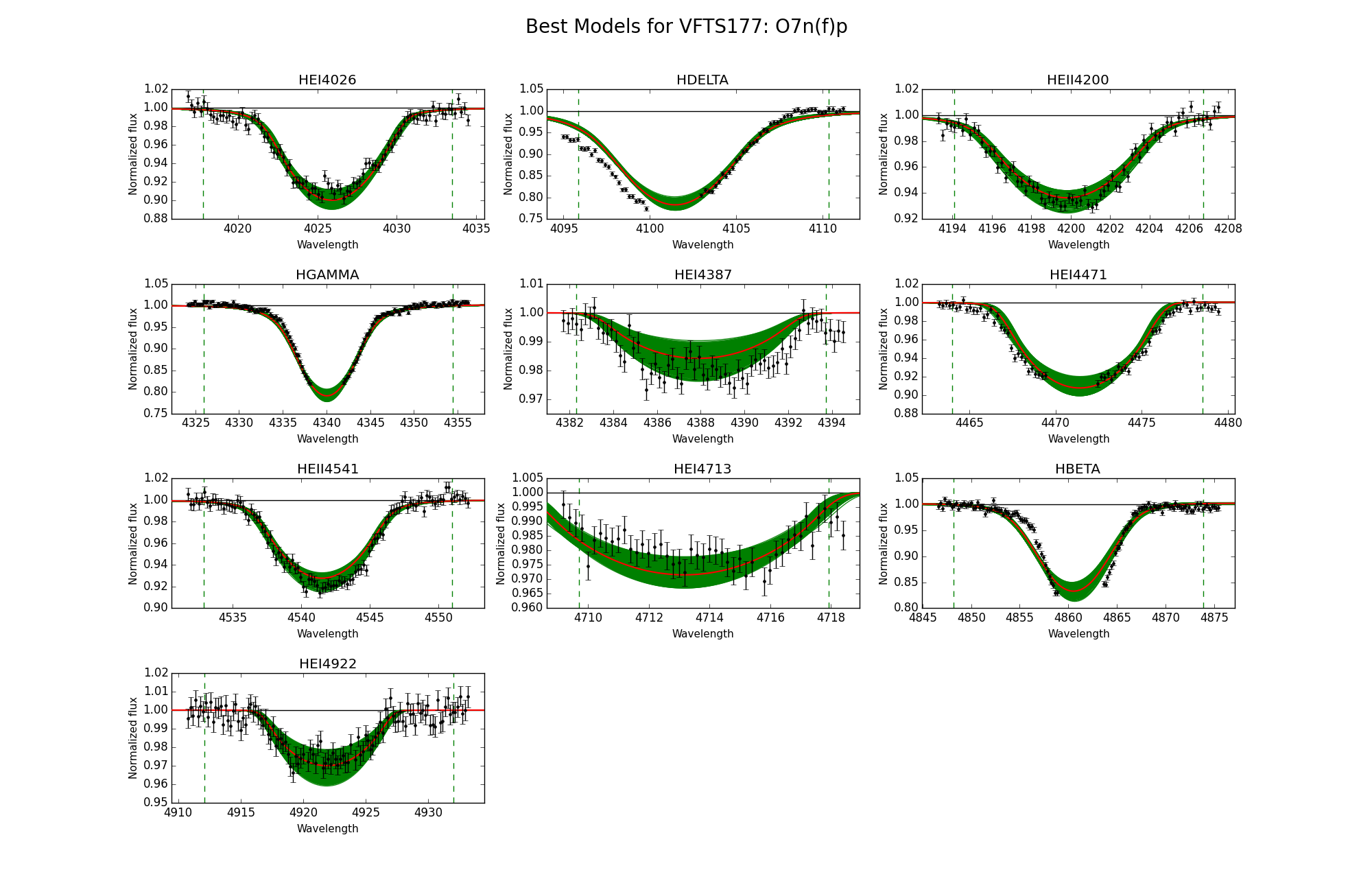}
\caption{Observed spectra, the 95\%\ probability models (green) and the best fit model (red) for VFTS\,145 and 177. The vertical dashed lines indicate 
the wavelength range used to fit the corresponding diagnostic line. [Color version available online.]}
\end{figure}

\end{appendix}

\end{document}